\documentclass[useAMS, usegraphicx, usenatbib]{mn2e}
\usepackage{aas_macros}
\usepackage{amsmath, amssymb}
\usepackage{graphicx}
\usepackage{natbib}
\usepackage[british]{babel}
\newcommand{\kunit}{\,h\,\mathrm{Mpc}^{-1}}
\newcommand{\runit}{\,h^{-1}\,\mathrm{Mpc}}
\newcommand{\runitk}{\,h^{-1}\,\mathrm{kpc}}
\newcommand{\munit}{\,h^{-1}\,\mathrm{M}_{\sun}}
\newcommand{\munitnoh}{\,\mathrm{M}_{\sun}}
\newcommand{\vunit}{\,\mathrm{km}\,\mathrm{s}^{-1}}
\newcommand{\Tunit}{\,\mathrm{K}}
\newcommand{\nunit}{\,\mathrm{eV}}
\newcommand{\bahamas}{\textsc{bahamas}}
\setcitestyle{authoryear,round,comma,aysep={},yysep={,},notesep={}}
\title[Linking the halo baryon fraction with $P(k)$]{Exploring the effects of galaxy formation on matter clustering through a library of simulation power spectra}
\author[M. P. van Daalen, I. G. McCarthy and J. Schaye]{Marcel P. van Daalen$^{1,2}$\thanks{E-mail: daalen@strw.leidenuniv.nl}, Ian G. McCarthy$^{3}$ and Joop Schaye$^{1}$\\\\
$^1$Leiden Observatory, Leiden University, P.O. Box 9513, 2300 RA Leiden, The Netherlands\\
$^2$Institute for Astronomy, University of Edinburgh, Royal Observatory, Blackford Hill, Edinburgh, EH9 3HJ, UK\\
$^3$Astrophysics Research Institute, Liverpool John Moores University, 146 Brownlow Hill, Liverpool L3 5RF, UK}
\begin{document}
\pagerange{\pageref{firstpage}--\pageref{lastpage}} \pubyear{2019}
\maketitle
\label{firstpage}
\begin{abstract}
Upcoming weak lensing surveys require a detailed theoretical understanding of the matter power spectrum in order to derive accurate and precise cosmological parameter values. While galaxy formation is known to play an important role, its precise effects are currently unknown. We present a set of $92$ matter power spectra from the OWLS, cosmo-OWLS and \bahamas{} simulation suites, including different $\Lambda$CDM cosmologies, neutrino masses, subgrid prescriptions and AGN feedback strengths. We conduct a detailed investigation of the dependence of the relative difference between the total matter power spectra in hydrodynamical and collisionless simulations on the effectiveness of stellar and AGN feedback, cosmology and redshift. The strength of AGN feedback can greatly affect the power on a range of scales, while a lack of stellar feedback can greatly increase the effectiveness of AGN feedback on large scales. We also examine differences in the initial conditions of hydrodynamic and N-body simulations that can lead to a $\sim 1\%$ discrepancy in the large-scale power, and furthermore show our results to be insensitive to cosmic variance. We present an empirical model capable of predicting the effect of galaxy formation on the matter power spectrum at $z=0$ to within $1\%$ for $k<1\kunit$, given only the mean baryon fraction in galaxy groups. Differences in group baryon fractions can also explain the quantitative disagreement between predictions from the literature. All total and dark matter only power spectra in this library will be made publicly available at \texttt{powerlib.strw.leidenuniv.nl}.
\end{abstract}
\begin{keywords}
cosmology: theory, large-scale structure of Universe -- galaxies: formation -- gravitational lensing: weak, surveys
\end{keywords}

\section{Introduction}
\label{sec:introduction}
Current and near-future weak lensing surveys like DES\footnote{\texttt{darkenergysurvey.org}}, LSST\footnote{\texttt{lsst.org}}, Euclid\footnote{\texttt{euclid-ec.org}} and WFIRST\footnote{\texttt{wfirst.gsfc.nasa.gov}} face a significant challenge when attempting to interpret their measurements: they require predictions of the matter power spectrum with a precision better than $1\%$ \citep{HutererTakada2005,Ivezic2008,Laureijs2009}. Presently, making predictions at this level down to sufficiently small scales is challenging even in a dark matter only Universe \citep[e.g.][]{Schneider2016} -- but unfortunately, the presence of baryons causes large additional complications. As first shown in \citet[][, hereafter VD11]{vanDaalen2011}, based on the OWLS suite of simulations \citep{Schaye2010}, stellar feedback and feedback from active galactic nuclei (AGN) in particular has a strong effect on the power out to relatively large scales, reducing the power by $1\%$ at a Fourier scale of $k=0.3\kunit$ to $28\%$ at $k=10\kunit$, relative to a dark matter only universe. This large-scale suppression in power primarily comes about by feedback heating and ejecting gas out to large distances, which is required in order to match X-ray observations of groups and clusters \citep[e.g.][]{McCarthy2010,McCarthy2011,LeBrun2014}. Secondary to this is the resulting change in the clustering of cold dark matter itself, dubbed the back-reaction. Follow-up work has shown that galaxy formation, when ignored, may result in biases of cosmological parameters that exceed the statistical errors of upcoming surveys by an order of magnitude \citep[e.g.][]{Semboloni2011,Zentner2013}.

The publicly available power spectra from VD11 only included one model with AGN feedback. Many authors have used the VD11 OWLS AGN power spectra and others to inform their models and test whether the effects of galaxy formation can be marginalized over \citep[e.g.][]{Semboloni2013,Zentner2013,MohammedSeljak2014,Harnois-Deraps2015,Eifler2015,SchneiderTeyssier2015,Foreman2016,MohammedGnedin2018,Huang2019,SchneiderTeyssier2019}. While other authors besides VD11 also find that AGN feedback has a significant effect on the power spectrum \citep[e.g.][]{Vogelsberger2014a,Hellwing2016,Peters2018,Springel2018,Chisari2018}, there is no consensus at anywhere near the $1\%$ level required -- which, itself, is an issue worth addressing.

The lack of (publicly available) simulations that test the effects of galaxy formation including AGN feedback (and, ideally, simultaneously cosmology) means that the simulated power spectra used in the literature do not fully reflect the theoretical uncertainties in the field of galaxy formation that still exist today. In addition, \citet{MohammedGnedin2018} have shown that methods aiming to mitigate the effects of baryons on weak lensing observables benefit from including models that are more extreme than is realistic in their training sets. Currently, the most realistic simulations, i.e.\ those including AGN, are often the most extreme as well, which is undesirable from a modelling perspective. The work of \citet{Huang2019} supports these findings, showing that adding more simulations with AGN feedback to the training set of a mitigating scheme allows for a stronger reduction of the bias in cosmological parameters.

The parametrization of the effects of galaxy formation on the matter power spectrum as formulated by \citet{Mead2015,Mead2016} is particularly widely used in clustering observations to marginalize over the effects of baryons, either directly or through its implementation in the model of \citet{Joudaki2017} \citep[e.g.][]{Hildebrandt2017,Copeland2018,vanUitert2018,Planck2018,Yoon2019}. Importantly, this model was calibrated solely to power spectra presented in VD11. Therefore, if the modifications to the dark matter only power spectrum are sufficiently different from those considered by VD11, they may not be captured by this model -- or not within the parameter range probed -- which may impact the interpretation of the observed data.

In this work, we take a step towards remedying some of these problems by presenting a large library of power spectra from OWLS, cosmo-OWLS \citep{LeBrun2014} and \bahamas{} \citep{McCarthy2017}, the latter containing -- for the first time -- AGN feedback that was calibrated to observations, with different cosmologies and neutrino masses. All power spectra presented here will become publicly available with the publication of this paper. While the underlying model in simulations with AGN is the same in each of the simulations presented here, many variations with different feedback strengths and/or scalings are explored, including some that go beyond what is expected to be realistic in order to allow sufficient flexibility for emulators and marginalization schemes. Using these power spectra, we attempt to deepen our understanding of how feedback influences the clustering of matter, and how this depends on some of the choices made when running the simulations. Most importantly, using these simulations we are able to present a model which is able to highly accurately predict the suppression of power due to galaxy formation at $z=0$ for $k<1\kunit$, as a function of only the baryon fraction at the galaxy group scale.

The simulations and methods used to calculate the power spectra are described in \S\ref{sec:simspow}, with Table~\ref{tab:sims} showing a list of all simulations with power spectra. In \S\ref{sec:results} we present a selection of these power spectra and investigate the effect of e.g.\ feedback strength, neutrino mass, redshift, cosmology and cosmic variance on the total matter and cold dark matter power spectrum. We also compare to power spectra from simulations including AGN from the literature and consider the reasons for the quantitative differences in the effects of galaxy formation on clustering found. At the end of \S\ref{sec:results}, we present and discuss our model for the large-scale suppression of power based on the baryon fraction of groups. Finally, we summarize and discuss our findings in \S\ref{sec:discussion}.

\section{Simulations and power spectra}
\label{sec:simspow}
\subsection{Simulation sets}
\label{subsec:sims}
In this work we present power spectra for three related sets of cosmological, hydrodynamical simulations: OWLS \citep{Schaye2010}, cosmo-OWLS \citep{LeBrun2014} and \bahamas{} \citep{McCarthy2017,McCarthy2018}. Since many of the OWLS power spectra were already presented in VD11, we focus on the latter two sets here. For some of the simulations in this set, power spectra were independently calculated and considered in \citet{Mummery2017}.

Cosmo-OWLS is an extension of OWLS that is aimed at studying the properties of groups and clusters, and to this end it includes simulations with larger boxes compared to OWLS ($200$ and $400\,\runit$ on a side versus at most $100\runit$ for OWLS), as well as variations in the strength of AGN feedback for those simulations that include it. \bahamas{} (BAryons and HAloes of MAssive Systems) is in turn a follow-up to cosmo-OWLS and even better suited for cosmological tests, as it includes more accurate initial conditions (using the Boltzmann code \textsc{camb} in combination with a modified version of N-GenIC that uses 2LPT), massive neutrinos \citep[following][]{Ali-HaimoudBird2013}, AGN in all hydrodynamical simulations, and more recent cosmologies. In addition, \bahamas{} is the first of these sets to calibrate the subgrid parameters for feedback from supernovae and active galactic nuclei to the observed present-day galaxy stellar mass function (SMF) and the hot gas mass fractions of groups and clusters, placing them among the most realistic cosmological simulations yet. In particular, the \bahamas{} simulations provide an excellent match to the galaxy SMF for all $M_*>10^{10}\munit$ \citep[see][]{McCarthy2017}. Relative to the standard OWLS and cosmo-OWLS AGN simulations, the subgrid physics prescriptions are unmodified, but the parameters are not. The subgrid parameters that were changed in \bahamas{} were the supernova-driven galactic wind velocity (from $600\vunit$ to $300\vunit$), the number of particles heated by each AGN event (from $1$ to $20$) and the AGN minimum heating temperature (from $\geq 10^8\Tunit$ to $10^{7.8}\Tunit$). The weaker stellar feedback is necessary to find agreement with observations at low stellar masses, as both OWLS AGN and cosmo-OWLS formed too few galaxies below $M_*=10^{11}\munit$ \citep{LeBrun2014,McCarthy2017}. Because of this change, the baryon fraction in stars goes up and the fraction in hot gas goes down, but the cold gas available for accretion by the supermassive black holes also increases. The lower minimum AGN heating temperature compensates for these shifts and brings the X-ray gas fractions in massive haloes back in agreement with observations. As a consequence of these adjustments, the relative role of strong feedback in \bahamas{} is somewhat smaller than it was in the previous models, and as we show in \S\ref{sec:results} this affects the power spectrum as well.

The most realistic simulation in the VD11 set of power spectra -- that is, the simulation in simultaneous agreement with the most observables -- is the WMAP7 OWLS AGN model with \textsc{camb} initial conditions, while the most realistic simulation in the current set is the \bahamas{} simulation with the low but non-zero total neutrino mass of $\sum m_\nu=0.06\nunit$ and an up-to-date Planck 2015 cosmology. We will therefore often use one of these simulations as a baseline for comparison -- though we stress again that no currently available simulation is expected to match the real Universe at the $1\%$ level. We will prefix the name of each hydrodynamical simulation with the set it belongs to; the former model we will dub OWLS\_AGN\_WMAP7\_CAMB (named AGN\_WMAP7 in VD11) and the latter BAHAMAS\_nu0.06\_Planck2015. All \bahamas{} simulations contain AGN, and therefore do not explicitly contain ``AGN'' in their name. Dark matter only simulations are given suffixes indicating box size and resolution, e.g.\ L100N512 for simulations with boxes $100\runit$ on a side and $512^3$ particles per type (fiducial values for OWLS), and L400N1024 for $400\runit$ boxes with $1024^3$ particles per type (fiducial values for cosmo-OWLS and \bahamas{}). To keep the names of the hydrodynamical simulations relatively short, their box size and resolution are only included when they differ from the fiducial values of the set. While the simulations from OWLS have an $8\times$ higher mass resolution, the \bahamas{} simulations are calibrated to observations and probe $64\times$ larger volumes. 

For a few of the models power spectra are available for different mass resolutions and/or box sizes. We investigate the effects these have on the matter clustering in detail in Appendix~\ref{app:restests}. The main conclusions are that the limited resolution of the simulations ($\sim 10^9\munit$ and $4\runitk$ at $z=0$ for the $400\runit$ boxes) mainly plays a role for $k\gtrsim 10\kunit$, and that calibrating simulations to observables at a fixed resolution is of greater importance than increasing said resolution.

A list of all simulations that we provide power spectra for can be found in Table~\ref{tab:sims}. The cosmological parameters corresponding to the different cosmologies of these simulations are listed in Table~\ref{tab:cosms}. The cosmologies probed here are based on the WMAP3 \citep{Spergel2007}, WMAP5 \citep{Komatsu2009}, WMAP7 \citep{Komatsu2011}, WMAP9 \citep{Hinshaw2013}, Planck 2013 \citep{Planck2014} and Planck 2015 \citep{Planck2016} data, with an additional cosmology (``BAO'') taking cosmological parameter values roughly in between those of WMAP and Planck. Note that the Planck 2015 cosmological parameters depend on the neutrino mass in such a way as to preserve the fit to CMB data, while for WMAP9 the density of CDM, $\Omega_\mathrm{c}$, was reduced with increasing neutrino mass so as to preserve the total matter density $\Omega_\mathrm{m}$; see \citet{McCarthy2018} for more information. Models for which power spectra were included in the VD11 release are marked with an asterisk. Below, we briefly expand on a few of the new physical models.

\begin{table*}
\caption{A list of the simulations that we provide total matter power spectra for, along with their dark matter counterparts. All simulations have $z=0$ power spectra, and most have power spectra that cover $z\leq 3$. A brief explanation of models that were not considered in VD11 can be found in \S\ref{subsec:sims}. For more details on the different models we refer to the papers which introduced them: \citet{Schaye2010} for OWLS, \citet{LeBrun2014} for cosmo-OWLS and \citet{McCarthy2017} for \textsc{bahamas}. The total mass in different neutrino species, $M_\nu=\sum m_\nu$, is listed where applicable. Simulations with power spectra that were made publicly available by VD11 are marked with a star. Simulation names start with the set they belong to (``C-OWLS'' referring to cosmo-OWLS). The fiducial box size and particle number identifier for cosmo-OWLS and \textsc{bahamas} is L400N1024, that for OWLS is L100N512.}
\centering
\tiny
\begin{tabular}{l c c l l}
\hline
\!\!\!\!Simulation & Cosmology & $M_\nu \mathrm{[eV]}$ & DMO counterpart & Comments \\ [0.5ex]
\hline
\!\!\!\!\textit{BAHAMAS\_nu0.06\_Planck2015} & Planck '15 & 0.06 & \textit{DMONLY\_2fluid\_nu0.06\_Planck2015\_L400N1024} & AGN/SN feedback calibrated to obs.\!\!\!\! \\ [1ex]
\!\!\!\!\textit{BAHAMAS\_nu0.06\_WMAP9} & WMAP9 & 0.06 & \textit{DMONLY\_2fluid\_nu0.06\_WMAP9\_L400N1024} & AGN/SN feedback calibrated to obs.\!\!\!\! \\ [1ex]
\!\!\!\!\textit{BAHAMAS\_nu0.12\_Planck2015} & Planck '15 & 0.12 & \textit{DMONLY\_2fluid\_nu0.12\_Planck2015\_L400N1024} & AGN/SN feedback calibrated to obs.\!\!\!\! \\ [1ex]
\!\!\!\!\textit{BAHAMAS\_nu0.12\_WMAP9} & WMAP9 & 0.12 & \textit{DMONLY\_2fluid\_nu0.12\_WMAP9\_L400N1024} & AGN/SN feedback calibrated to obs.\!\!\!\! \\ [1ex]
\!\!\!\!\textit{BAHAMAS\_nu0.24\_Planck2015} & Planck '15 & 0.24 & \textit{DMONLY\_2fluid\_nu0.24\_Planck2015\_L400N1024} & AGN/SN feedback calibrated to obs.\!\!\!\! \\ [1ex]
\!\!\!\!\textit{BAHAMAS\_nu0.24\_WMAP9} & WMAP9 & 0.24 & \textit{DMONLY\_2fluid\_nu0.24\_WMAP9\_L400N1024} & AGN/SN feedback calibrated to obs.\!\!\!\! \\ [1ex]
\!\!\!\!\textit{BAHAMAS\_nu0.48\_Planck2015} & Planck '15 & 0.48 & \textit{DMONLY\_2fluid\_nu0.48\_Planck2015\_L400N1024} & AGN/SN feedback calibrated to obs.\!\!\!\! \\ [1ex]
\!\!\!\!\textit{BAHAMAS\_nu0.48\_WMAP9} & WMAP9 & 0.48 & \textit{DMONLY\_2fluid\_nu0.48\_WMAP9\_L400N1024} & AGN/SN feedback calibrated to obs.\!\!\!\! \\ [1ex]
\!\!\!\!\textit{BAHAMAS\_nu0\_BAO\_L200N512} & BAO & 0 & \textit{DMONLY\_nu0\_BAO\_L200N512} & AGN/SN feedback calibrated to obs.\!\!\!\! \\ [1ex]
\!\!\!\!\textit{BAHAMAS\_nu0\_Planck2013} & Planck '13 & 0 & \textit{DMONLY\_2fluid\_nu0\_Planck2013\_L400N1024} & AGN/SN feedback calibrated to obs.\!\!\!\! \\ [1ex]
\!\!\!\!\textit{BAHAMAS\_nu0\_WMAP9} & WMAP9 & 0 & \textit{DMONLY\_2fluid\_nu0\_WMAP9\_L400N1024} & AGN/SN feedback calibrated to obs.\!\!\!\! \\ [1ex]
\!\!\!\!\textit{BAHAMAS\_nu0\_WMAP9\_L100N512} & WMAP9 & 0 & \textit{DMONLY\_2fluid\_nu0\_WMAP9\_L100N512} & Using the L400N1024 calib. params.\!\!\!\! \\ [1ex]
\!\!\!\!\textit{BAHAMAS\_nu0\_WMAP9\_v2} & WMAP9 & 0 & \textit{DMONLY\_2fluid\_nu0\_v2\_WMAP9\_L400N1024} & AGN/SN feedback calibrated to obs.\!\!\!\! \\ [1ex]
\!\!\!\!\textit{BAHAMAS\_nu0\_WMAP9\_v3} & WMAP9 & 0 & \textit{DMONLY\_2fluid\_nu0\_v3\_WMAP9\_L400N1024} & AGN/SN feedback calibrated to obs.\!\!\!\! \\ [1ex]
\!\!\!\!\textit{BAHAMAS\_Theat7.6\_nu0\_WMAP9} & WMAP9 & 0 & \textit{DMONLY\_2fluid\_nu0\_WMAP9\_L400N1024} & As calibr., but lower AGN heating\!\!\! \\ [1ex]
\!\!\!\!\textit{BAHAMAS\_Theat8.0\_nu0\_WMAP9} & WMAP9 & 0 & \textit{DMONLY\_2fluid\_nu0\_WMAP9\_L400N1024} & As calibr., but higher AGN heating\!\!\!\! \\ [1ex]
\!\!\!\!\textit{C-OWLS\_AGN\_Mseed800\_Theat8.5\_WMAP7\_L100N512} & WMAP7 & - & \textit{DMONLY\_WMAP7\_L100N512} & BHs seeded for $\geq\! 800$ DM particles\!\!\!\!\!\!\!\!\!\! \\ [1ex]
\!\!\!\!\textit{C-OWLS\_AGN\_Mseed800\_Theat8.7\_WMAP7\_L100N512} & WMAP7 & - & \textit{DMONLY\_WMAP7\_L100N512} & BHs seeded for $\geq\! 800$ DM particles\!\!\!\!\!\!\!\!\!\! \\ [1ex]
\!\!\!\!\textit{C-OWLS\_AGN\_Mseed800\_WMAP7\_L100N512} & WMAP7 & - & \textit{DMONLY\_WMAP7\_L100N512} & BHs seeded for $\geq\! 800$ DM particles\!\!\!\!\!\!\!\!\!\! \\ [1ex]
\!\!\!\!\textit{C-OWLS\_AGN\_Planck2013} & Planck '13 & - & \textit{DMONLY\_Planck2013\_L400N1024} & - \\ [1ex]
\!\!\!\!\textit{C-OWLS\_AGN\_Theat8.3\_WMAP7} & WMAP7 & - & \textit{DMONLY\_WMAP7\_L400N1024} & - \\ [1ex]
\!\!\!\!\textit{C-OWLS\_AGN\_Theat8.5\_Planck2013} & Planck '13 & - & \textit{DMONLY\_Planck2013\_L400N1024} & - \\ [1ex]
\!\!\!\!\textit{C-OWLS\_AGN\_Theat8.5\_WMAP7} & WMAP7 & - & \textit{DMONLY\_WMAP7\_L400N1024} & - \\ [1ex]
\!\!\!\!\textit{C-OWLS\_AGN\_Theat8.5\_WMAP7\_L100N256} & WMAP7 & - & \textit{DMONLY\_WMAP7\_L100N256} & - \\ [1ex]
\!\!\!\!\textit{C-OWLS\_AGN\_Theat8.5\_WMAP7\_L100N512} & WMAP7 & - & \textit{DMONLY\_WMAP7\_L100N512} & - \\ [1ex]
\!\!\!\!\textit{C-OWLS\_AGN\_Theat8.7\_Planck2013} & Planck '13 & - & \textit{DMONLY\_Planck2013\_L400N1024} & - \\ [1ex]
\!\!\!\!\textit{C-OWLS\_AGN\_Theat8.7\_WMAP7} & WMAP7 & - & \textit{DMONLY\_WMAP7\_L400N1024} & - \\ [1ex]
\!\!\!\!\textit{C-OWLS\_AGN\_WMAP7} & WMAP7 & - & \textit{DMONLY\_WMAP7\_L400N1024} & - \\ [1ex]
\!\!\!\!\textit{C-OWLS\_AGN\_WMAP7\_L200N1024} & WMAP7 & - & \textit{DMONLY\_WMAP7\_L200N1024} & - \\ [1ex]
\!\!\!\!\textit{C-OWLS\_AGN\_WMAP7\_L200N512} & WMAP7 & - & \textit{DMONLY\_WMAP7\_L200N512} & - \\ [1ex]
\!\!\!\!\textit{C-OWLS\_NOCOOL\_UVB\_Planck2013} & Planck '13 & - & \textit{DMONLY\_Planck2013\_L400N1024} & No AGN feedback \\ [1ex]
\!\!\!\!\textit{C-OWLS\_NOCOOL\_UVB\_WMAP7} & WMAP7 & - & \textit{DMONLY\_WMAP7\_L400N1024} & No AGN feedback \\ [1ex]
\!\!\!\!\textit{C-OWLS\_NOCOOL\_UVB\_WMAP7\_L100N256} & WMAP7 & - & \textit{DMONLY\_WMAP7\_L100N256} & No AGN feedback \\ [1ex]
\!\!\!\!\textit{C-OWLS\_NOCOOL\_UVB\_WMAP7\_L100N512} & WMAP7 & - & \textit{DMONLY\_WMAP7\_L100N512} & No AGN feedback \\ [1ex]
\!\!\!\!\textit{C-OWLS\_NOCOOL\_UVB\_WMAP7\_L200N512} & WMAP7 & - & \textit{DMONLY\_WMAP7\_L200N512} & No AGN feedback \\ [1ex]
\!\!\!\!\textit{C-OWLS\_REF\_Planck2013} & Planck '13 & - & \textit{DMONLY\_Planck2013\_L400N1024} & No AGN feedback \\ [1ex]
\!\!\!\!\textit{C-OWLS\_REF\_WMAP7} & WMAP7 & - & \textit{DMONLY\_WMAP7\_L400N1024} & No AGN feedback \\ [1ex]
\!\!\!\!\textit{C-OWLS\_REF\_WMAP7\_L200N1024} & WMAP7 & - & \textit{DMONLY\_WMAP7\_L200N1024} & No AGN feedback \\ [1ex]
\!\!\!\!\textit{C-OWLS\_REF\_WMAP7\_L200N512} & WMAP7 & - & \textit{DMONLY\_WMAP7\_L200N512} & No AGN feedback \\ [1ex]
\!\!\!\!\textit{OWLS\_AGN$^*$} & WMAP3 & - & \textit{DMONLY\_L100N512$^*$} & - \\ [1ex]
\!\!\!\!\textit{OWLS\_AGN\_LOBETA} & WMAP3 & - & \textit{DMONLY\_L100N512$^*$} & Alternative AGN model \\ [1ex]
\!\!\!\!\textit{OWLS\_AGN\_LOBETA\_NOSN} & WMAP3 & - & \textit{DMONLY\_L100N512$^*$} & No SN feedback, alt. AGN model \\ [1ex]
\!\!\!\!\textit{OWLS\_AGN\_LOBETA\_Theat7.0} & WMAP3 & - & \textit{DMONLY\_L100N512$^*$} & Alternative AGN model \\ [1ex]
\!\!\!\!\textit{OWLS\_AGN\_WMAP7} & WMAP7 & - & \textit{DMONLY\_WMAP7\_L100N512} & - \\ [1ex]
\!\!\!\!\textit{OWLS\_AGN\_WMAP7\_CAMB$^*$} & WMAP7 & - & \textit{DMONLY\_WMAP7\_CAMB\_L100N512$^*$} & Uses CAMB initial power spectrum\!\! \\ [1ex]
\!\!\!\!\textit{OWLS\_AGN\_WMAP7\_L100N256} & WMAP7 & - & \textit{DMONLY\_WMAP7\_L100N256} & - \\ [1ex]
\!\!\!\!\textit{OWLS\_DBLIMFCONTSFML14} & WMAP3 & - & \textit{DMONLY\_L100N512$^*$} & No AGN feedback \\ [1ex]
\!\!\!\!\textit{OWLS\_DBLIMFCONTSFV1618} & WMAP3 & - & \textit{DMONLY\_L100N512$^*$} & No AGN feedback \\ [1ex]
\!\!\!\!\textit{OWLS\_DBLIMFV1618$^*$} & WMAP3 & - & \textit{DMONLY\_L100N512$^*$} & No AGN feedback \\ [1ex]
\!\!\!\!\textit{OWLS\_EOS1p0} & WMAP3 & - & \textit{DMONLY\_L100N512$^*$} & No AGN feedback \\ [1ex]
\!\!\!\!\textit{OWLS\_IMFSALP} & WMAP3 & - & \textit{DMONLY\_L100N512$^*$} & No AGN feedback \\ [1ex]
\!\!\!\!\textit{OWLS\_NOAGB\_L100N256} & WMAP3 & - & \textit{DMONLY\_L100N256} & No AGN feedback \\ [1ex]
\!\!\!\!\textit{OWLS\_NOAGB\_NOSNIa} & WMAP3 & - & \textit{DMONLY\_L100N512$^*$} & No AGN feedback \\ [1ex]
\!\!\!\!\textit{OWLS\_NONRAD} & WMAP3 & - & \textit{DMONLY\_L100N512$^*$} & No AGN feedback \\ [1ex]
\!\!\!\!\textit{OWLS\_NOSN$^*$} & WMAP3 & - & \textit{DMONLY\_L100N512$^*$} & No AGN or SN feedback \\ [1ex]
\!\!\!\!\textit{OWLS\_NOSN\_NOZCOOL$^*$} & WMAP3 & - & \textit{DMONLY\_L100N512$^*$} & No AGN or SN feedback \\ [1ex]
\!\!\!\!\textit{OWLS\_NOZCOOL$^*$} & WMAP3 & - & \textit{DMONLY\_L100N512$^*$} & No AGN feedback \\ [1ex]
\!\!\!\!\textit{OWLS\_REF$^*$} & WMAP3 & - & \textit{DMONLY\_L100N512$^*$} & No AGN feedback \\ [1ex]
\!\!\!\!\textit{OWLS\_REF\_L100N256} & WMAP3 & - & \textit{DMONLY\_L100N256} & No AGN feedback \\ [1ex]
\!\!\!\!\textit{OWLS\_REF\_WMAP7} & WMAP7 & - & \textit{DMONLY\_WMAP7\_L100N512} & No AGN feedback \\ [1ex]
\!\!\!\!\textit{OWLS\_REF\_WMAP7\_L100N256} & WMAP7 & - & \textit{DMONLY\_WMAP7\_L100N256} & No AGN feedback \\ [1ex]
\!\!\!\!\textit{OWLS\_SNIaGAUSS} & WMAP3 & - & \textit{DMONLY\_L100N512$^*$} & No AGN feedback \\ [1ex]
\!\!\!\!\textit{OWLS\_WDENS$^*$} & WMAP3 & - & \textit{DMONLY\_L100N512$^*$} & No AGN feedback \\ [1ex]
\!\!\!\!\textit{OWLS\_WML1V848$^*$} & WMAP3 & - & \textit{DMONLY\_L100N512$^*$} & No AGN feedback \\ [1ex]
\!\!\!\!\textit{OWLS\_WML4$^*$} & WMAP3 & - & \textit{DMONLY\_L100N512$^*$} & No AGN feedback \\ [1ex]
\!\!\!\!\textit{OWLS\_WPOTNOKICK} & WMAP3 & - & \textit{DMONLY\_L100N512$^*$} & No AGN feedback \\ [1ex]
\!\!\!\!\textit{OWLS\_WTHERMAL\_WMAP5} & WMAP5 & - & \textit{DMONLY\_WMAP5\_L100N512} & No AGN feedback \\ [1ex]
\!\!\!\!\textit{OWLS\_WVCIRC} & WMAP3 & - & \textit{DMONLY\_L100N512$^*$} & No AGN feedback \\ [1ex]
\hline
\end{tabular}
\label{tab:sims}
\end{table*}

\begin{table*}
\caption{The cosmological parameters of the simulations in Table~\ref{tab:sims}. The ``BAO'' parameters are chosen to be roughly in between those of WMAP9 and Planck. Numbers in parentheses are total neutrino masses in eV.}
\centering
\setlength{\tabcolsep}{12pt}
\begin{tabular}{l l l l l l l l}
\hline
Cosmology & $\Omega_\mathrm{m}$ & $\Omega_\Lambda$ & $\Omega_\mathrm{b}$ & $\Omega_\nu$ & $\sigma_8$ & $n_\mathrm{s}$ & $h$ \\ [0.5ex]
\hline
WMAP3 & 0.238 & 0.762 & 0.0418 & 0 & 0.74 & 0.951 & 0.73 \\ [1ex]
WMAP5 & 0.258 & 0.742 & 0.0441 & 0 & 0.796 & 0.963 & 0.719 \\ [1ex]
WMAP7 & 0.272 & 0.728 & 0.0455 & 0 & 0.81 & 0.967 & 0.704 \\ [1ex]
WMAP9 ($0$) & 0.2793 & 0.7207 & 0.0463 & 0 & 0.8211 & 0.972 & 0.700 \\ [1ex]
WMAP9 ($0.06$) & 0.2793 & 0.7207 & 0.0463 & 0.0013 & 0.8069 & 0.972 & 0.700 \\ [1ex]
WMAP9 ($0.12$) & 0.2793 & 0.7207 & 0.0463 & 0.0026 & 0.7924 & 0.972 & 0.700 \\ [1ex]
WMAP9 ($0.24$) & 0.2793 & 0.7207 & 0.0463 & 0.0053 & 0.7600 & 0.972 & 0.700 \\ [1ex]
WMAP9 ($0.48$) & 0.2793 & 0.7207 & 0.0463 & 0.0105 & 0.7001 & 0.972 & 0.700 \\ [1ex]
BAO & 0.3 & 0.7 & 0.05 & 0 & 0.8 & 0.96 & 0.7 \\ [1ex]
Planck 2013 ($0$) & 0.3175 & 0.6825 & 0.049 & 0 & 0.8341 & 0.9624 & 0.6711 \\ [1ex]
Planck 2015 ($0.06$) & 0.3067 & 0.6933 & 0.0482 & 0.0014 & 0.8085 & 0.9701 & 0.6787 \\ [1ex]
Planck 2015 ($0.12$) & 0.3091 & 0.6909 & 0.0488 & 0.0028 & 0.7943 & 0.9693 & 0.6768 \\ [1ex]
Planck 2015 ($0.24$) & 0.3129 & 0.6871 & 0.0496 & 0.0057 & 0.7664 & 0.9733 & 0.6723 \\ [1ex]
Planck 2015 ($0.48$) & 0.3197 & 0.6803 & 0.0513 & 0.0117 & 0.7030 & 0.9811 & 0.6643 \\ [1ex]
\hline
\end{tabular}
\label{tab:cosms}
\end{table*}

\subsubsection{Models with AGN}
\label{subsubsec:AGN}
As in the OWLS AGN model, cosmo-OWLS and \bahamas{} use the \citet{Springel2005} prescription for black hole seeding, and the \citet{BoothSchaye2009} prescriptions for black hole merging, accretion and AGN feedback. This model has several free parameters, although its authors have shown the model to be insensitive to some of these due to self-regulation. The most important parameter for the effect of feedback on large scales is the minimum heating temperature for AGN feedback, $\Delta T_\mathrm{heat}$ \citep[see][]{LeBrun2014,Pike2014}.\footnote{We note here that the parameters that the large-scale gas properties are most sensitive to may be different for simulations using a hydro solver different from SPH. See e.g.\ \citet{Dubois2012,Hahn2017}.} Black holes in this model store a fraction of the energy gained from accretion until they are able to heat a fixed number of particles ($n_\mathrm{heat}$) by $\Delta T_\mathrm{heat}$, to ensure that the heated gas does not cool in an artificially short time for numerical reasons and that the time between feedback events is shorter than the Salpeter time for Eddington-limited accretion. The fiducial value of this parameter is $10^8\Tunit$ in OWLS and cosmo-OWLS and $10^{7.8}\Tunit$ in \bahamas{}. For some simulations in the current set, a different value than the fiducial one is adopted; in this case the simulation name includes a suffix ``TheatX.Y'' for $\Delta T_\mathrm{heat}=10^{X.Y}\Tunit$.

Besides the minimum heating temperature several other parameters concerning the AGN can be varied. By default, seed black holes are placed in any halo with at least $100$ dark matter particles in its Friends-of-Friends group, which makes the black hole seeding resolution-dependent. To investigate the effect of this, several simulations were re-run with a black hole seeding criterion of $800$ dark matter particles, thus having the same halo mass threshold as a simulation with $8\times$ worse mass resolution. The names of these simulations include the suffix ``Mseed800''.

In the fiducial AGN model the accretion efficiency scales as a power law of the density in the high-density regime. The power law slope is $\beta=2$ by default. In some models, a shallower slope of $\beta=1$ is explored. Such simulations have the suffix ``LOBETA'' in their name.

Finally, note that all \bahamas{} simulations adopt stellar and AGN feedback parameters calibrated to observations. In \bahamas{} simulations that include the suffix ``Theat'' only the AGN heating parameter adopts a non-calibrated value. The calibrated values are based on the L400N1024 simulations, but to test for strong convergence a L100N512 simulation with the same parameters was also run, and power spectra for it are presented here. Care was taken that the \emph{physical} parameters were kept fixed with the change in box size and resolution, for example by keeping the minimum halo mass rather than the number of DM particles for black hole seeding fixed -- we refer to Appendix C of \citet{McCarthy2017} for more information.

\subsubsection{Models with physical processes turned off}
\label{subsubsec:NO}
Many OWLS and cosmo-OWLS simulations in the current set do not include AGN (indicated by the lack of ``AGN'' in their name), but in some the effect of switching off (additional) subgrid physics was tested. Examples include ``NOSN'' (no SN feedback) and ``NOZCOOL'' (no metal-line cooling), both of which featured in VD11. New in the current set are ``NOAGB'', in which mass loss from Asymptotic Giant Branch stars is turned off; ``NOSNIa'', in which there is no mass loss from type Ia supernovae; ``NONRAD'', a non-radiative simulation which includes no radiative cooling or heating at all (and hence no star formation etc. either); and ``NOCOOL\_UVB'', which also does not include radiative cooling, although net photoheating is allowed. While such simulations are not seen as realistic, they can still offer interesting extremes for modelling the effects of baryons on the clustering of matter. Simulations without SNe but with AGN are particularly interesting to examine the interplay between the two types of feedback, which we consider in \S\ref{subsec:SNAGN}.

\subsubsection{Models with non-standard star formation or stellar winds}
\label{subsubsec:IMFSN}
In VD11 power spectra were presented for several OWLS models with subgrid prescriptions for star formation or stellar winds that differed from those used in the reference simulations. In the current set several more are included, which we will briefly explain in alphabetical order here. We do not focus on the effects of these models in this work, but they are still useful to gauge the impact of theoretical uncertainties on the matter power spectrum. For all of these models more information can be found in \citet{Schaye2010}.

Models named ``DBLIMF'' use a top-heavy stellar initial mass function (IMF) in high-pressure environments. This increases the number of SNe per unit stellar mass, and this additional available energy can be applied in subgrid models in various ways. One such simulation, ``DBLIMFV1618'', was included in the VD11 release; in it, the additional energy was used to increase the wind speed from $600$ to $1618\vunit$, which had a similar effect on the matter clustering as including AGN. An underlying assumption of this model is that the rate of formation of massive stars is continuous with the gas pressure as the IMF changes suddenly. Here we include two more variations on ``DBLIMF'', for both of which a continuous star formation law is assumed instead (``CONTSF''). One of these still puts the additional energy from a top-heavy IMF into a faster supernova-driven wind (``V1618''), but the other instead puts the additional energy into increasing the wind's mass loading (``ML14''). While we do not show so here, assuming a continuous star formation law somewhat diminishes the effect that the SNe have on the power spectrum (though there is still an up to $10\%$ decrease in power for $k<10\kunit$ compared to dark matter only), and increasing the mass loading instead of the wind speed barely changes the matter clustering at $z=0$ compared to the fiducial model (``REF'').

To model the interstellar medium a polytropic equation of state with slope $\gamma_\mathrm{eff}=4/3$ is imposed; in ``EOS1p0'', this slope is instead $1$. While this somewhat diminishes star formation at high redshift, the effect on matter clustering at $z=0$ is negligible.

The fiducial stellar initial mass function is that of \citet{Chabrier2003}, but ``IMFSALP'' uses the \citet{Salpeter1955} IMF instead. This causes the amount of metals and thereby the amount of star formation to decrease, and SNe are less frequent. As a consequence, matter clusters $\sim 1\%$ stronger for $1\leq k \leq 10\kunit$ compared to the fiducial model.

The time between the formation of its progenitor and a type Ia supernova depends on binary evolution, and the distribution of delay times has some uncertainty. The fiducial model assumes an exponential decline with time, but in ``SNIaGAUSS'' a Gaussian distribution is assumed instead \citep[see][]{Wiersma2009}. The difference in clustering compared to the fiducial model is quite small (roughly half that of assuming a Salpeter IMF).

Finally, we include three more models in which the implementation of SN-driven winds is varied, in addition to those included in VD11 (which were ``WDENS'', ``WML1V848'' and ``WML4''). ``WPOTNOKICK'' and ``WVCIRC'' are both approximations of momentum-driven winds, i.e.\ galactic outflows driven not primarily by SN explosions but by radiation pressure. ``WPOTNOKICK'' is based on the \citet{OppenheimerDave2006} model (though without hydrodynamical decoupling) and assumes wind velocities $v_\mathrm{w}=3\sigma$ with mass loading $\eta=150\vunit/\sigma$, where $\sigma$ is the galaxy velocity dispersion as estimated from the local potential. ``WVCIRC'' is the same, except that the velocity dispersion is estimated by first running an on-the-fly halo finder and then calculating the circular velocity, $v_\mathrm{c}$, from the resulting halo mass and virial radius, setting $\sigma=\sqrt{2}v_\mathrm{c}$. Both models diminish the amount of clustering on scales $k\lesssim 40\kunit$ by up to $30\%$, with a magnitude and scale-dependence that is highly similar to that of the OWLS AGN model. We note however that in these kind of implementations of momentum-driven winds the total amount of energy is not limited, and may exceed that available from radiation. Lastly, in ``WTHERMAL'' the fiducial kinetic SN feedback model is replaced by the energy-driven (thermal) model of \citet{DallaVecchiaSchaye2012}, which, like the fiducial model, injects only $40\%$ of the available SN energy to drive winds. The thermal feedback model is more effective at driving winds and the simulation is less sensitive to its parameters compared to the kinetic SN feedback model. Its effect on the power spectrum is also larger, on average only a factor of two below that of AGN feedback for $k<10\kunit$, in terms of the suppression relative to dark matter only.

\subsection{Modified dark matter only simulations}
\label{subsec:dmonly}
OWLS and cosmo-OWLS use the common approximation of initializing the particles using the total matter transfer function, while for the \bahamas{} simulations the dark matter and baryons are initialized instead with their own respective transfer functions. As \citet{Valkenburg2017} have shown, this can create percent-level differences on all scales of the matter power spectrum at redshift zero. A consequence of the change in initialization between \bahamas{} and its predecessors is that it introduces a $1-2\%$ offset in clustering between a hydrodynamical \bahamas{} simulation and its dark matter only counterpart, which is especially noticeable on large scales. As our goal is to probe the effect of galaxy formation, rather than initial conditions, we have run a second set of dark matter only simulations which, like the hydrodynamical simulations, contain $2\times 1024^3$ particles with mass ratios $\Omega_\mathrm{b}:\Omega_\mathrm{c}$. While both particle species act like dark matter, the lighter (baryon-mass) particle species is initialized with the baryon transfer function instead of the cold dark matter one, the end result of which is a $1-2\%$ stronger clustering on all scales. We find that the large-scale power in these simulations agree with their hydrodynamical counterparts to $<0.1\%$. Simulations run in this way are dubbed ``DMONLY\_2fluid''.

Even though the total matter transfer function is used in both dark matter only and hydrodynamical simulations in OWLS and cosmo-OWLS, for some of them we still observe $0.1-0.2\%$ offsets on large scales between the two. This is due to several related effects, all consequences of how the initial conditions were set up: having twice as many particles in one versus the other leading to numerical differences in their evolution; not including a phase shift when splitting the initial particles in offset dark matter and baryon particles which creates artificial power; and spurious clumping between dark matter and baryon particles when the force softening is smaller than the interparticle spacing, as it is here, even with staggered initial conditions \citep[see e.g.][]{OLearyMcQuinn2012,Angulo2013,Valkenburg2017}. These issues do not arise in the comparison of the \bahamas{} 2-fluid simulations and their hydrodynamical counterparts, as there the first of these effects is absent while the others are present in equal measure in both runs.

We further explore the differences between the 1-fluid and 2-fluid simulations in Appendix~\ref{app:2fluid}. As we conclude there, replacing the DMONLY counterparts of the (cosmo\discretionary{-)}{}{)}OWLS simulations with 2-fluid runs that, unlike those for \bahamas{}, use the same (total) transfer function for both components, would remove the $\sim 0.1\%$ large-scale offsets currently present for some of these simulations. However, since the effect is small and re-running many simulations would be computationally expensive, we have chosen not to do so. For the results of \S\ref{subsec:barfrac}, we instead correct the power spectra of cosmo-OWLS DMONLY simulations by multiplying with a constant so as to bring them into $<0.1\%$ agreement with their hydrodynamical counterparts on the largest scales measured, motivated by the results of Appendix~\ref{app:2fluid}.

\begin{figure}
\begin{center}
\includegraphics[width=1.0\columnwidth, trim=16mm 8mm 10mm -4mm]{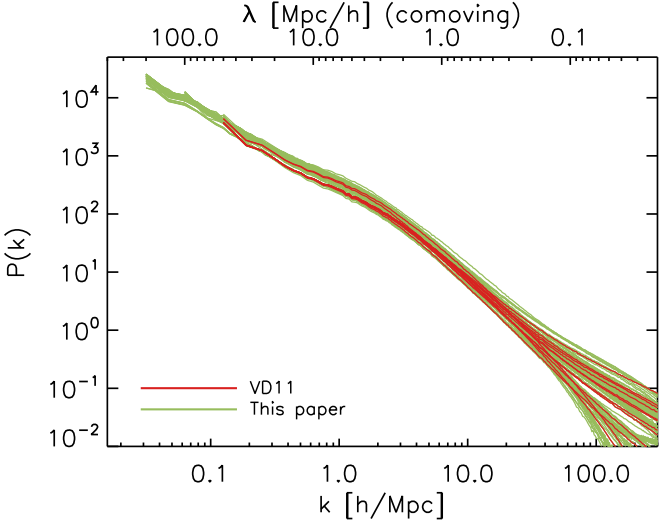}
\caption{All matter power spectra provided with the current paper are shown in green. The power spectra released by VD11, which are also part of this set, are shown in red. The current set expands on the 2011 release with more cosmologies, non-zero neutrino masses and different strengths of AGN feedback. The bottom x-axis shows the comoving Fourier scale $k$ while the top axis shows the corresponding comoving physical scale $\lambda=2\pi/k$.}
\label{fig:z0_power}
\end{center}
\end{figure}

\begin{figure*}
\begin{center}
\includegraphics[width=1.0\columnwidth, trim=21mm 8mm 5mm -4mm]{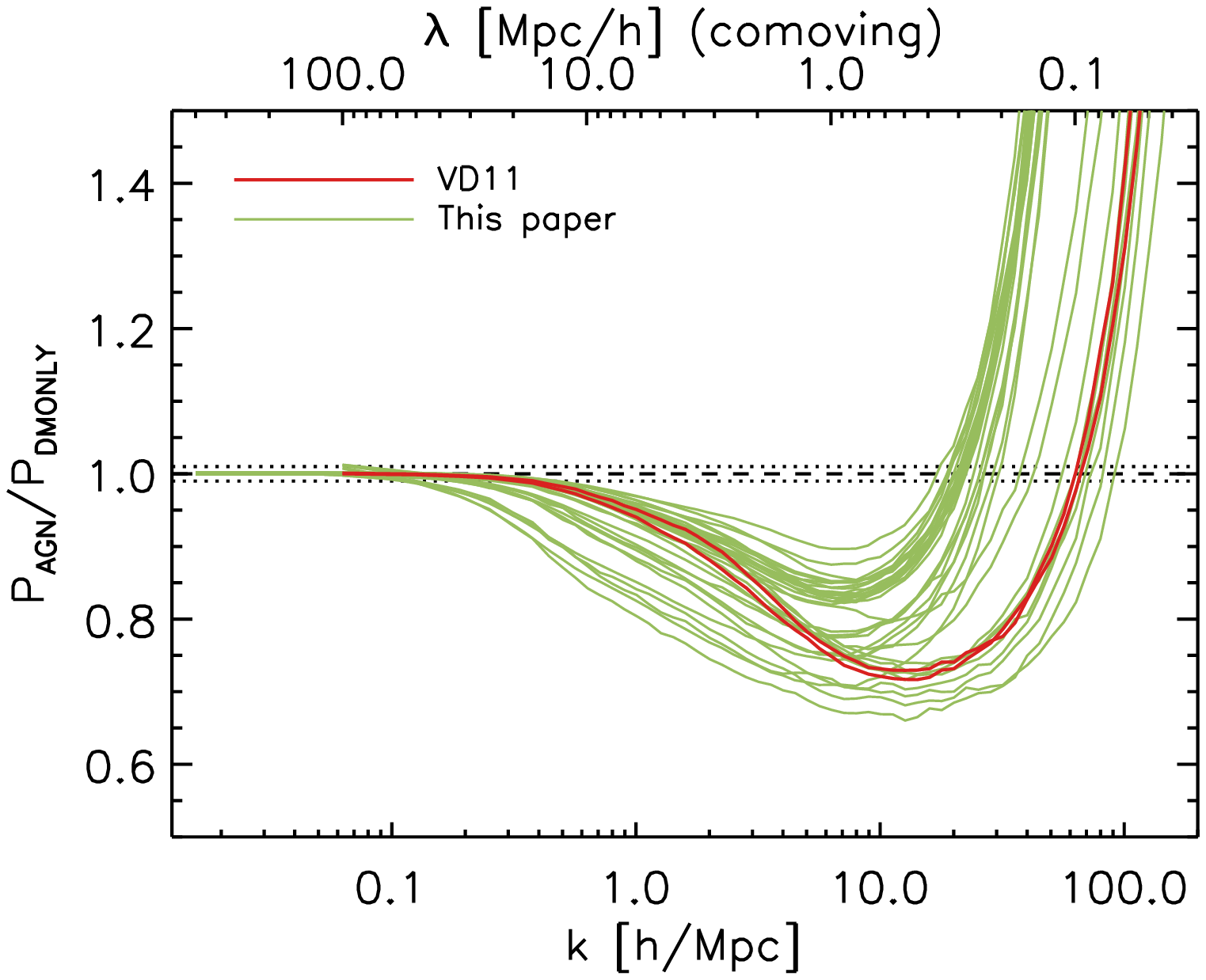}
\includegraphics[width=1.0\columnwidth, trim=11mm 8mm 15mm -4mm]{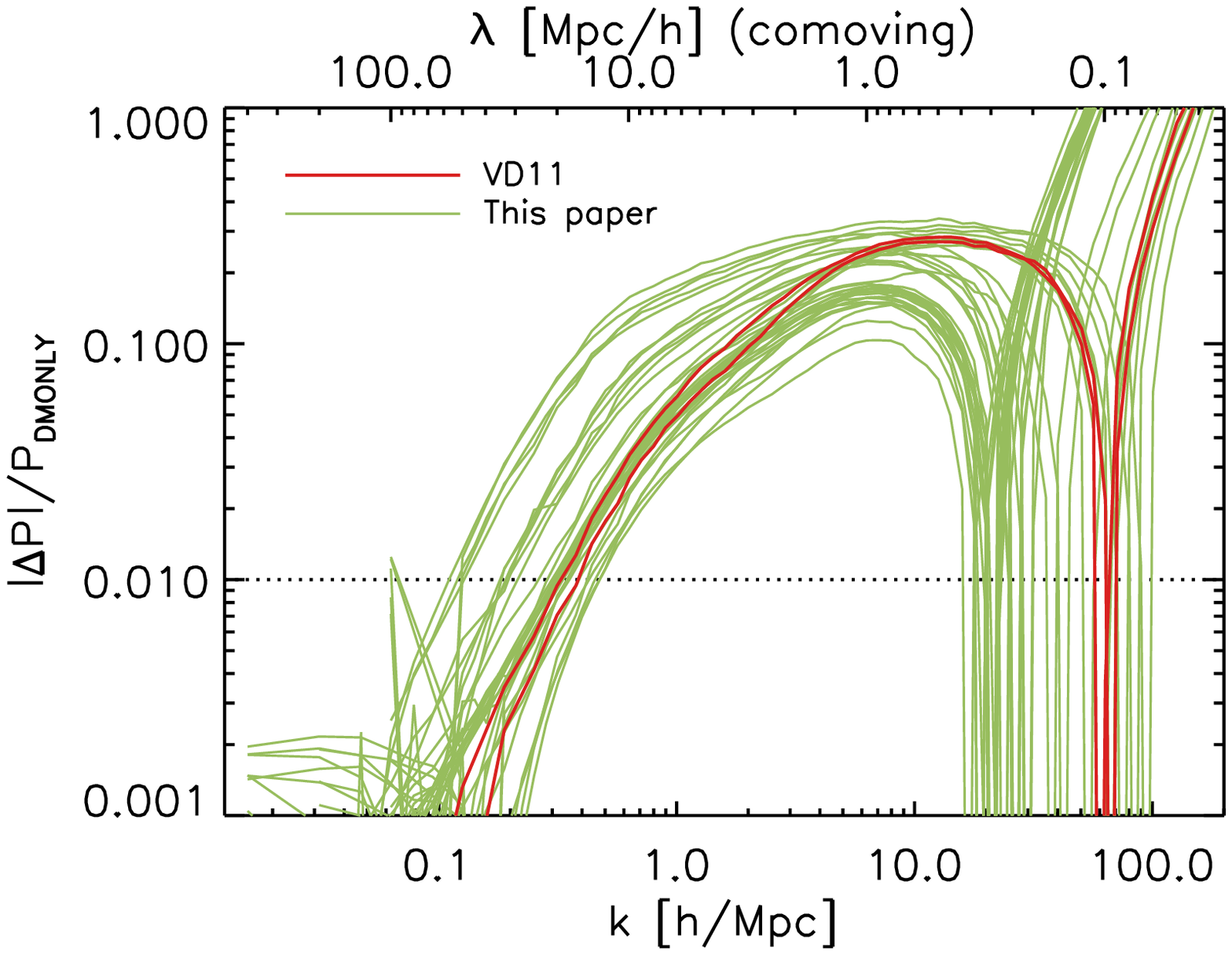}
\caption{The effect of galaxy formation including AGN feedback on the matter power spectrum. The y-axis shows the change in power relative to a simulation with only dark matter, but otherwise identical initial conditions, linear on the left and logarithmic on the right. Each line shows a simulation with a different feedback strength, cosmology, and/or neutrino mass. The power spectra from the two AGN simulations included in VD11 are highlighted in red (both have identical physics but a slightly different cosmology). The relative effect of galaxy formation is only weakly dependent on cosmology or neutrino mass, and the range in effects seen here is therefore mainly due to changes in the strength of AGN (or stellar) feedback. All subsequent figures will use a logarithmic axis, as we are interested in small changes to the power.}
\label{fig:z0_power_pairs}
\end{center}
\end{figure*}

\subsection{Power spectra}
\label{subsec:pow}
Like VD11, we have used a modified version of \textsc{powmes} \citep{Colombi2009} to calculate highly accurate power spectra for each of the simulations in our sample from their particle data, down to $k\approx 500\kunit$. We refer to these publications for more information on the method and its accuracy. We have used the same method to calculate power spectra for EAGLE, which we compare to in \S\ref{subsec:complit}. For the \bahamas{} simulations with massive neutrinos an additional step is needed, as the neutrinos themselves are not included as particles but using the method outlined in \citet{Ali-HaimoudBird2013}. A useful by-product of this method is the neutrino-only power spectrum, which is included with every simulation output. Since the neutrino overdensities can be assumed to be in phase with those of the remaining (non-relativistic) matter in the simulation \citep[see][]{Ali-HaimoudBird2013}, we can write:
\begin{equation}
\hat{\delta}_\nu(\mathbf{k}) = \left(\frac{P_\nu(k)}{P_\mathrm{m}(k)}\right)^{1/2}\hat{\delta}_\mathrm{m}(\mathbf{k}),
\label{eq:deltanu}
\end{equation}
where $\hat{\delta}(\mathbf{k})$ is the Fourier transform of the density contrast $\delta(\mathbf{x})$, $P(k)=\left<|\hat{\delta}(\mathbf{k})|^2\right>_k$ is the power on Fourier scale $k$ and the subscripts $\nu$ and $\mathrm{m}$ denote the neutrinos and the remaining matter, respectively. Denoting the fraction of matter in neutrinos as $f_\nu=\Omega_\nu/\Omega_\mathrm{m}$, we can write the density contrast field of all matter in Fourier space as:
\begin{eqnarray}
\nonumber
\quad\quad\hat{\delta}_\mathrm{tot}(\mathbf{k})\!\!\!&=&\!\!\!(1-f_\nu)\hat{\delta}_\mathrm{m}(\mathbf{k})+f_\nu\hat{\delta}_\nu(\mathbf{k})\\
\!\!\!&=&\!\!\!\left[(1-f_\nu)+f_\nu\left(\frac{P_\nu(k)}{P_\mathrm{m}(k)}\right)^{1/2}\right]\hat{\delta}_\mathrm{m}(\mathbf{k}).
\label{eq:deltatot}
\end{eqnarray}
For the \bahamas{} simulations with massive neutrinos we can therefore combine the neutrino-only and particle power spectra to find a total matter power spectrum through:
\begin{equation}
P_\mathrm{tot}(k)=\left[(1-f_\nu)+f_\nu\left(\frac{P_\nu(k)}{P_\mathrm{m}(k)}\right)^{1/2}\right]^2P_\mathrm{m}(k).
\label{eq:Ptot}
\end{equation}
All power spectra are normalized to the total matter density in the simulated volume and shot-noise subtracted.

When showing a ratio of power spectra, we re-bin our power spectra in bins of minimum size $0.05\,\mathrm{dex}$ in $k$ to reduce visual noise.

\section{A comparison of power spectra}
\label{sec:results}
In this section we use power spectra from the current set to investigate the range of relative effects on the total and CDM power spectra that can be brought about with AGN feedback, which changes in the models have the largest impact on the clustering of matter, and how these impacts change with cosmology and redshift. We also compare to power spectra in the literature, starting with the power spectra released by VD11.

\subsection{Comparison to VD11}
\label{subsec:compVD11}
In Figure~\ref{fig:z0_power} we show all power spectra in the current set, highlighting those previously released by VD11. The vertical range spanned on large scales is mostly an indication of the range in cosmology probed, while that on the smallest scales indicates the range in galaxy formation models. The simulations in the current set allow us to probe larger scales and provide both a denser and broader sampling of parameter space.

Since AGN feedback has previously been shown to have the largest impact on the matter power spectrum out of all investigated aspects of galaxy formation, we show the difference in the power spectrum relative to the dark matter only prediction for all simulations that include AGN feedback in Figure~\ref{fig:z0_power_pairs}. Once again the AGN simulations from VD11 are highlighted. This figure shows the wide range in AGN feedback impacts on the matter power spectrum tested by the current set, including both models with more extreme and milder effects than the AGN simulations of VD11. As previous authors have shown, the relative effect of galaxy formation is only weakly dependent on cosmology or neutrino mass \citep[e.g.\ VD11,][]{Mead2016,Mummery2017}, and the range in effects seen here is therefore mainly due to changes in the strength of AGN (or -- as we show in \S\ref{subsec:SNAGN} -- stellar) feedback. In some cases, the AGN feedback is so strong as to even affect the power spectrum on the largest scales probed, but in most cases the offsets of $0.1-0.2\%$ on the largest scales (visible in the logarithmic right-hand panel) have numerical origins (see \S\ref{subsec:dmonly}).

We note here that the power spectrum is changed by feedback rearranging matter -- primarily gas --around galaxies, in some cases out to beyond the virial radius. While the positions of galaxies and haloes may change as well when baryons are added or feedback is varied, \citet{vanDaalen2014} have shown that this does not drive the shifts in the power spectrum. Furthermore, note that the power spectrum changes on scales larger than the maximum scale over which matter is displaced in real space. In the language of the halo model, feedback changes the large-scale power by decreasing the 2-halo term in a mass-dependent way.

\begin{figure}
\begin{center}
\includegraphics[width=1.0\columnwidth, trim=16mm 8mm 10mm -4mm]{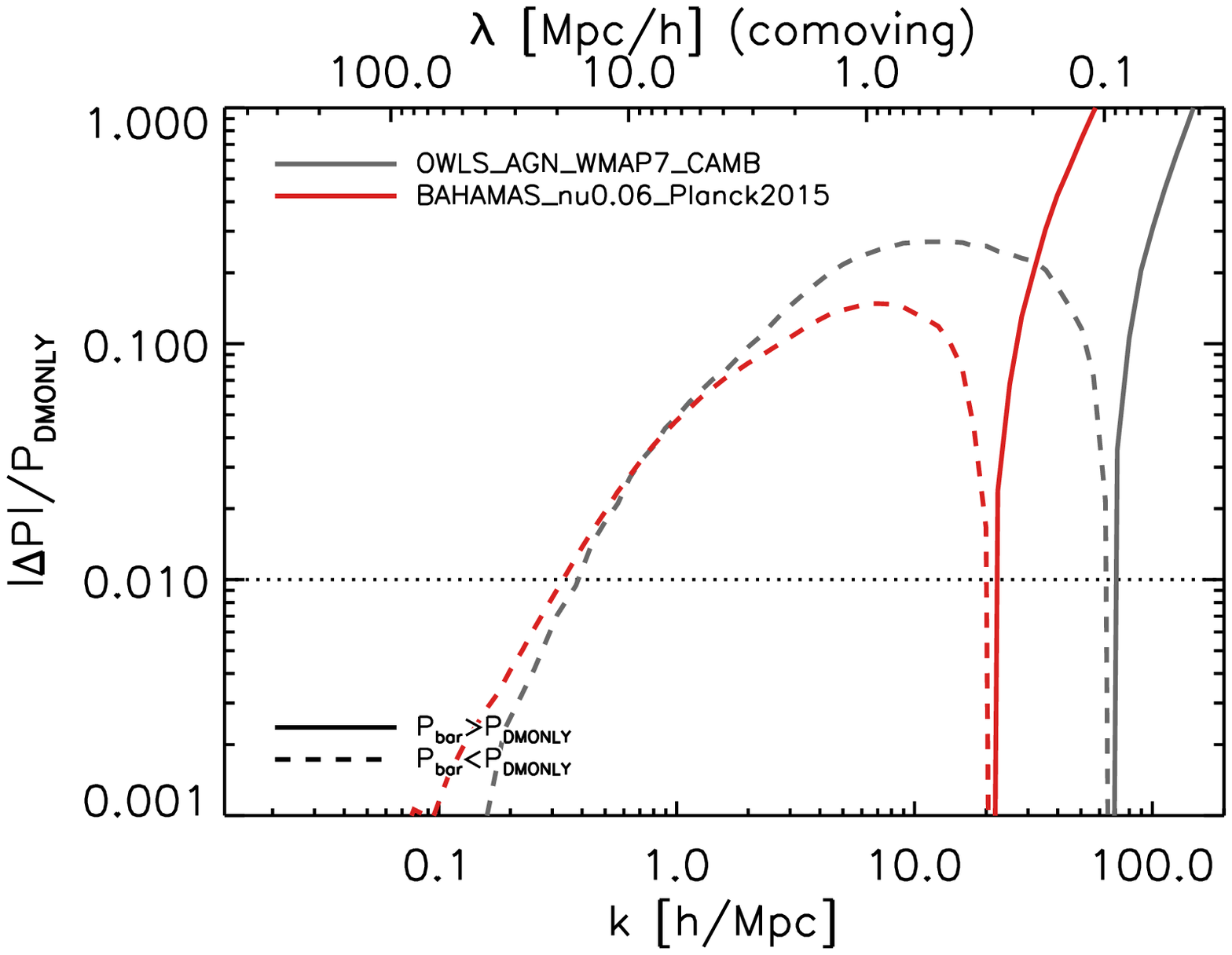}
\caption{A comparison of the relative effect of galaxy formation between the most realistic simulation included in the VD11 power spectra (grey, OWLS) and the most realistic simulation in the current set (red, \bahamas{}). In the new model, the feedback lowers the power less overall, although the effect extends to slightly larger scales. While the latter can be attributed to the larger box size, the former is a combined result of changes in the feedback prescription and a lower resolution (see main text). However, the new model is in much better agreement with observables, including the galaxy stellar mass function, the stellar-to-halo mass relation and the hot gas fraction as a function of halo mass for groups and clusters.}
\label{fig:diff_oldvsnew}
\end{center}
\end{figure}

We compare the most realistic simulation from VD11 (a WMAP7 OWLS AGN simulation, in grey) to that of the current set (a Planck 2015 \bahamas{} simulation, in red) in Figure~\ref{fig:diff_oldvsnew}. Overall, the feedback in the newer simulation is weaker, though it extends to larger scales and still reduces the power by $>10\%$ for $k<10\kunit$.\footnote{We note here that the differences between the two simulations on the largest scales, $k\lesssim 0.5\kunit$, can be largely attributed to the effects described in Appendix~\ref{app:2fluid}. Using a 2-fluid DMONLY run -- which mimics the initial conditions of the AGN simulation more closely -- to compute the relative OWLS power spectrum brings the two results into much closer agreement.} The transition from suppressed to enhanced clustering, relative to dark matter only, has shifted from $k\approx 70\kunit$ to $k\approx 20\kunit$. We will refer to this as the cross-over scale.

As discussed in \citet{McCarthy2017}, the feedback in the \bahamas{} simulations was calibrated to the galaxy stellar mass function and the gas fractions in groups and clusters. Relative to the (cosmo\discretionary{-)}{}{)}OWLS AGN model, the SN feedback wind velocity is lower in \bahamas{}, reducing its effectiveness and allowing more low-mass galaxies to form. Since this means less gas is ejected than before, and therefore more is available to the supermassive black holes, the heating temperature of AGN feedback is slightly lowered in order to bring the hot gas fractions back in agreement with observations. This may explain the reduction in the peak of the effect of AGN feedback seen in Figure~\ref{fig:diff_oldvsnew}.

\begin{figure}
\begin{center}
\includegraphics[width=1.0\columnwidth, trim=16mm 8mm 10mm -4mm]{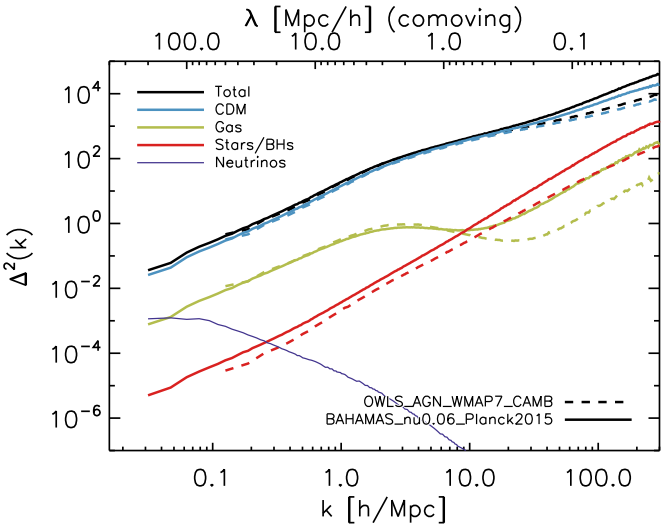}
\caption{A comparison of the power spectra of the different components for the most realistic simulation included in the VD11 power spectra (dashed lines) and the most realistic simulation in the current set (solid lines, \bahamas{}). The power spectra of OWLS AGN have been renormalized to the Planck 2015 cosmology through the square of the linear growth factor. Despite this, the \bahamas{} simulations generally show stronger clustering in all components except the gas on scales $k\lesssim 10\kunit$. The primary reason for this is the weaker feedback in \bahamas{} compared to OWLS, allowing more low-mass galaxies to form. On smaller scales, the differences seen here are driven by the change in resolution.}
\label{fig:components}
\end{center}
\end{figure}

\begin{figure*}
\begin{center}
\includegraphics[width=1.0\columnwidth, trim=16mm 8mm 21mm -4mm]{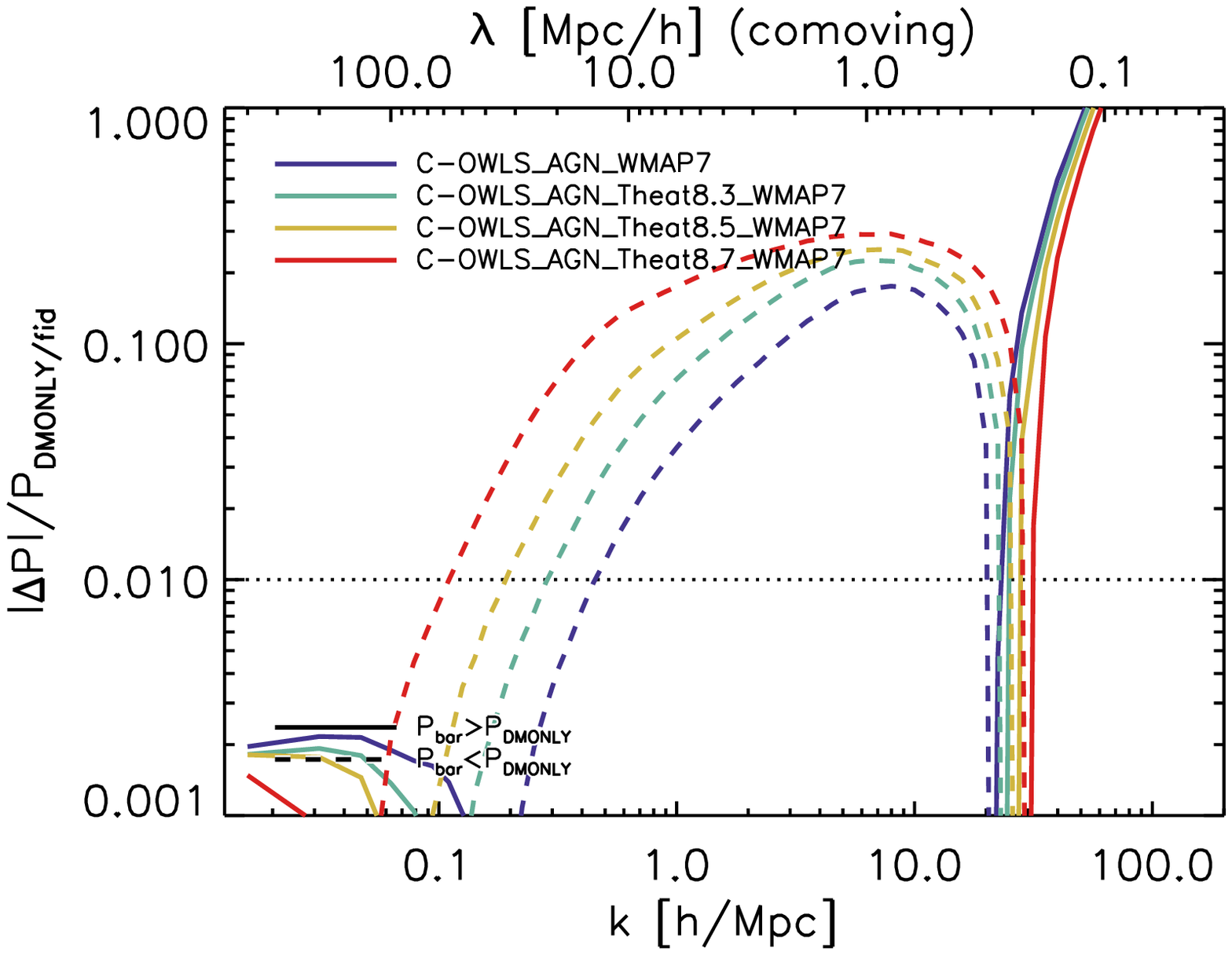}
\includegraphics[width=1.0\columnwidth, trim=27mm 8mm 10mm -4mm]{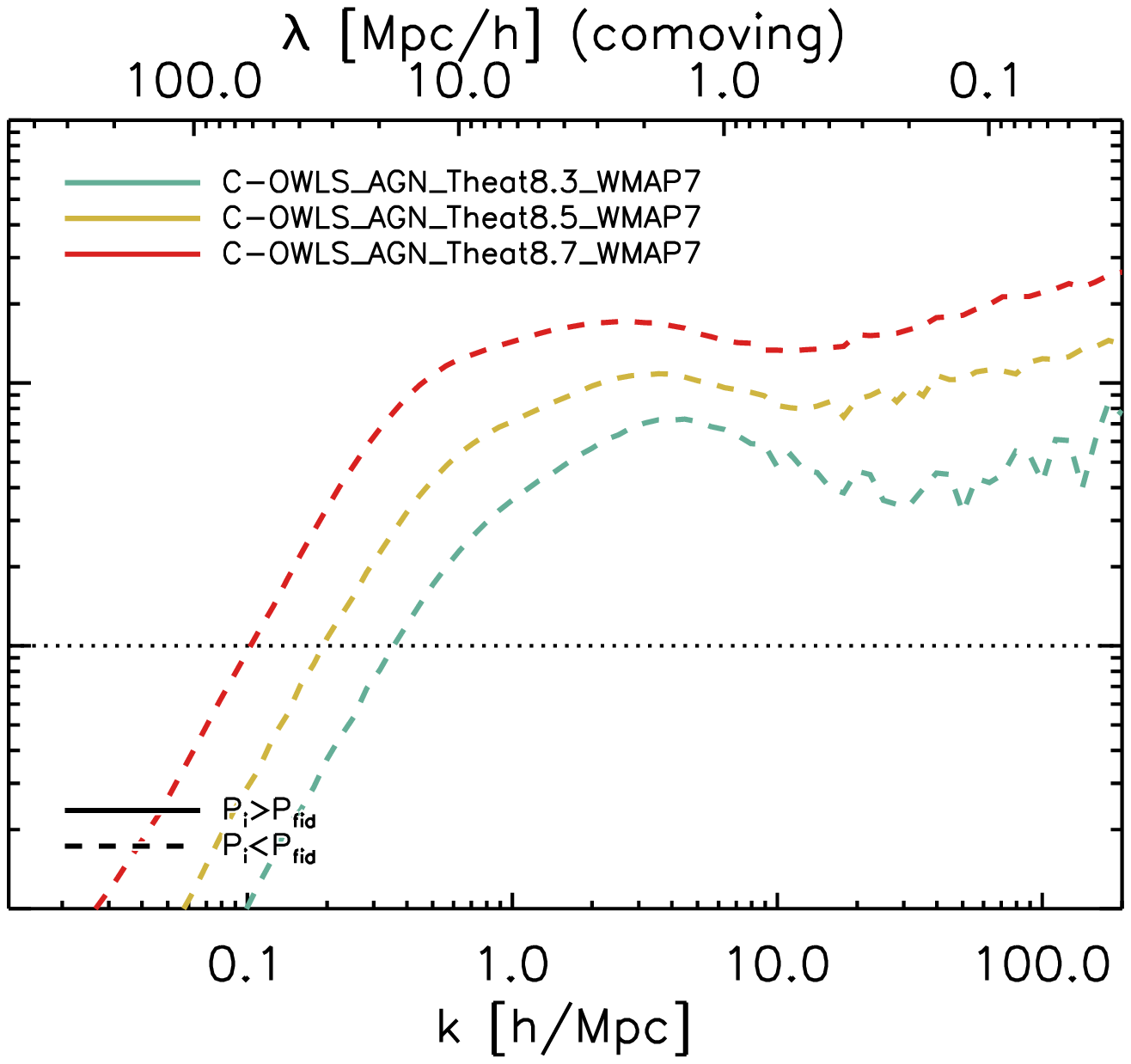}
\caption{The effect of the strength of AGN feedback. Shown are relative power spectra for four WMAP7 cosmo-OWLS simulations with different values for the heating temperature $\Delta T_\mathrm{heat}$ of AGN feedback. From blue to red, the values are $\Delta T_\mathrm{heat}=10^{8.0}\Tunit$ (fiducial), $10^{8.3}\Tunit$, $10^{8.5}\Tunit$ and $10^{8.7}\Tunit$. \textit{Left:} The effect of galaxy formation relative to a dark matter only simulation. Increasing the heating temperature generally increases the effectiveness of AGN feedback. For cosmo-OWLS, the best observational agreement with the properties of groups and clusters is found for $\Delta T_\mathrm{heat}=10^{8.3}\Tunit$ \citep[green line, see][]{LeBrun2014}. \textit{Right:} The effect of increasing the effectiveness of AGN feedback relative to the fiducial cosmo-OWLS AGN simulation ($\Delta T_\mathrm{heat}=10^{8.0}\Tunit$). As the effectiveness of AGN feedback is increased the power is reduced on all scales.}
\label{fig:diff_Theat}
\end{center}
\end{figure*}

AGN feedback -- at least as implemented in these simulations -- is not independent of resolution. This is in part because only haloes resolved with at least 100 dark matter particles are seeded with a black hole, but more importantly, because of the expected lack of convergence when subgrid prescriptions and parameters are held fixed as the physically resolved scales change \citep[see e.g.][]{Bourne2015,McCarthy2017}. We explore the effects of resolution in Appendix~\ref{app:restests}. One of our findings is that lowering the mass resolution causes the large-scale decrease in power due to feedback to diminish, as is also seen in Figure~\ref{fig:diff_oldvsnew}. In addition, the cross-over scale is sensitive to resolution. However, as shown in \citet{McCarthy2017}, the \bahamas{} simulations are in much better agreement with observables than the OWLS AGN model, including the galaxy stellar mass function, the stellar-to-halo mass relation and the hot gas fraction as a function of halo mass for groups and clusters. Despite its lower resolution -- due to its larger box size -- the effects of galaxy formation as seen in the simulation shown in red in Figure~\ref{fig:diff_oldvsnew} should therefore be viewed as the most realistic, at least up to $k\approx 10\kunit$.

Considering the large changes in e.g.\ the galaxy SMF in \bahamas{} versus that in (cosmo\discretionary{-)}{}{)}OWLS, it is perhaps surprising that the relative effect on the matter power spectrum is similar on large scales. As shown by \citet{vanDaalenSchaye2015}, the dominant contribution to the power spectrum on scales $k\lesssim 20\kunit$ comes from groups and clusters ($M\gtrsim 10^{13.5}\munit$), which provide almost all the signal on scales $k\approx 1\kunit$ -- and the properties of groups and clusters are well reproduced by the AGN feedback in both OWLS and \bahamas{} (by construction in the latter). This also explains why the differences between the two simulations shown in Figure~\ref{fig:diff_oldvsnew} are smallest around $k\approx 1\kunit$. We come back to this point in \S\ref{subsec:complit}.

Finally, we consider the power spectra of the different mass components from the most realistic simulations in VD11 and the current set in Figure~\ref{fig:components}: cold dark matter (blue), gas (yellow), stars (red), and for the latter simulation, neutrinos (purple). The OWLS AGN components are shown as dashed lines, the \bahamas{} components as solid. All components are normalized to the total mass in the simulated volume, and the total power is shown in black. To allow for easier comparison we show the dimensionless power for each, $\Delta^2(k)\equiv k^3P(k)/(2\pi^2)$, with the VD11 power spectra renormalized to the Planck 2015 cosmology through a factor $D(\mathrm{P15})^2/D(\mathrm{W7})^2$, where $D(\mathrm{X})$ is the linear growth factor for cosmology X. Significant differences between the power spectra for the two simulations remain in spite of this renormalization, which are partly due to variations in galaxy formation, and partly due to differences in resolution (and on the very largest scales, box size).

The clustering in all components in \bahamas{} is stronger on almost all scales, with the exception of the gas for $k\lesssim 8\kunit$. The differences are especially large on small scales, $k\gtrsim 20\kunit$ -- but these, unlike the changes seen for $k<10\kunit$, are almost entirely driven by the difference in resolution. Furthermore, the larger box size contributes to a slightly increased clustering in all components due to the presence of more massive objects. The effects of the differences in galaxy formation are to decrease the clustering of gas on scales $k\lesssim 10\kunit$, and to increase the clustering of stars. This is expected, as the weaker SN feedback in \bahamas{}, relative to (cosmo\discretionary{-)}{}{)}OWLS, allows more gas to cool and form stars. At fixed box size and resolution, the main effects of increasing the strength of AGN feedback are to suppress the clustering of gas on scales $0.5\lesssim k\lesssim 10\kunit$, and to slightly decrease the clustering of stars on all scales (not shown here).

\subsection{Variations in AGN feedback}
\label{subsec:varAGN}
By comparing power spectra of cosmo-OWLS simulations that use different AGN heating temperatures but are otherwise identical, we can examine how the strength of AGN feedback impacts the total matter power spectrum. We do so in Figure~\ref{fig:diff_Theat}, where the AGN heating temperature increases from blue (fiducial) to red. In the left-hand panel, we consider the effects relative to the dark matter only counterpart, while in the right-hand panel we compare the simulations to the one with the lowest (the fiducial) heating temperature of $10^8\Tunit$. Increasing the temperature increases the duty cycle of the AGN, as it takes longer for the black holes to reach the threshold energy for feedback, but the individual events are more powerful. The latter effect dominates, making the impact of galaxy formation larger when the heating temperature goes up and more gas is being blown out of the galaxies.

Looking at the left-hand panel of Figure~\ref{fig:diff_Theat}, we see that the matter power spectrum is indeed more suppressed as the heating temperature increases. The differences caused by increasing the heating temperature can be better appreciated when looking at the changes relative to the fiducial simulations, as in the right-hand panel of Figure~\ref{fig:diff_Theat}. Here we see that increasing the heating temperature has the largest relative impact for $k\gtrsim 1\kunit$, increasing the small-scale suppression by an amount nearly independent of scale. Larger heating temperatures suppress the power out to larger scales.

The fiducial \bahamas{} simulations, which have a heating temperature of $10^{7.8}\Tunit$ but also a larger reservoir of cold gas available for accretion, agree very well with the results for $10^8\Tunit$ (blue) on scales $k>1\kunit$, but are closer to those for $10^{8.3}\Tunit$ (cyan) for $0.1<k<1\kunit$.

\subsection{Variations in cosmology}
\label{subsec:varcosm}
\begin{figure}
\begin{center}
\includegraphics[width=1.0\columnwidth, trim=16mm 8mm 10mm -4mm]{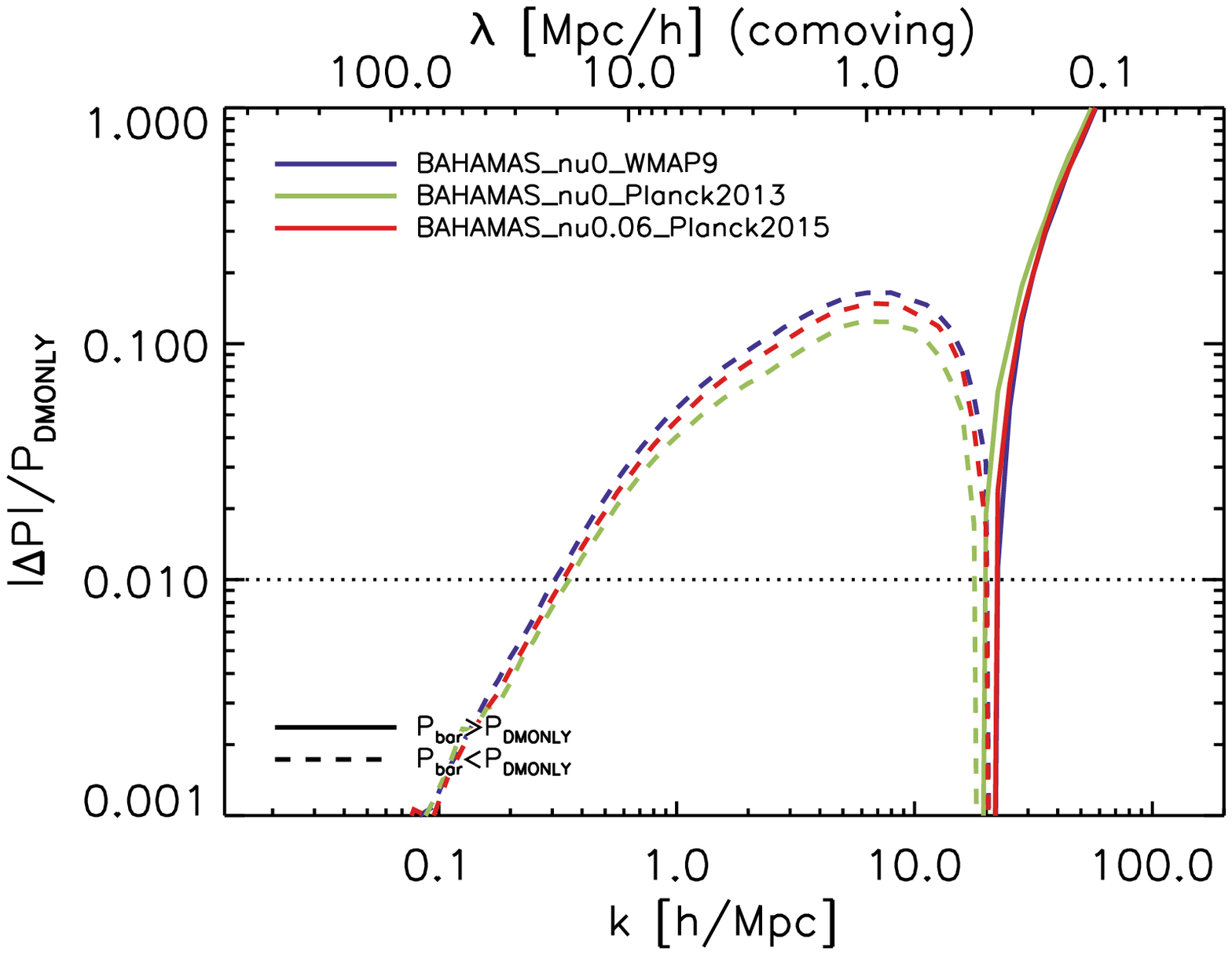}
\caption{The impact of changing the cosmology on the relative effect of galaxy formation on the matter power spectrum. From blue, to green, to red the relative effect of assuming a WMAP9, Planck 2013 or Planck 2015 cosmology are shown. The latter assumes a non-zero neutrino mass, but as Figure~\ref{fig:diff_nu} will show, this has little to no impact on the curve shown here. While the choice of cosmology -- at least in the range probed here -- does not affect the largest or smallest scales probed, there is a small but significant change in the strength of the suppression of power on scales $1\lesssim k\lesssim 10\kunit$, with WMAP9 predicting a larger suppression than Planck.}
\label{fig:diff_cosm}
\end{center}
\end{figure}

VD11 showed that the effect of galaxy formation was almost completely independent of (reasonably small) changes in cosmology by comparing the results a WMAP3 and a WMAP7 AGN simulation. In Figure~\ref{fig:diff_cosm} we conduct a similar investigation by comparing the baryonic effects for \bahamas{} simulations with a WMAP9 (blue), Planck 2013 (green) and Planck 2015 (red) cosmology. The latter is the only one that includes a non-zero neutrino mass, but as we will show shortly, the impact of this is negligible. While the differences are generally small, significant shifts (up to $4\%$ in absolute or $30\%$ in relative terms) in the suppression for $1<k<10\kunit$ can be seen here. The largest difference is between WMAP9 and Planck 2013, which are known to be in tension. However, the distance in parameter space between these cosmologies is similar to that of WMAP3 and WMAP7 (and significantly smaller for $\sigma_8$), so there is \textit{a priori} no reason to expect a larger change than found by VD11 for the cosmologies shown here. The fact that the simulations have a $64\times$ larger volume than those of VD11, and therefore include more massive objects, may well play a role here. The Planck 2015 results are in between those of WMAP9 and Planck 2013, as one might expect from Table~\ref{tab:cosms}.

Based on the limited cosmologies tested here, we cannot exclude the possibility that cosmological parameter variations of the same order could produce larger shifts in the effect of galaxy formation on matter clustering. However, variations in cosmology are separable from the effects of galaxy formation for at least some parameters, within current uncertainties, as \citet{Stafford2019} have recently shown for a running scalar spectral index, and Pfeiffer et al. (in prep.) will demonstrate for dynamical dark energy models.

\begin{figure*}
\begin{center}
\includegraphics[width=1.0\columnwidth, trim=16mm 8mm 21mm -4mm]{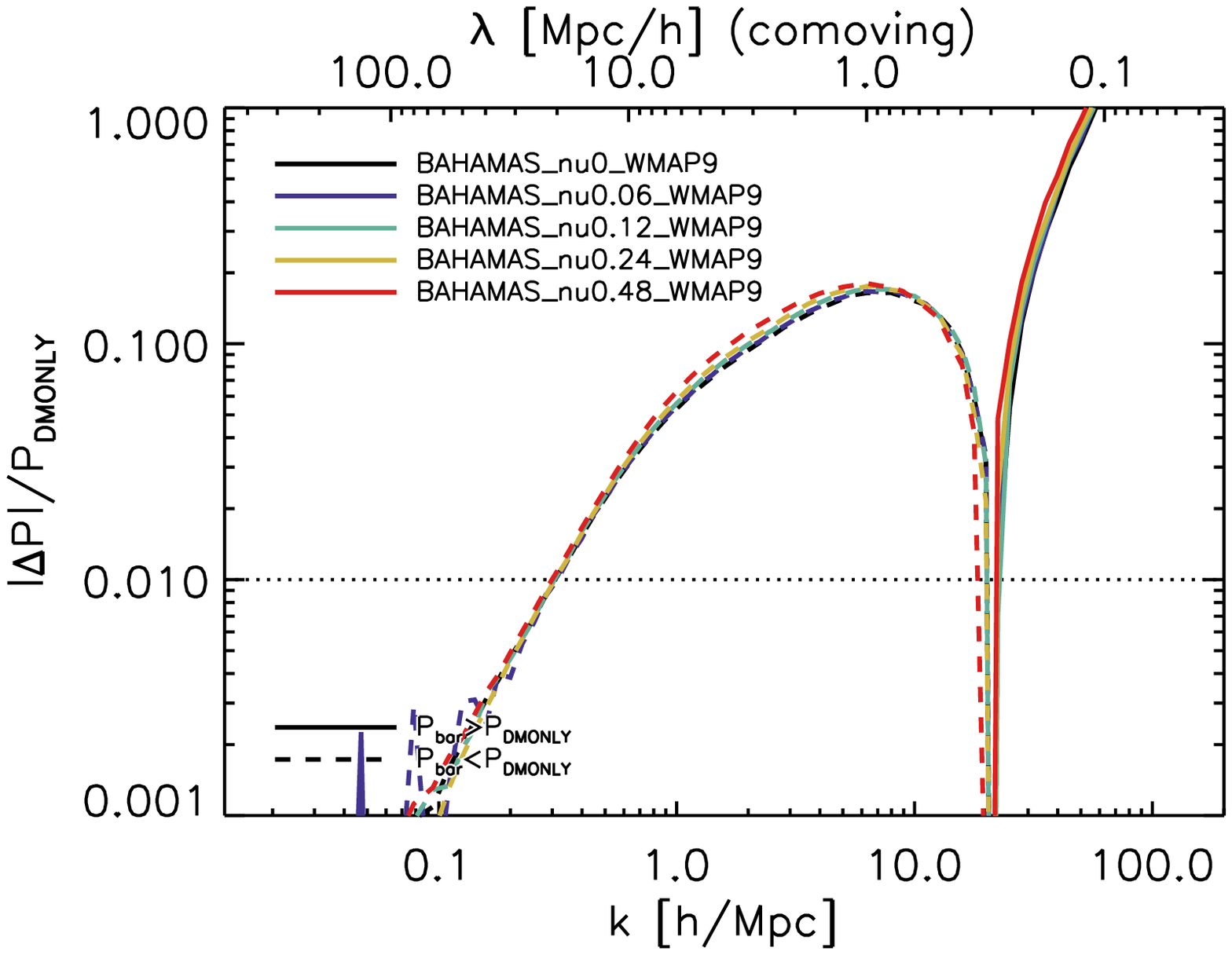}
\includegraphics[width=1.0\columnwidth, trim=27mm 8mm 10mm -4mm]{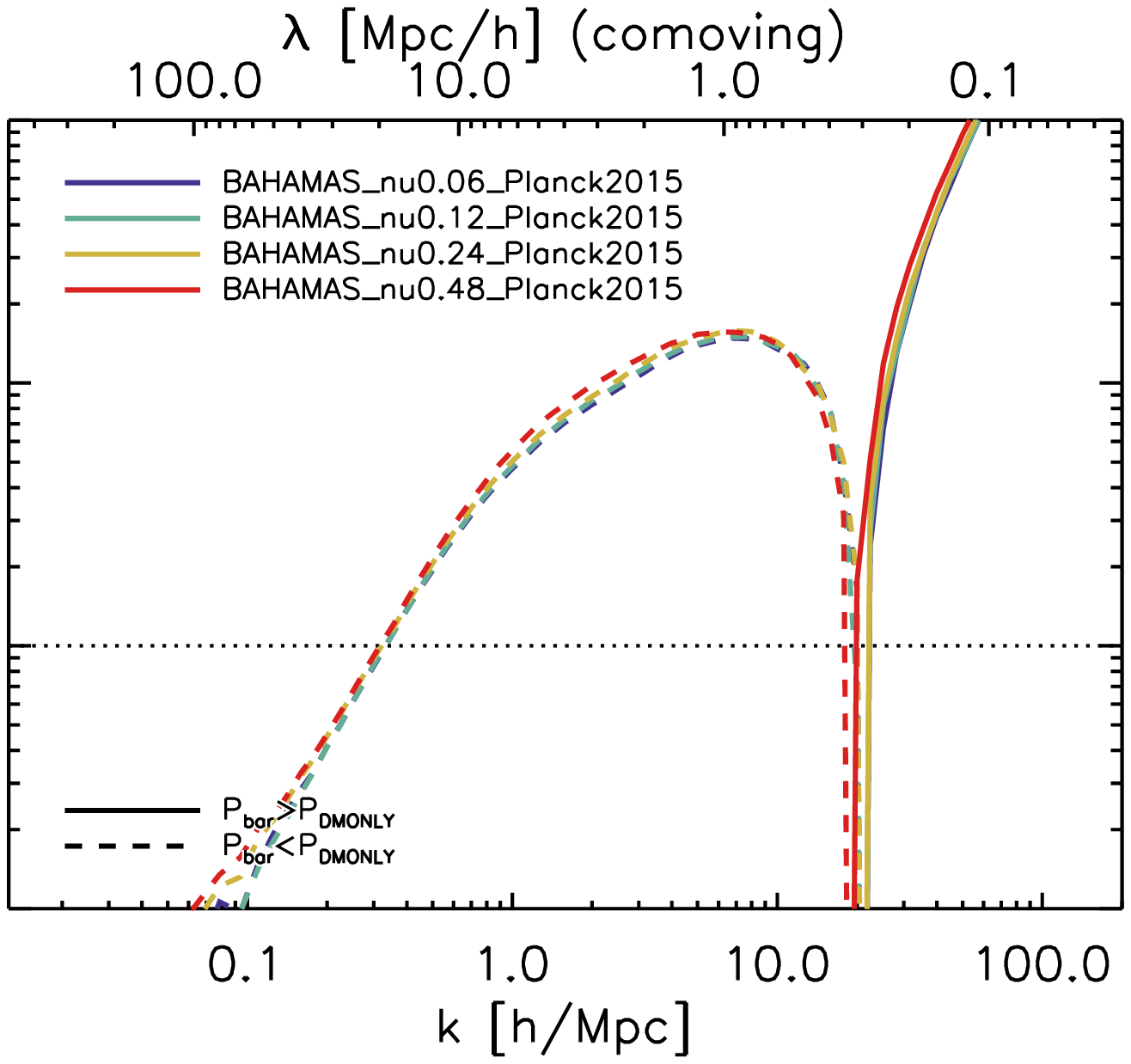}
\caption{The role of the neutrino mass in changing the relative effect of galaxy formation on the matter power spectrum. Each line corresponds to the absolute relative difference in power for a set of simulations with a different neutrino mass. From blue to red: $M_\nu=0.06\nunit$, $0.12\nunit$, $0.24\nunit$ and $0.48\nunit$. The left panel additionally shows $M_\nu=0$ in black. \textit{Left:} For the WMAP9 cosmology, the neutrino mass is increased at fixed cosmology, so effectively more dark matter is replaced by neutrinos as the neutrino mass increases. The effects of neutrinos and galaxy formation are almost completely independent, in particular for low neutrino masses, and so can be modelled separately. Note that in the $0.06\nunit$ simulation, some event between $z=2$ and $z=1.75$ causes the AGN simulation to go out of sync, which produces larger than expected differences even at $z=0$ (large-scale fluctuations). \textit{Right:} For a Planck 2015 cosmology, the neutrino mass is increased while allowing the other cosmological parameters to vary in such a way as to provide the best fit to CMB data. Evidently, this does not affect the relative effect of galaxy formation on the matter power spectrum, except perhaps at the very largest scales probed ($k\approx 0.1\kunit$).}
\label{fig:diff_nu}
\end{center}
\end{figure*}

In Figure~\ref{fig:diff_nu} we test the dependence of the effects of galaxy formation on another aspect of cosmology, namely the neutrino mass. In the left-hand panel we compare the WMAP9 \bahamas{} run with different neutrino masses, and in the right-hand panel we do the same for Planck 2015. Both could affect the power spectrum in different ways, as the cosmology is held fixed with increasing neutrino mass for the former but not the latter (see \S\ref{subsec:sims} and \citealt{McCarthy2018}). From blue to red, the total neutrino mass increases from $M_\nu=0.06\nunit$ to $0.12\nunit$, $0.24\nunit$ and $0.48\nunit$. The left panel additionally shows $M_\nu=0$ in black. We stress that we consider the power spectrum in each simulation relative to a dark matter only simulation with the \emph{same} neutrino mass, so as to scale out the direct effect of adding neutrinos on the clustering of matter.

As \citet{Mummery2017} previously showed for \bahamas{}, baryons act nearly independently of the neutrino mass -- in line with the findings of \citet{Mead2016} -- and the results shown here confirm this: the impact of galaxy formation on the power spectrum is nearly unchanged in all cases. Looking at the left panel in more detail, we see that increasing the neutrino mass from zero to $0.06\nunit$ has almost no effect on the \emph{relative} power spectrum, but the higher the neutrino mass, the stronger the change.\footnote{For the power spectrum, what matters most is the mass of the most massive neutrino. This is not so different for $M_\nu=0.06$ and $0.12\,\mathrm{eV}$ (assuming the normal hierarchy), so the neutrino power spectra will also be similar, though the normalization of the total power spectrum changes (and for the Planck 2015 cosmology there are additional parameter changes).} The shift in suppression is largest around $k\approx 2\kunit$, which, according to the results of \citet{vanDaalenSchaye2015}, is where the power spectrum is dominated by groups and clusters. Additionally, the WMAP9 $0.06\nunit$ simulation shows some unique features around $k\approx 0.1\kunit$, but these are likely numerical in origin, e.g.\ due to small shifts in positions.

Looking at the right-hand panel, we can draw the same conclusions, except that there is very slightly more evolution with neutrino mass on the very largest scales probed ($k\approx 0.1\kunit$). This is likely due to the large change in $\sigma_8$ between the Planck 2015 simulations with the smallest and largest neutrino mass, necessary in order to maintain agreement with CMB data.

\subsection{Back-reaction on CDM}
\label{subsec:backreaction}
As shown by previous authors (e.g.\ VD11, see also \S\ref{subsec:complit}), galaxy formation and its associated feedback events do not only change the clustering of gas and stars but also of the cold dark matter, an effect dubbed simply the back-reaction. To consider this back-reaction we compare the CDM-only power spectrum of hydrodynamical simulations to the matter power spectrum in the DMONLY simulation, multiplying the former by a factor $[(\Omega_\mathrm{c}+\Omega_\mathrm{b})/\Omega_\mathrm{c}]^2=[(\Omega_\mathrm{m}-\Omega_\nu)/(\Omega_\mathrm{m}-\Omega_\mathrm{b}-\Omega_\nu)]^2$ to compensate for the difference in normalization.

\begin{figure}
\begin{center}
\includegraphics[width=1.0\columnwidth, trim=16mm 8mm 10mm -4mm]{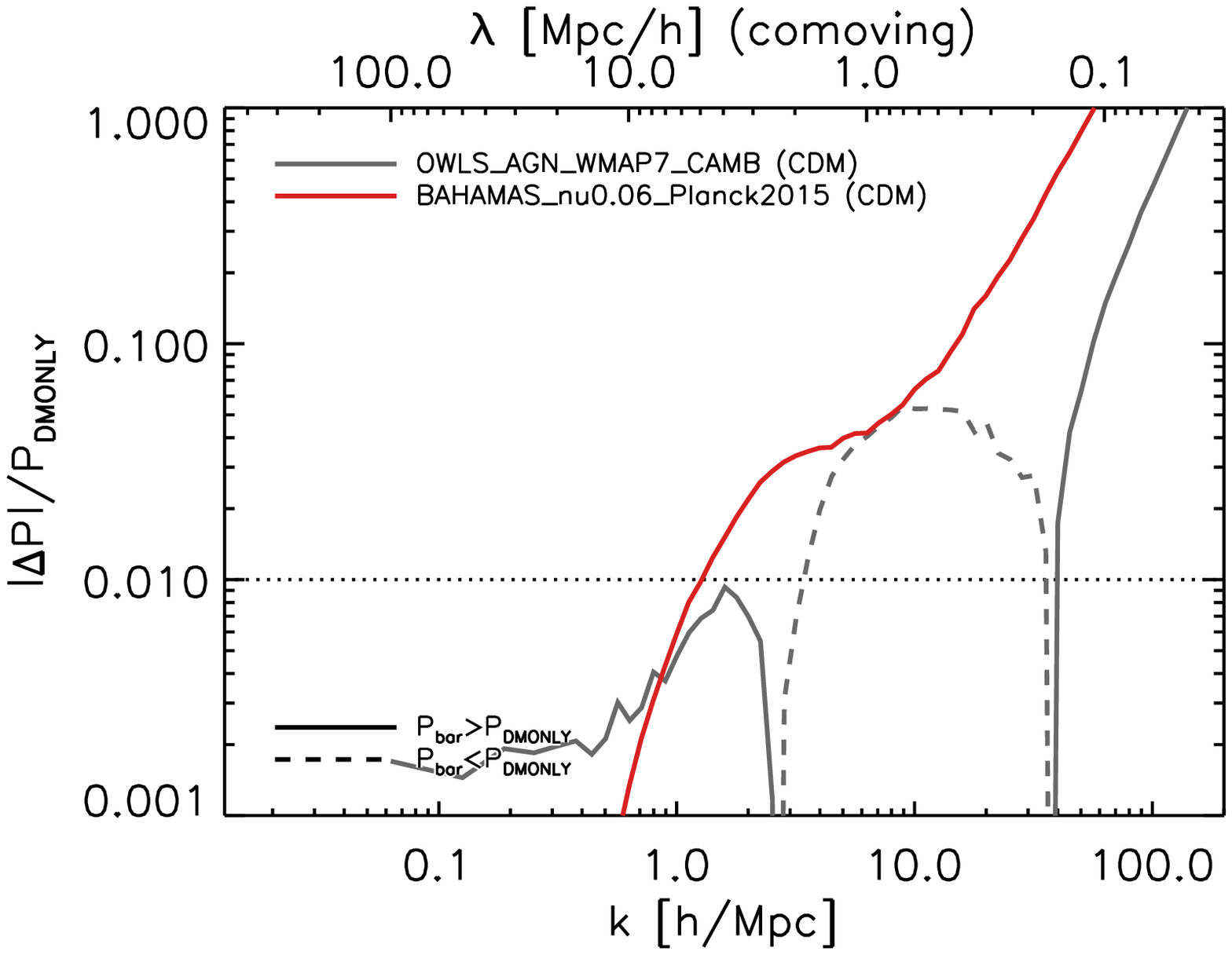}
\caption{The back-reaction of galaxy formation on the power spectrum of cold dark matter. The back-reaction of OWLS AGN presented in VD11 is shown in grey while the back-reaction of the current most realistic simulation is shown in red. The generally less effective AGN feedback in \bahamas{} means that the reduction in CDM power relative to dark matter only around $k\sim 10\kunit$ is no longer seen here. This makes the back-reaction somewhat easier to model as the increase in CDM power towards smaller scales is associated with halo contraction.}
\label{fig:back_oldvsnew}
\end{center}
\end{figure}

We first compare the back-reaction for the most realistic simulation in the current set to that of VD11, in Figure~\ref{fig:back_oldvsnew}. The results of VD11, shown in grey, predicted that the CDM power spectrum around $k=10\kunit$ is suppressed by several percent, in line with halo expansion, while being enhanced on the smallest scales probed, in line with halo contraction. They also saw a corresponding increase in power of up to $1\%$ for $k\approx 1\kunit$. The results for \bahamas{}, in red, are different: the power in cold dark matter is increased on all scales, relative to a dark matter only simulation, monotonically increasing towards smaller scales. Interestingly, agreement is found only for $k\approx 1\kunit$, just as for the total matter power spectrum, which happens to correspond to scales where the relative contribution of groups and clusters is maximized.\footnote{Here, too, we note that compensating for the large-scale $\sim 0.1\%$ offset in power seen for OWLS would bring the simulation into slightly better agreement with \bahamas{} on the largest scales. For more information we again refer to Appendix~\ref{app:2fluid}.}

\begin{figure*}
\begin{center}
\includegraphics[width=1.0\columnwidth, trim=16mm 8mm 21mm -4mm]{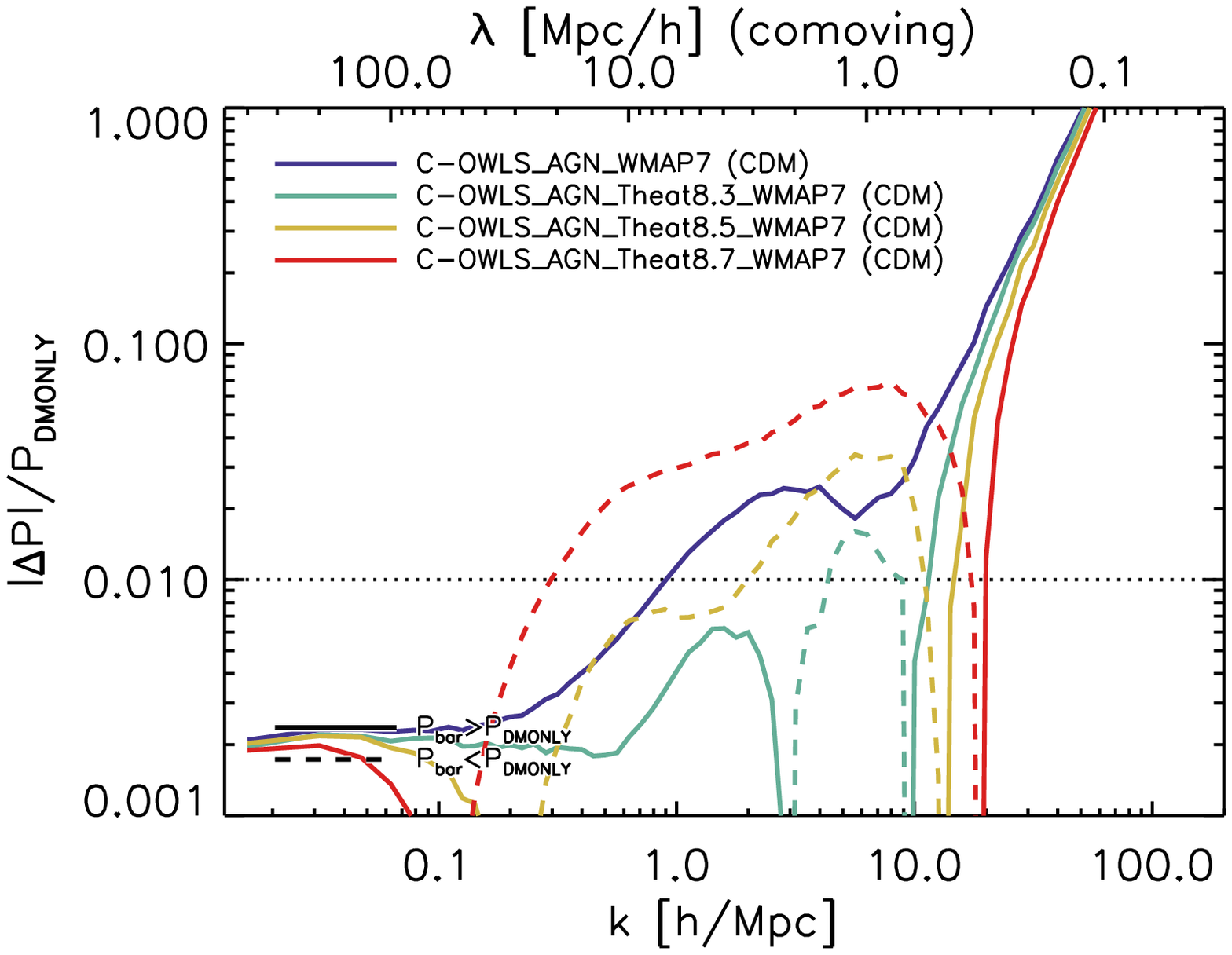}
\includegraphics[width=1.0\columnwidth, trim=27mm 8mm 10mm -4mm]{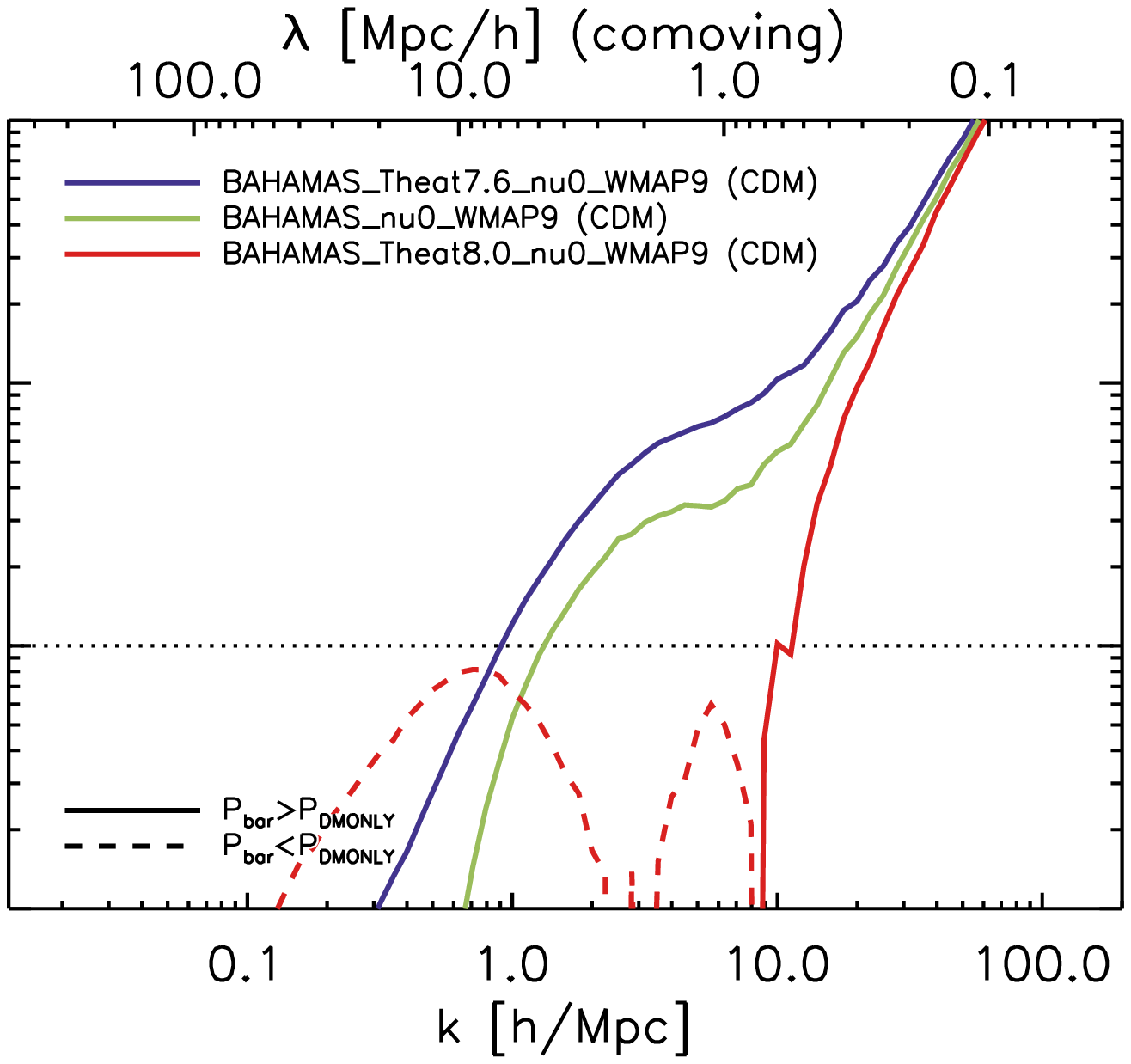}
\caption{The effect of the strength of AGN feedback on the back-reaction on the CDM power spectrum. \textit{Left:} As in the left panel of Figure~\ref{fig:diff_Theat}, we consider a set of four cosmo-OWLS simulations that differ only in their AGN heating temperature. From blue to red, the values are $\Delta T_\mathrm{heat}=10^{8.0}\Tunit$ (fiducial), $10^{8.3}\Tunit$, $10^{8.5}\Tunit$ and $10^{8.7}\Tunit$. At higher temperatures (i.e.\ more effective feedback, towards red), an increasingly strong reduction in CDM power on scales $k\lesssim 10\kunit$ can be seen, due to gas ejection causing the halo to expand relative to a dark matter only scenario. \textit{Right:} The dependence of the back-reaction on the AGN heating temperature in \bahamas{}. From blue to red, the values are $\Delta T_\mathrm{heat}=10^{7.6}\Tunit$, $10^{7.8}\Tunit$ (fiducial) and $10^{8.0}\Tunit$. As the strength of feedback increases, the enhancement of power on large scales diminishes. Because of the lower effectiveness of stellar feedback in \bahamas{} versus cosmo-OWLS, a suppression in the CDM power spectrum is seen already for $\Delta T_\mathrm{heat}=10^8\Tunit$.}
\label{fig:back_Theat}
\end{center}
\end{figure*}

In part, the differences are due to a change in resolution: by lowering the mass resolution, the effectiveness of AGN feedback is decreased and the cross-over scale (if there is one) moves to $\sim 2\times$ larger scales, leaving less room for suppression. At the same time, two physical changes can play a significant role as well: the larger box size provides more massive objects (and lowers the effect of cosmic variance, see \S\ref{subsec:cosvar}), and the lower AGN heating temperature in \bahamas{} allows for more halo contraction. By comparing with other power spectra in our library (not shown), we can isolate the effect of running the same simulation with a different minimum BH seed mass, a different particle resolution and/or a different box size. We find that, at the fiducial heating temperature, increasing the minimum seed mass has a very minor effect, while decreasing the particle resolution at fixed box size leads to a positive back-reaction (i.e.\ contraction) on all scales, though generally weaker than that seen in \bahamas{}. Increasing the box size from $100$ to $400\runit$ as well further increases the CDM clustering on scales $k \lesssim 8\kunit$, but by at most $1\%$. From there, switching to the \bahamas{} model adds another $\leq 2\%$ to the back-reaction on scales $2\lesssim k \lesssim 10\kunit$. Therefore, going from OWLS to \bahamas{} the particle resolution at fixed box size has the largest effect on the back-reaction, at least for $k\gtrsim 3\kunit$. For the total signal, this is only the case for $k\gtrsim 10\kunit$, meaning the back-reaction has a relatively strong dependence of the back-reaction on box size and resolution.

In Figure~\ref{fig:back_Theat} we show the dependence of the back-reaction on the AGN heating temperature for cosmo-OWLS (left) and \bahamas{} (right). Focussing first on the left-hand panel, we see that the simulation with the fiducial heating temperature of $\Delta T_\mathrm{heat}=10^8\Tunit$ shows only enhancement, similar to \bahamas{} in Figure~\ref{fig:back_oldvsnew}, in line with our findings above. Increasing the heating temperature (cyan, yellow and red) suppresses the CDM power spectrum on increasingly large scales, as more and more material is blown to large scales by feedback, causing the outer halo to expand (or contract less) relative to the dark matter only counterparts.

Looking at the panel on the right, we see that increasing the heating temperature of \bahamas{} from the fiducial value of $10^{7.8}\Tunit$ (green) to $10^8\Tunit$ (red) is enough to cause a small but significant suppression of the CDM power for all scales $k\lesssim 8\kunit$ (not shown). Lowering the heating temperature by the same factor (blue) instead allows the CDM power spectrum to be enhanced on all scales.

We have checked that the impact of changing the cosmology or neutrino mass of the simulations on the back-reaction is comparable to that on the total power, and we do not show it here.

\subsection{Evolution}
\label{subsec:redshift}
In this section we examine how the effects of galaxy formation on the total and CDM power spectra evolve in the most realistic simulation of the current set. We first consider the total matter, in Figure~\ref{fig:diff_z}. From blue to red, we show the relative power spectrum for $z=3$ down to $z=0$. Note that below $z=0.5$ the output frequency is doubled. The evolution of the impact of galaxy formation on the power spectrum is largely monotonic with time, with the large-scale suppression increasing down to redshift zero while the cross-over scale moves to smaller (co-moving) scales (the down-turn for $k>20\kunit$ is numerical and should be ignored). The exception is that below $z=0.5$, the suppression on scales $0.8\lesssim k\lesssim 8\kunit$ diminishes somewhat, due to the growth of the most massive haloes. In simulations with weaker AGN feedback (not shown here), the late-time decrease in suppression happens at the same redshifts and scales as shown in Figure~\ref{fig:diff_z}, but is more substantial. Conversely, when the feedback is stronger than in BAHAMAS\_nu0.06\_Planck2015, it is able to counter the increased dark matter clustering somewhat and the suppression of power diminishes less or not at all, although even for the highest heating temperatures in our simulation set ($10^{8.7}\,\mathrm{K}$), for $z<0.5$ the suppression does not proceed to grow on these scales either.

Comparing our result to those of VD11 for OWLS AGN (their Figure~8), we see that they are very similar, even down to the slight decrease in suppression for $z<0.5$. The same is true for the evolution of the matter power spectrum of other simulations in our set containing AGN.\footnote{We note here that other authors find very different results for the redshift evolution of the relative matter power spectrum. We plan to explore these differences in future work.}

In Figure~\ref{fig:back_z} we consider the evolution of the back-reaction. The effect of galaxy formation on the CDM clustering is roughly constant for $k\approx 10\kunit$. On smaller scales, the enhancement relative to dark matter only diminishes somewhat, but this may be an effect of resolution, like the downturn seen on these scales (which should be disregarded for that reason). The evolution is particularly strong on scales $k\approx 2\kunit$, again corresponding to the scales dominated by groups and clusters.

\subsection{Interplay between stellar and AGN feedback}
\label{subsec:SNAGN}
Interesting differences between the OWLS/cosmo-OWLS and \bahamas{} simulations arise because of the interplay between stellar and AGN feedback. With the set of simulations presented in this work, we can examine the impact that changes in stellar feedback have on the effectiveness of AGN feedback.

We first consider the OWLS AGN model in addition to two variations that were not included in VD11, in Figure~\ref{fig:diff_SNAGN}. In the model shown in green, the accretion model's dependence on gas density is changed. Comparing the results for this model to the fiducial one, in blue, we see that this has almost no impact on the effectiveness of AGN feedback. However, if we now also disable stellar feedback (in red), the power spectra change drastically. Without winds driven by stellar feedback, the (self-regulated) AGN has a larger reservoir of cold gas at its disposal, and both accretes and heats far more gas (leading to much more massive supermassive black holes as well). Consequently, it is able to suppress the matter clustering to a much larger degree, largely compensating for the strong increase in star formation in low-mass galaxies. This agrees with \citet{BoothSchaye2013}, who used OWLS to demonstrate that stellar feedback diminishes the effectiveness of AGN feedback. This result also shows why the \bahamas{} heating temperature needed to go down to match observations after reducing the stellar feedback: less gas ejected by stars means more gas is available to the AGN, making it more effective at a fixed heating temperature.

Stellar feedback primarily quenches star formation in low-mass galaxies (above the knee of the SMF), and AGN primarily in high-mass galaxies (below the knee). It therefore seems reasonable to assume that if the strength of the AGN feedback is such that it quenches high-mass galaxies in such a way as to agree with observations, then the predictions for its effects on the matter power spectrum are realistic as well. These results show that this is not the case: stellar feedback determines what gas remains in a galaxy, to be heated by an AGN once the galaxy is sufficiently massive to host it \citep[e.g.][]{Bower2017}. It is important, in order to predict the right amount of power suppression due to AGN, that the strength of stellar feedback is correctly calibrated to observations as well. Calibrating only AGN feedback using hot gas fractions in addition to the high-mass SMF may or may not be enough; further research is needed for this.

\subsection{Effects of cosmic variance}
\label{subsec:cosvar}
The effects of galaxy formation on the matter power spectrum depend on mass and environment, with the large-scale suppression of power being dominated by the strongest AGN. Not only are these AGN only found in very overdense environments, but the descendants of the haloes that host them are themselves the dominant contributor to the power on all scales $k\lesssim 10\kunit$, as shown by \citet{vanDaalenSchaye2015}. Since high-mass haloes are rare, the predicted suppression of power due to galaxy formation could be susceptible to cosmic variance and may depend on the size of the volume probed.

Several approaches can be taken to assess the importance of cosmic variance. Ideally, one would simulate a larger volume at fixed resolution and check for convergence, but this is often computationally prohibitive. Instead, one could take the reverse approach and compare the results of the fiducial simulations to that of smaller volumes, although drawing conclusions from this about the larger volume is difficult, even more so if the smaller volumes do not probe linear scales and/or do not contain highly overdense or underdense regions. Recently, \citet{Chisari2018} avoided the latter issue by considering instead eight sub-volumes drawn from their fiducial $100\runit$ simulation, finding significant variation between them, depending on whether a massive object was present in a sub-volume. A similar study was performed by \citet{Peters2018}, who performed $60$ zoom-in simulations of randomly selected sub-volumes from a $(3.2\,\mathrm{Gpc})^3$ parent volume, each $40.16\runit$ on a side and resimulated using the hydrodynamical code and resolution of \bahamas{}. They found large variation between the predicted relative power in each sub-volume, although the median effect of galaxy formation on the power spectrum provides an excellent match to that of the full $400\runit$ \bahamas{} simulations.

\begin{figure}
\begin{center}
\includegraphics[width=1.0\columnwidth, trim=16mm 8mm 10mm -4mm]{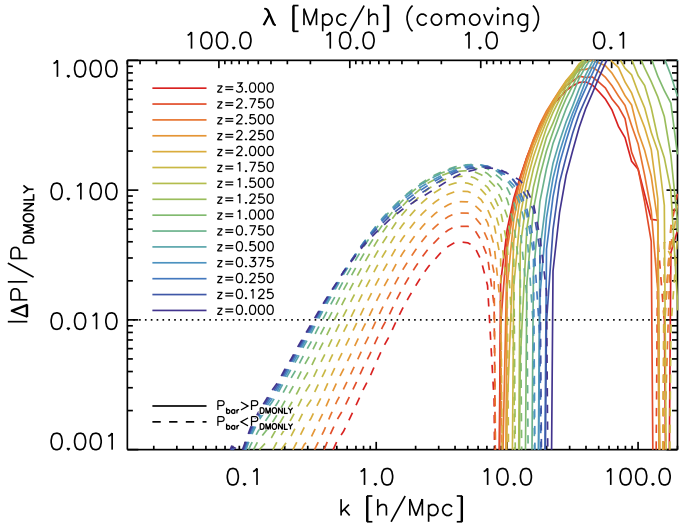}
\caption{The redshift evolution of the effect of galaxy formation on the power spectrum for the most realistic simulation in the current set, BAHAMAS\_nu0.06\_Planck2015. The large-scale reduction in power due to AGN feedback steadily increases from high redshift down to $z\approx 0.5$, while the shift from a power reduction to a power increase (relative to dark matter only) keeps moving to smaller (co-moving) scales all the way down to redshift zero. The downturn seen on the smallest scales ($k\gtrsim 40\kunit$) for the highest redshifts is due to a lack of resolution.}
\label{fig:diff_z}
\end{center}
\end{figure}

Here, we avoid some of the issues mentioned above by taking a different approach, instead performing two additional runs of a calibrated WMAP9 \bahamas{} simulation at fixed box size and resolution, but with different initial conditions. All resimulations use the same subgrid parameters and differ only in the random phases of their initial conditions. Matching dark matter only runs were also performed. The results are shown in Figure~\ref{fig:diff_cosmicvar}: all three simulations predict a nearly identical effect of galaxy formation on the total matter power spectrum, suggesting that the effects of cosmic variance can be ignored for these simulations. We note that this does not apply to the power spectra themselves: while not shown here, the matter power spectra of each resimulated volume show random variations which can reach $\sim 10\%$ even on large scales -- however, the \emph{ratio} of power spectra with and without baryons in the same volume is converged to high precision.

The $400\runit$ simulations shown here probe well into the linear regime. Still, one might wonder whether the power (and the effect of galaxy formation) is not suppressed on the largest scales probed due to additional linear modes that cannot be included. While it is computationally prohibitive to perform a much larger volume simulation at the same resolution to check this, we believe this is unlikely to make a significant difference, at least for the relative effect of galaxy formation. The difference in power is $\lesssim 0.1\%$ on scales larger than $k=0.1\kunit$, where the power is probed by hundreds of statistically independent modes already for a $400\runit$ volume. Any changes in the effects of galaxy formation due to including additional modes are therefore expected to be $\lesssim 0.1\%$ as well. Extremely rare overdensities may still be missed and could play a role for the power -- though, given that the effectiveness of AGN feedback drops off for the most massive haloes in the current volume, and since the fraction of the total mass in these haloes is very small, we don't expect these to play a significant role for the relative power spectrum.

Given this convergence, the ratio of hydrodynamical and dark matter only \bahamas{} simulations may be used to accurately correct matter power spectra from large-volume dark matter only runs, emulator predictions or analytical power spectra up to $k\approx 10\kunit$, at least for matching cosmologies, leaving only the uncertainty in galaxy formation to be accounted for.

\begin{figure}
\begin{center}
\includegraphics[width=1.0\columnwidth, trim=16mm 8mm 10mm -4mm]{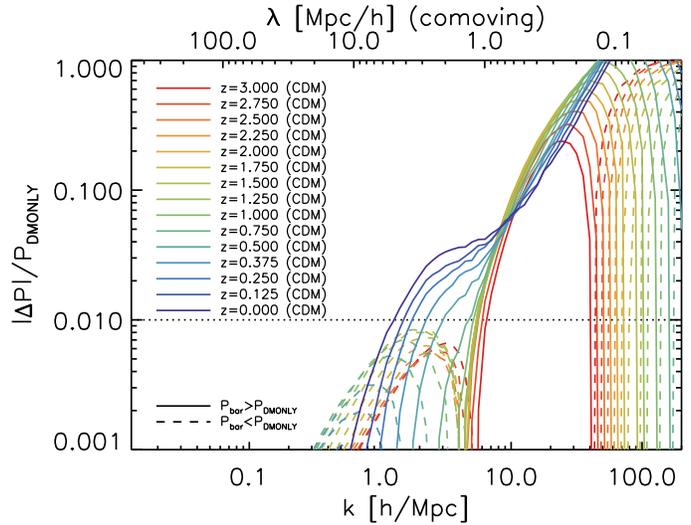}
\caption{As Figure~\ref{fig:diff_z}, but showing the redshift evolution of the back-reaction on the CDM power spectrum. Down to $z\approx 1$, the cold dark matter shows increased clustering for $k\gtrsim 5\kunit$ and slightly decreased clustering on larger scales. However, at lower redshifts the dark matter is able to contract on larger scales as well, increasing clustering on all scales $k\lesssim 8\kunit$.}
\label{fig:back_z}
\end{center}
\end{figure}

\subsection{Comparison to power spectra from the literature}
\label{subsec:complit}
In Figure~\ref{fig:diffback_literature} we compare the relative effect of galaxy formation on the total matter (left) and CDM power spectrum (right) of the most realistic \bahamas{} simulation (red) and power spectra from the literature. Included in the comparison are OWLS AGN (grey, VD11), EAGLE \citep[purple,][]{Hellwing2016}, Illustris \citep[blue,][]{Vogelsberger2014a}, IllustrisTNG \citep[cyan and green,][]{Springel2018} and Horizon-AGN \citep[orange,][]{Chisari2018}. All simulations considered here contain both stellar and AGN feedback, but employ different subgrid implementations, box sizes, resolutions and hydrodynamics solvers. We note that to reduce visual noise, we applied the re-binning mentioned in \S\ref{subsec:pow} to these power spectra as well where necessary, imposing a minimum bin size of $0.05\,\mathrm{dex}$ in $k$. For Horizon-AGN, data was not available on scales smaller than $k=32\kunit$.

As the left-hand panel of Figure~\ref{fig:diffback_literature} shows, not all simulations are in quantitative agreement, and they certainly do not agree at the level required for Stage IV weak lensing surveys (e.g.\ Euclid, LSST and WFIRST) -- however, there is qualitative agreement. All simulations shown here agree that the total matter power spectrum is suppressed on all scales $1<k<20\kunit$ relative to dark matter only, with a $10-30\%$ suppression at $k=10\kunit$. If we assume that the large-scale $\approx 0.7\%$ offset in power in Horizon-AGN is due to differences in the initial conditions as explored in Appendix~\ref{app:2fluid}, then compensating for these differences (not shown) results in a prediction that is close to that of Illustris TNG100, and therefore qualitative agreement on all scales $k\lesssim 10\kunit$. We note, however, that \citet{Huang2019} showed that the MassiveBlack-II simulation \citep[][, not shown here]{Khandai2015,Tenneti2015}, which also contains both stellar and AGN feedback, predicts suppression only for $k\lesssim 2\kunit$.

\begin{figure}
\begin{center}
\includegraphics[width=1.0\columnwidth, trim=16mm 8mm 10mm -4mm]{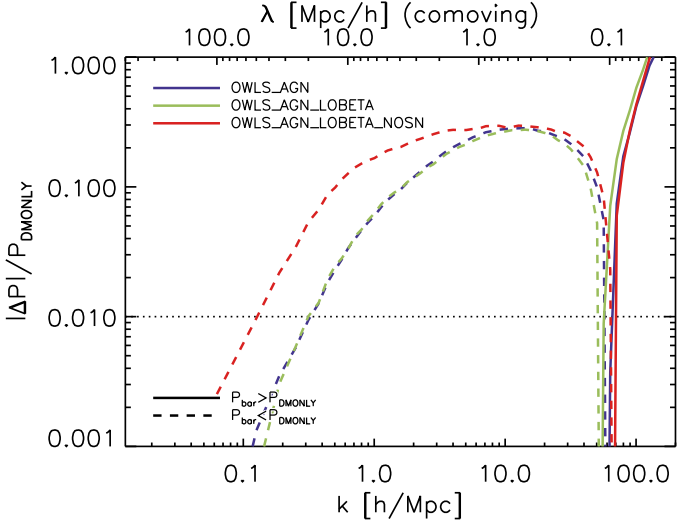}
\caption{Here we show the effect of changing the AGN accretion model and removing stellar feedback on the matter power spectrum, providing insight into the interplay between the two forms of feedback. The fiducial WMAP3 AGN model is shown in blue. In the ``LOBETA'' model, shown in green, the density dependence of gas accretion by the black holes is shallower (see \S\ref{subsubsec:AGN}), but this has almost no effect on the power spectra. However, removing stellar feedback (red lines) has a tremendous impact on the power, greatly increasing the suppression of power due to AGN on large scales.}
\label{fig:diff_SNAGN}
\end{center}
\end{figure}

The right-hand panel compares the predictions for the back-reaction of galaxy formation on dark matter clustering. With the exception of Illustris, all simulations predict an enhancement of power on scales $k\lesssim 1\kunit$, or $k\lesssim 5\kunit$ when additionally excluding OWLS AGN. With the exception of \bahamas{}, all simulations predict a $\sim 1\%$ suppression of power for $k>10\kunit$ with a cross-over scale at $k\approx 40-80\kunit$. This is likely connected to the fact that the \bahamas{} simulations predict a relatively large cross-over scale for the total matter power spectrum, $k\approx 20\kunit$, while all other simulations shown here still show suppression of power on this scale (see left-hand panel). As shown in \S\ref{subsec:backreaction}, the amount of suppression predicted for the dark matter power depends on the strength of the AGN feedback, which would explain the large suppression seen here for Illustris and the relatively large suppression for $k\approx 10\kunit$ seen for OWLS AGN. \citet{Tenneti2015} showed that the back-reaction in the MassiveBlack-II simulation is in qualitative agreement with the results for \bahamas{} shown here, although they find a stronger enhancement, which implies that feedback is less effective.

However, a closer look at both panels reveals that not all differences can be as easily explained, even qualitatively. For example, the \bahamas{} simulation predicts a larger total matter suppression on large scales than e.g.\ IllustrisTNG, implying stronger feedback, yet \bahamas{} is the only simulation shown to not predict any dark matter suppression at all, implying weaker feedback. We will now compare these predictions in more detail, taking into account observational constraints, box size, and resolution.

\begin{figure}
\begin{center}
\includegraphics[width=1.0\columnwidth, trim=16mm 8mm 10mm -4mm]{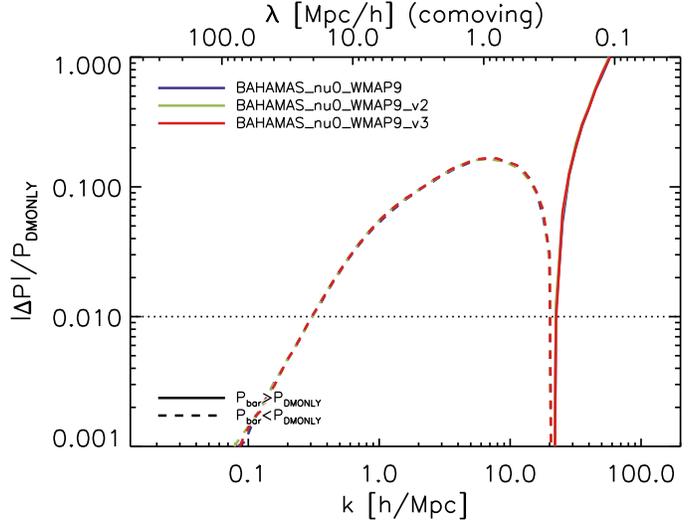}
\caption{Here we show the impact of cosmic variance on our results, by comparing three pairs of simulations with identical box size, resolution and physics, but different initial conditions. The relative total matter power spectra are virtually identical in all three cases on all scales probed here, suggesting that cosmic variance can safely be ignored for these volumes.}
\label{fig:diff_cosmicvar}
\end{center}
\end{figure}

\begin{figure*}
\begin{center}
\includegraphics[width=1.0\columnwidth, trim=16mm 8mm 21mm -4mm]{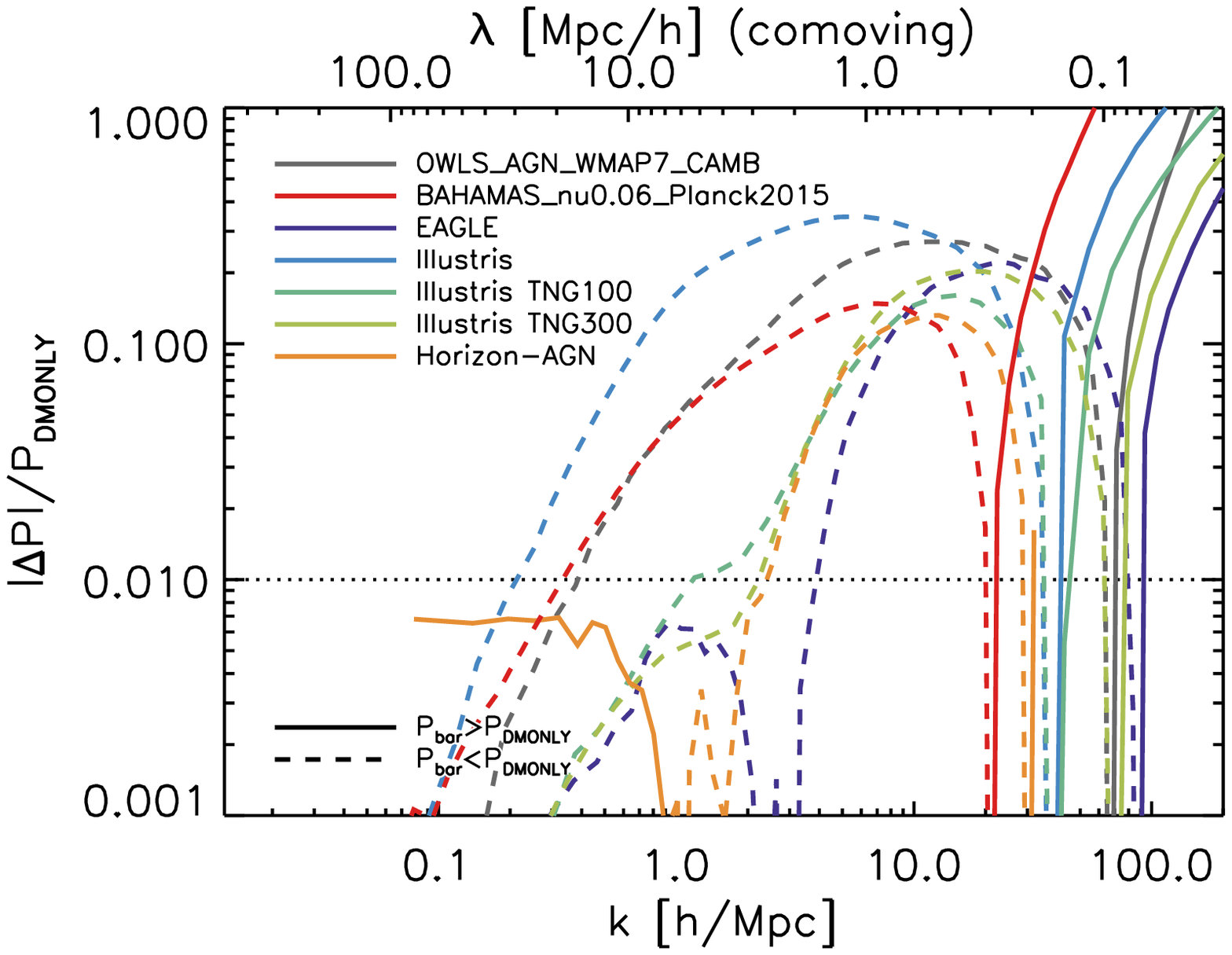}
\includegraphics[width=1.0\columnwidth, trim=27mm 8mm 10mm -4mm]{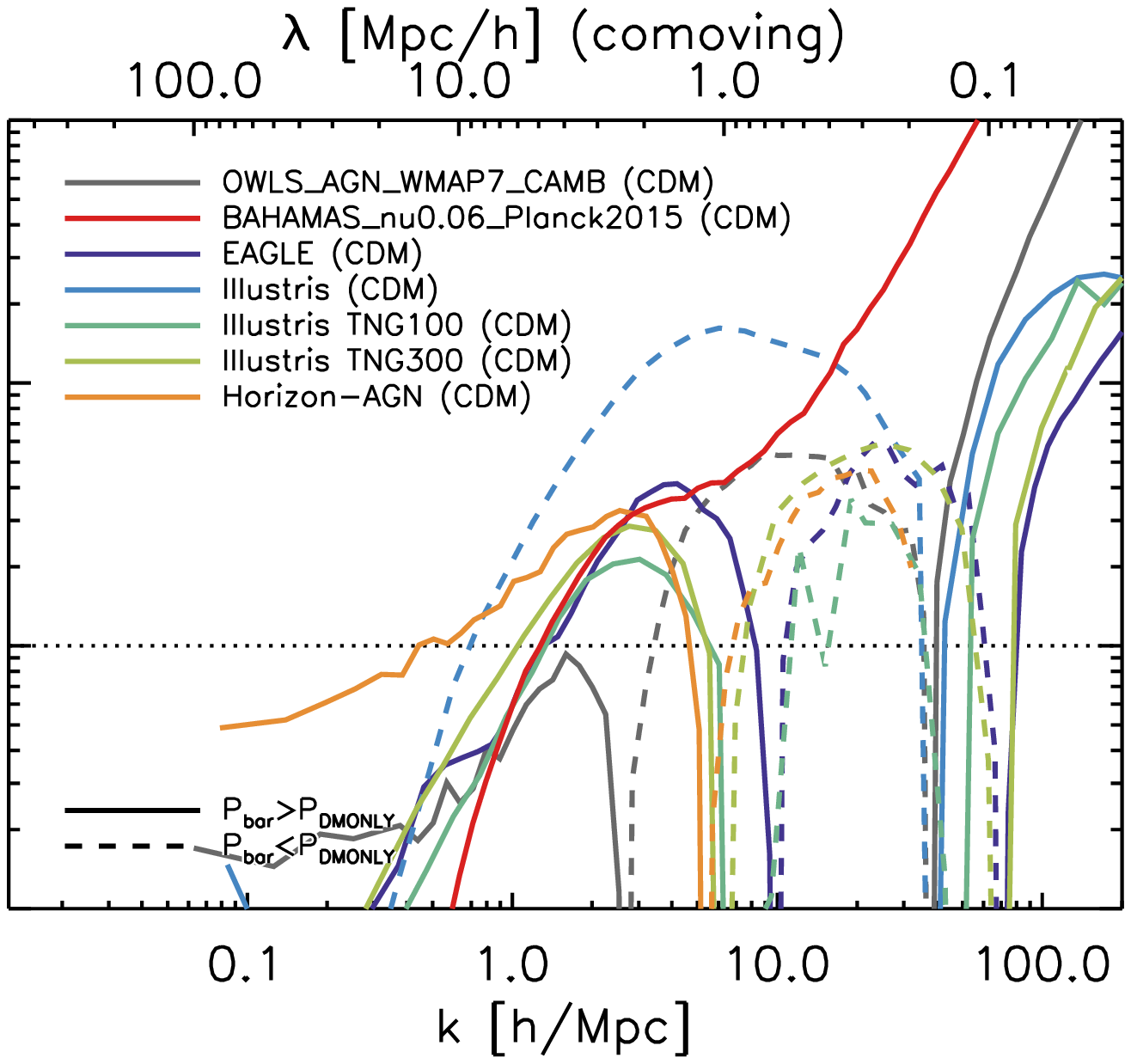}
\caption{A comparison of the relative effect of galaxy formation on the matter power spectrum between the most realistic simulation in the current set (red, \bahamas{}) and power spectra from the literature. We again show the relative effect for the most realistic simulation from VD11 (grey), and compare to that of EAGLE \citep[purple,][]{Hellwing2016}, Illustris \citep[blue,][]{Vogelsberger2014a}, IllustrisTNG \citep[cyan and green,][]{Springel2018} and Horizon-AGN \citep[orange,][]{Chisari2018}. All simulations include stellar and AGN feedback. \textit{Left:} A comparison of the total matter power spectra. There is currently no complete consensus on the exact effect of galaxy formation, though all simulations shown here agree that the total matter power spectrum is suppressed for $1<k<20\kunit$ relative to dark matter only, with a $10-30\%$ suppression at $k=10\kunit$. All simulations predict a cross-over scale between $20$ and $100\kunit$. \textit{Right:} A comparison of the predictions for the back-reaction of galaxy formation on the CDM power spectrum. All models predict enhancement on the smallest scales (except Horizon-AGN, whose power spectrum does not probe small enough scales to tell) and, with the exception of the original Illustris simulation, predict some enhancement for $k\approx 2\kunit$ as well. All simulations except \bahamas{} additionally predict a suppression of several percent on scales $k\approx 20\kunit$.}
\label{fig:diffback_literature}
\end{center}
\end{figure*}

\subsubsection{Understanding the range of predicted effects: total matter}
\label{subsubsec:understandlitdiff}
It is interesting to consider what causes the quantitatively different predictions for the effects of galaxy formation on the matter power spectrum seen in Figure~\ref{fig:diffback_literature}. The predicted suppression is highly dependent on the effectiveness of AGN, and to a lesser extent, stellar feedback. The strength of these processes is \textit{a priori} unknown and must be constrained using observables. A reasonable approach to understanding the amount of power suppression predicted by a simulation would therefore be to look at which observables are used and how they compare to the numerical results. We start by considering the effect of galaxy formation on the total matter power spectrum (left-hand panel of Figure~\ref{fig:diffback_literature}) on large scales, $k<10\kunit$.

We first consider the simulation predicting the largest suppression on scales $0.1<k<10\kunit$: Illustris (blue). The Illustris simulation was constrained using the observed global star formation efficiency \citep{Vogelsberger2014b}, but had several issues which were summarized by \citet{Nelson2015}. Among these is an underestimated gas fraction within $R_\mathrm{500c}$ in group-size haloes due to too-violent radio-mode AGN feedback, which explains the large suppression predicted by this simulation. \citet{Nelson2015} also mention that galaxies below the knee of the stellar mass function are insufficiently quenched, which implies that stellar feedback is not efficient enough -- this, in turn, may provide the AGN with too much gas to heat and eject to large scales (see \S\ref{subsec:SNAGN}).

These shortcomings were addressed with the IllustrisTNG simulations \citep[e.g.][]{Springel2018}, one with a $75\runit$ box (TNG100, matching Illustris; cyan) and another with a $205\runit$ box (TNG300; green). TNG indeed predicts a much lower large-scale suppression of power than Illustris. While the two simulations differ in box size and resolution, their predictions for the relative power agree for $k<10\kunit$, with only a slight deviation around $k\approx 1\kunit$. The Horizon-AGN simulation, described in \citet{Chisari2018}, agrees with these results as well (after a renormalization of the dark matter only power spectrum as mentioned above).

As mentioned above, one of the shortcomings of Illustris addressed with IllustrisTNG was the underestimated gas fraction in groups. \citet{Weinberger2017} and \citet{Pillepich2018a} show that the IllustrisTNG model indeed produces gas fractions that are in better agreement with observations -- however, these studies used simulation volumes $(30\runit)^3$ and $(25\runit)^3)$ in size, respectively, and were not able to probe haloes beyond $M\approx 10^{13.5}\munit$. As \citet{vanDaalenSchaye2015} showed, haloes above this mass limit provide the dominant contribution to the matter power spectrum on scales $k\lesssim 20\kunit$, supplying nearly all signal for $k\approx 1\kunit$ -- additionally, (the progenitors of) these haloes are where AGN feedback has the largest effect on the matter distribution.

\citet{Barnes2018} show that, unlike the smaller boxes using the same model, the full large-volume IllustrisTNG simulations over-predict the gas fractions of massive haloes at redshift zero with respect to observations. Likewise, \citet[][, their Figure 13]{Chisari2018} show that in Horizon-AGN the fraction of bound gas at $z=0$ is over-predicted for $M_{500}>10^{13}\munit$, relative to observations. This implies too-weak AGN feedback in these haloes, which could well explain the discrepancies between the predicted large-scale suppression in the matter power spectrum for IllustrisTNG/Horizon-AGN on the one hand and OWLS/\bahamas{} on the other as seen in Figure~\ref{fig:diffback_literature} ($k\lesssim 10\kunit$).

AGN feedback in Horizon-AGN is implemented following \citet{Dubois2012}, with a kinetic jet mode at low accretion rates and a thermal quasar mode at high accretion. For the latter mode the \citet{BoothSchaye2009} model is used, as in (cosmo\discretionary{-)}{}{)}OWLS and \bahamas{}. However, the minimum heating temperature, which in e.g.\ \citet{LeBrun2014} was found to be of large influence to the effectiveness of AGN feedback and needed to be sufficiently high for the feedback energy not to be immediately radiated away (see also \S\ref{subsec:varAGN}), is set to zero in Horizon-AGN. While the jet mode feedback dominates in Horizon-AGN for $z\lesssim 2$, Figure 13 of \citet{Chisari2018} shows that the gas fractions in massive haloes are already quite high at $z=2$, and either mode of feedback might therefore be responsible for the too-high gas fractions at $z=0$. A minimum heating temperature of zero could certainly account for ineffective quasar mode feedback in a particle-based simulation, but since Horizon-AGN is mesh-based, more work is needed before one can tell whether this parameter choice has the same effect there. IllustrisTNG also implements feedback from supermassive black holes in two different modes in a mesh-based code, and may therefore show some similarities with Horizon-AGN in the behaviour of its AGN feedback.

Finally, we consider the predictions of EAGLE for the large-scale matter power spectrum, presented in \citet{Hellwing2016} and reproduced in Figure~\ref{fig:diffback_literature} in purple. EAGLE predicts a suppression close to that of IllustrisTNG, though it is typically smaller, particularly around $k=3\kunit$. Given that EAGLE's simulation volume is the smallest presented here, this may well be due to it not probing sufficiently massive haloes. While the original Illustris simulation has the second-smallest volume of those considered, its extreme feedback compensated for the effect of its lack of massive haloes. \citet{Schaye2015} shows that EAGLE also over-predicts gas fractions at the massive end, $M\gtrsim 10^{14}\munit$, similar to Horizon-AGN. \citet{Hellwing2016} point to this as the main reason for the smaller suppression in power predicted for this simulation, compared to OWLS AGN. This, in addition to its small volume, may well explain why EAGLE generally predicts the smallest large-scale effect on the total matter power spectrum of the simulations considered in Figure~\ref{fig:diffback_literature}.

In short, we find that differences in the predicted suppression of power on scales $k\lesssim 10\kunit$ may in large part be explained by differences in the gas fractions of groups and clusters, with possibly a minor contribution from box size. Of the simulations considered here, the gas fractions in \bahamas{} are closest to observations, which is expected given that these were used as a constraint in choosing its subgrid parameters. It has the largest volume as well, which as shown in \S\ref{subsec:cosvar} is expected to be sufficient for cosmic variance to be small. We therefore expect the power spectrum of the \bahamas{} simulation shown in Figure~\ref{fig:diffback_literature} to be closest to that of the real Universe on scales $k\lesssim 10\kunit$.

However, the large volume of \bahamas{} comes at the expense of resolution. Indeed, all other simulations shown here are able to resolve $2-8\times$ smaller scales. In Appendix~\ref{app:restests} we demonstrate that an increase in resolution yields significantly different predictions for the power spectrum on scales $k\approx 10\kunit$ and smaller, which recalibration can only slightly compensate for. The cross-over scale predicted by \bahamas{} is too large, and the relative suppression of the total matter spectrum is expected to be $\sim 10\%$ to smaller scales than shown here. Besides resolution, the relative power spectrum on small scales is also set by the stellar fraction, as stars dominate the power spectrum on sufficiently small scales. Discrepancies in resolution and stellar fractions are therefore expected to explain the differences in the relative small-scale clustering seen in the left-hand panel of Figure~\ref{fig:diffback_literature}. However, as our primary interest is in predicting the power spectrum observed by Stage IV experiments, we leave a more detailed exploration of clustering on scales $k>10\kunit$ to future work.

\begin{figure*}
\begin{center}
\includegraphics[width=1.0\columnwidth, trim=16mm 8mm 10mm -4mm]{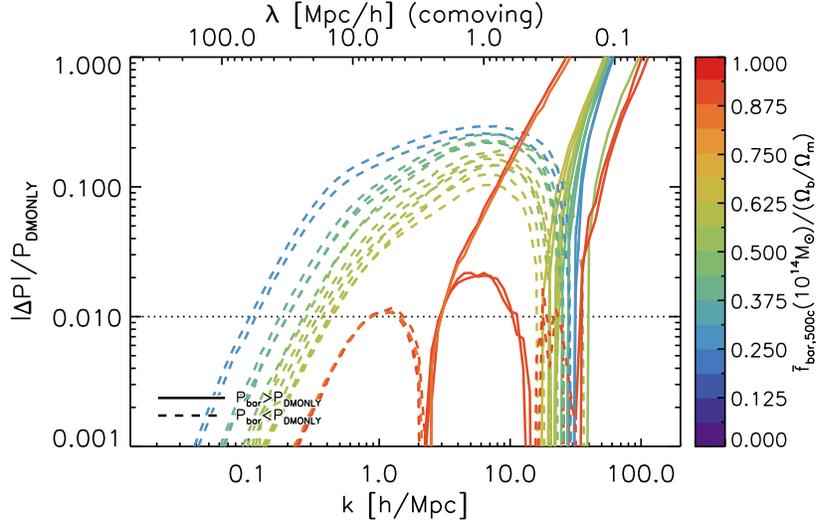}
\caption{The effect of galaxy formation on cosmo-OWLS and \bahamas{} simulations with $400\runit$ boxes. Each simulation is colour-coded by its mean baryon fraction within $R_\mathrm{500c}$ for $M_\mathrm{500c}\approx 10^{14}\munit$ haloes, rescaled by the cosmic baryon fraction. The power in the cosmo-OWLS simulations has been renormalized to remove the large-scale offset where applicable (see \S\ref{subsec:dmonly}). The effect of galaxy formation is strongly correlated with the group baryon fraction, a lower fraction corresponding to a stronger suppression of power.}
\label{fig:fbar500pow}
\end{center}
\end{figure*}

\subsubsection{Understanding the range of predicted effects: back-reaction}
\label{subsubsec:understandlitback}
Even though on large scales ($k\lesssim 2\kunit$) the \bahamas{} simulation shown here predicts a much larger suppression of the total power than EAGLE and both IllustrisTNG simulations, these simulations are in close agreement for the back-reaction on these scales (the same applying to Horizon-AGN assuming again a large-scale correction for initial conditions). Only the original Illustris simulation disagrees, predicting suppression instead of enhancement, but this can be explained by its extremely strong feedback and the results of \S\ref{subsec:varAGN}. The agreement of the other simulations implies that the clustering of dark matter is rather insensitive to changes in the baryonic matter distribution on large scales, which are mainly driven by the expulsion of gas. Differences in clustering on galactic scales ($k\gtrsim 10\kunit$), however, do seem to correlate with changes in the gas distribution, and here resolution plays a more significant role than for large-scale effects, in addition to the role played by the strength of feedback itself. These conclusions are supported by the results of Appendix~\ref{app:restests}, where we show that a recalibrated \bahamas{} run with a higher resolution predicts that the suppression of the total power extends to $1.4\times$ smaller scales and thereby significantly reduces the dark matter clustering around $k\approx 10\kunit$. Since these scales are quite small compared to those probed by Stage IV weak lensing surveys, we refrain from exploring the differences in the back-reaction for $k\gtrsim 10\kunit$ in more detail here.

\subsection{The baryon fraction as a predictor of power suppression}
\label{subsec:barfrac}
Given that changes to the total power due to galaxy formation on sufficiently large scales seem to correlate with even small shifts in the amount of gas in groups and clusters, it is interesting to consider whether the gas fraction of massive haloes can be used to predict the amount of power suppression seen in these simulations. If so, then the large-scale power spectrum might be fixed by cosmology and just one other (importantly, measurable) quantity, independently of how feedback was implemented in a hydrodynamical simulation.

Since most of the large-scale suppression of power is brought about by AGN removing gas from primarily groups of galaxies, a natural measure to correlate with would be the gas fraction within the virial radius of $M\sim 10^{14}\munit$ haloes. The lower the gas fraction, the more gas has been evacuated and the larger the expected suppression of power. However, the problem with this is that in the absence of strong feedback, the galaxies overcool and convert the gas into stars instead, leading to a low gas fraction but no suppression of power on large scales. We therefore instead use the \emph{baryon} fraction within the virial radius, which is lowered only through the removal of gas from the halo. To compensate for differences in cosmology, we normalize these fractions for each simulation by the cosmic baryon fraction, $\Omega_\mathrm{b}/\Omega_\mathrm{m}$. In what follows, we will focus on the mean renormalized baryon fractions within $R_\mathrm{500c}\equiv R_\mathrm{500,crit}$ for haloes with masses $M_{500c}$ in the range $[6\times 10^{13},2\times 10^{14}]\munitnoh$, which we refer to as:
\begin{equation}
\label{eq:fbar}
\tilde{f}_\mathrm{bar,500c}(10^{14}\munitnoh)\equiv \bar{f}_\mathrm{bar,500c}(10^{14}\munitnoh)/(\Omega_\mathrm{b}/\Omega_\mathrm{m}).
\end{equation}

In Figure~\ref{fig:fbar500pow}, we show the effect of galaxy formation on those cosmo-OWLS and \bahamas{} simulations that have $400\runit$ boxes and for which baryon fractions are available. This includes most simulations from this set, but not those with non-zero neutrino masses save BAHAMAS\_nu0.06\_Planck2015. The cosmo-OWLS simulations shown here have had their DMONLY counterparts renormalized where applicable to account for any large-scale offset (see \S\ref{subsec:dmonly}). The relative power spectra have been colour-coded by $\tilde{f}_\mathrm{bar,500c}(10^{14}\munitnoh)$. It is clear immediately that, at least for $k\lesssim 3\kunit$, the suppression of power is monotonic with the baryon fraction, with higher suppression corresponding to a lower baryon fraction in group-sized haloes. This is expected: the lower the baryon fraction, the more mass was ejected from the halo by feedback, and hence the stronger the effect on the power spectrum.

We note again that the suppression on scales larger than the size of haloes comes about through feedback lowering the mass of haloes, thereby lowering the 2-halo term contribution of these haloes. Where this gas is distributed \emph{to} should not be relevant at the $1\%$ level on large scales, just that it is removed from clustered regions -- after all, unless feedback moves matter over distances $\sim 10\runit$, changing the mass of clustered regions is the only way to significantly change the power for $k\lesssim 1\kunit$. Because of this, we might expect to be able to model the power suppression on large scales as a function of only the baryon fraction in group-sized haloes, which dominate the change in clustering.

\begin{figure}
\begin{center}
\includegraphics[width=1.0\columnwidth, trim=16mm 8mm 10mm 9mm]{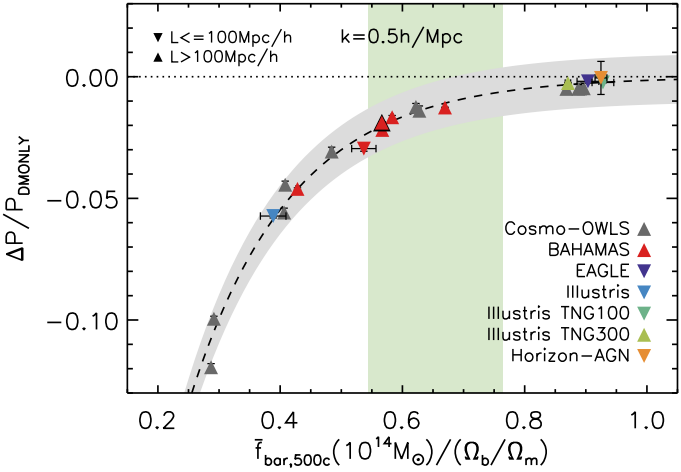}
\caption[The effect of galaxy formation as a function of baryon fraction.]{The effect of galaxy formation on the matter power spectrum at the scale $k=0.5\kunit$ as a function of the (renormalized) mean baryon fraction in $\sim 10^{14}\munitnoh$ haloes. Shown are simulations from cosmo-OWLS and \bahamas{} (grey and red), EAGLE (purple), Illustris (blue), TNG100 and TNG300 (cyan and green) and Horizon-AGN (orange). Where necessary, dark matter only power spectra were corrected to agree with their hydrodynamical counterparts on the largest scale available (see \S\ref{subsec:dmonly}). Baryon fractions were calculated within $R_\mathrm{500c}$ for haloes in the mass range $M_\mathrm{500c}=[6\times 10^{13},2\times 10^{14}]\munitnoh$. The dashed curve shows that at this $k$, a simple exponential function of the renormalized baryon fraction, $-\exp(-5.990\tilde{f}_\mathrm{bar,500c}-0.5107)$, fits the predictions for the suppression of power of all simulations to within $1\%$ (grey band). The vertical green band shows a range of renormalized mean group-scale baryon fractions roughly consistent with observations. BAHAMAS\_nu0.06\_Planck2015 is outlined in black.}
\label{fig:fbar500pow_lin}
\end{center}
\end{figure}

In Figure~\ref{fig:fbar500pow_lin} we show the change in the matter power spectrum due to feedback relative to the dark matter only power spectrum at $k=0.5\kunit$, as a function of the renormalized mean group baryon fraction $\tilde{f}_\mathrm{bar,500c}(10^{14}\munitnoh)$. Besides the simulations shown in Figure~\ref{fig:fbar500pow}, shown here as upward grey and red triangles, we also show results for several simulations with smaller volumes, namely the uncalibrated \bahamas{} re-run BAHAMAS\_nu0\_WMAP9\_L100N512 (red downward triangle, see Appendix~\ref{app:restests}), EAGLE (purple), Illustris (blue), TNG100 and TNG300 (cyan and green), and Horizon-AGN (orange). The dashed line is given by a simple exponential, $-\exp(-5.990\tilde{f}_\mathrm{bar,500c}-0.5107)$, the grey band around it denoting a deviation of $1\%$. For all simulations, the results fall within this band. The vertical green band shows a range of renormalized mean group-scale baryon fractions roughly consistent with observations\footnote{The observation range for the mean is based on the combined data sets of \citet{Vikhlinin2006}, \citet{Maughan2008}, \citet{Sun2009}, \citet{Pratt2009}, \citet{Rasmussen2009}, \citet{Lin2012}, \citet{Sanderson2013}, \citet{Gonzalez2013}, \citet{Budzynski2014}, \citet{Lovisari2015}, \citet{Pearson2017} and \citet{Kravtsov2018}.}. The calibrated \bahamas{} simulations all lie within this band, and BAHAMAS\_nu0.06\_Planck2015 is outlined in black.

While they are too small to see for many of these points, both horizontal and vertical error bars were calculated. The statistical uncertainty in the individual power spectra largely drops out when taking the ratio of power spectra with identical initial conditions, though systematic uncertainties like cosmic variance remain. Since all residual uncertainties in the power ratio are difficult to estimate, we take a conservative stance and only use $400\runit$ boxes (for which cosmic variance is assumed to play a negligible role) in fitting our model, with an uncertainty of $10^{-3}$ at any $k$. For (cosmo\discretionary{-)}{}{)}OWLS boxes, for which no 2-fluid dark matter runs were performed, we apply a correction factor to the DMONLY power spectra on all scales such that the power measured on the largest scales is approximately identical to that of the hydrodynamical runs. Since this correction is approximate, we use the maximum of $10^{-3}$ and the original large-scale offset (typically $2\times 10^{-3}$) as the uncertainty on the relative power. The same was done for the other simulations shown, though the only significant correction was for Horizon-AGN. Note the vertical error bars shown in Figure~\ref{fig:fbar500pow_lin} do not include systematics due to small box size even for simulations where they would be significant (i.e.\ most simulations not included in the fit). The uncertainties in the baryon fractions are calculated directly as the standard error on the mean.

At every individual value of $k$, the power decrement as a function of the baryon fraction of $\sim 10^{14}\munitnoh$ haloes is fit surprisingly well by a simple two-parameter exponential, as seen in Figure~\ref{fig:fbar500pow_lin}.\footnote{\label{modelnote}Note that for $k=0.5\kunit$, our model, by construction, reduces to exactly such an exponential. The model shown in this figure is the full model with parameters as given in Table~\ref{tab:modelparams}, but is virtually identical to the best-fit two-parameter exponential.} This is also the basis of our full (empirical) fitting function, given by:
\begin{equation}
\label{eq:powbarmodel}
\frac{P_\mathrm{bar}}{P_\mathrm{DMO}}=1-\frac{2^a+2^b(c\tilde{f}_\mathrm{bar})^{b-a}}{k^{-a}+(c\tilde{f}_\mathrm{bar})^{b-a}k^{-b}}\exp{\left(d\tilde{f}_\mathrm{bar}+e\right)}.
\end{equation}
Here the additional parameters $a$, $b$ and $c$ facilitate a fit to the relative power spectra at fixed baryon fraction. The effect of galaxy formation on the power spectrum up to $k=1\kunit$ is approximated as two power laws in $k$ with a smooth transition between them at a scale that depends linearly on the baryon fraction, which was empirically determined to give accurate results. Beyond $k=1\kunit$, however, this approximation fails. This is because on these scales the power spectrum is sensitive to the changes in the internal structure of group-sized haloes, which requires additional modelling. In halo model terms, for $k>1\kunit$ the power spectrum is sensitive to the 1-halo term of $\sim 10^{14}\munitnoh$ haloes, while its contribution is constant for $k<1\kunit$ \citep[e.g.][]{Debackere2019}. We therefore do not attempt to apply our current model to these scales.

\begin{table}
\caption{The best-fit parameter values of the model in equation~\eqref{eq:powbarmodel}, which gives the total matter power spectrum relative to a dark matter only power spectrum for $k\leq 1\kunit$. $\tilde{f}_\mathrm{bar}$ is the mean baryon fraction relative to the cosmic value, i.e.\ $\tilde{f}_\mathrm{bar}\equiv \bar{f}_\mathrm{bar}/(\Omega_\mathrm{b}/\Omega_\mathrm{m})$. The baryon fractions used in fitting the model are measured within either $R_\mathrm{500c}$ or $R_\mathrm{200c}$ for haloes with masses (either $M_\mathrm{500c}$ or $M_\mathrm{200c}$) in the range $[6\times 10^{13},2\times 10^{14}]\munitnoh$.}
\centering
\setlength{\tabcolsep}{7.5pt}
\begin{tabular}{l l l l l l}
\hline
Definition & $a$ & $b$ & $c$ & $d$ & $e$ \\ [0.5ex]
\hline
$\tilde{f}_\mathrm{bar,500c}$ & 2.215 & 0.1276 & 1.309 & -5.990 & -0.5107 \\ [1ex]
$\tilde{f}_\mathrm{bar,200c}$ & 2.111 & 0.0038 & 1.371 & -5.816 & -0.4005 \\ [1ex]
\hline
\end{tabular}
\label{tab:modelparams}
\end{table}

\begin{figure*}
\begin{center}
\includegraphics[width=0.69\columnwidth, trim=16mm 18mm 25mm 9mm]{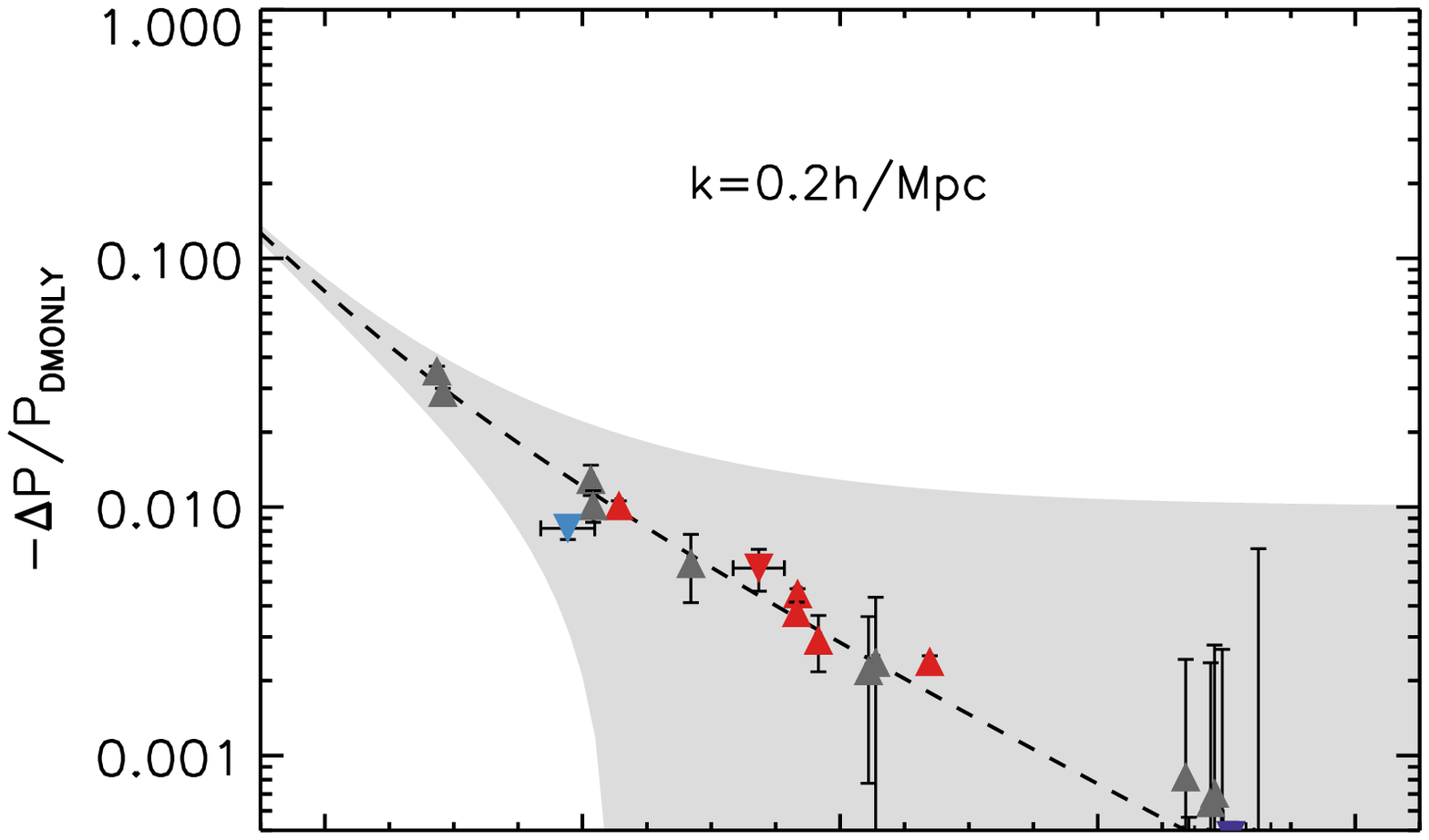}
\includegraphics[width=0.69\columnwidth, trim=24mm 18mm 17mm 9mm]{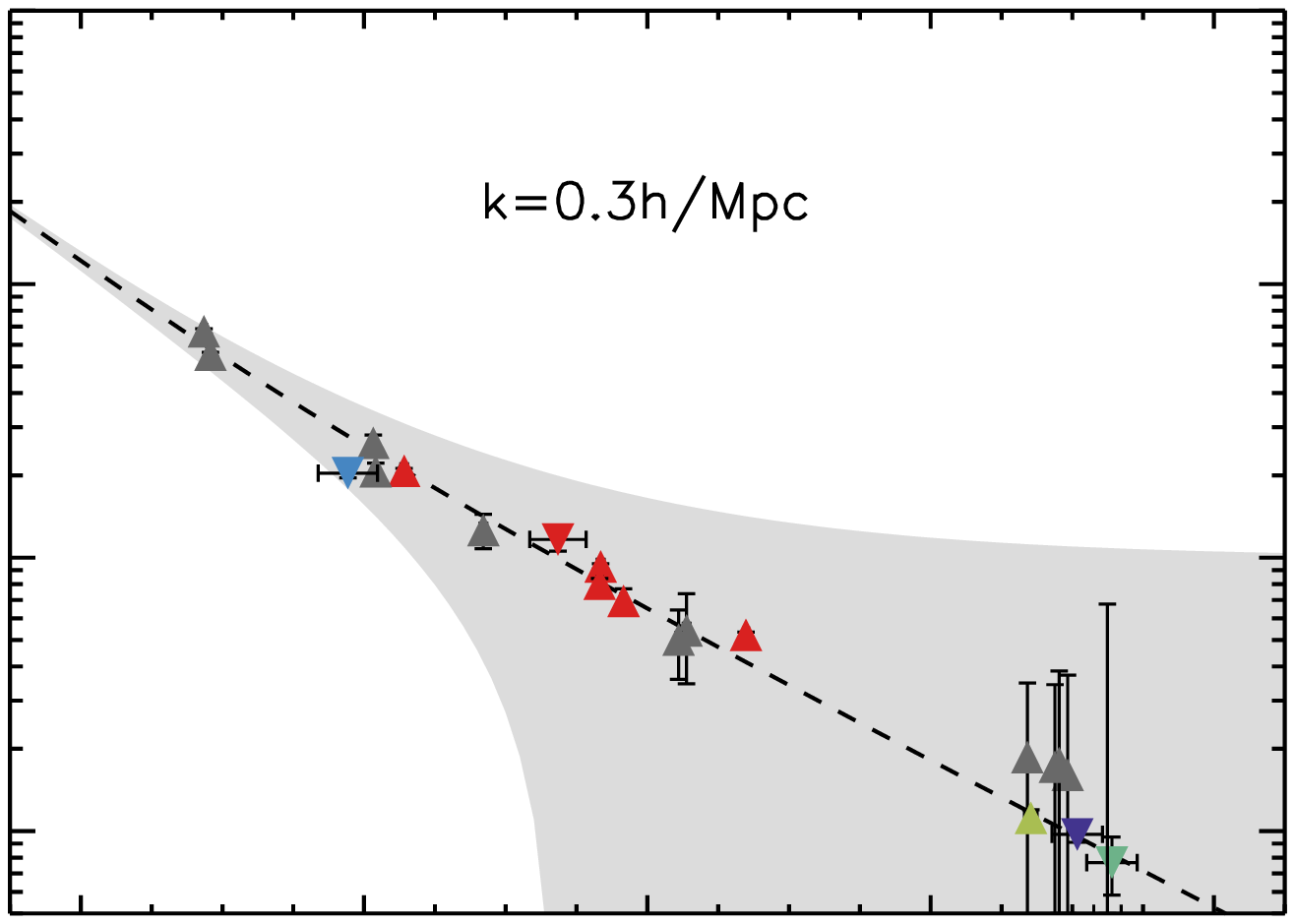}
\includegraphics[width=0.69\columnwidth, trim=32mm 18mm 9mm 9mm]{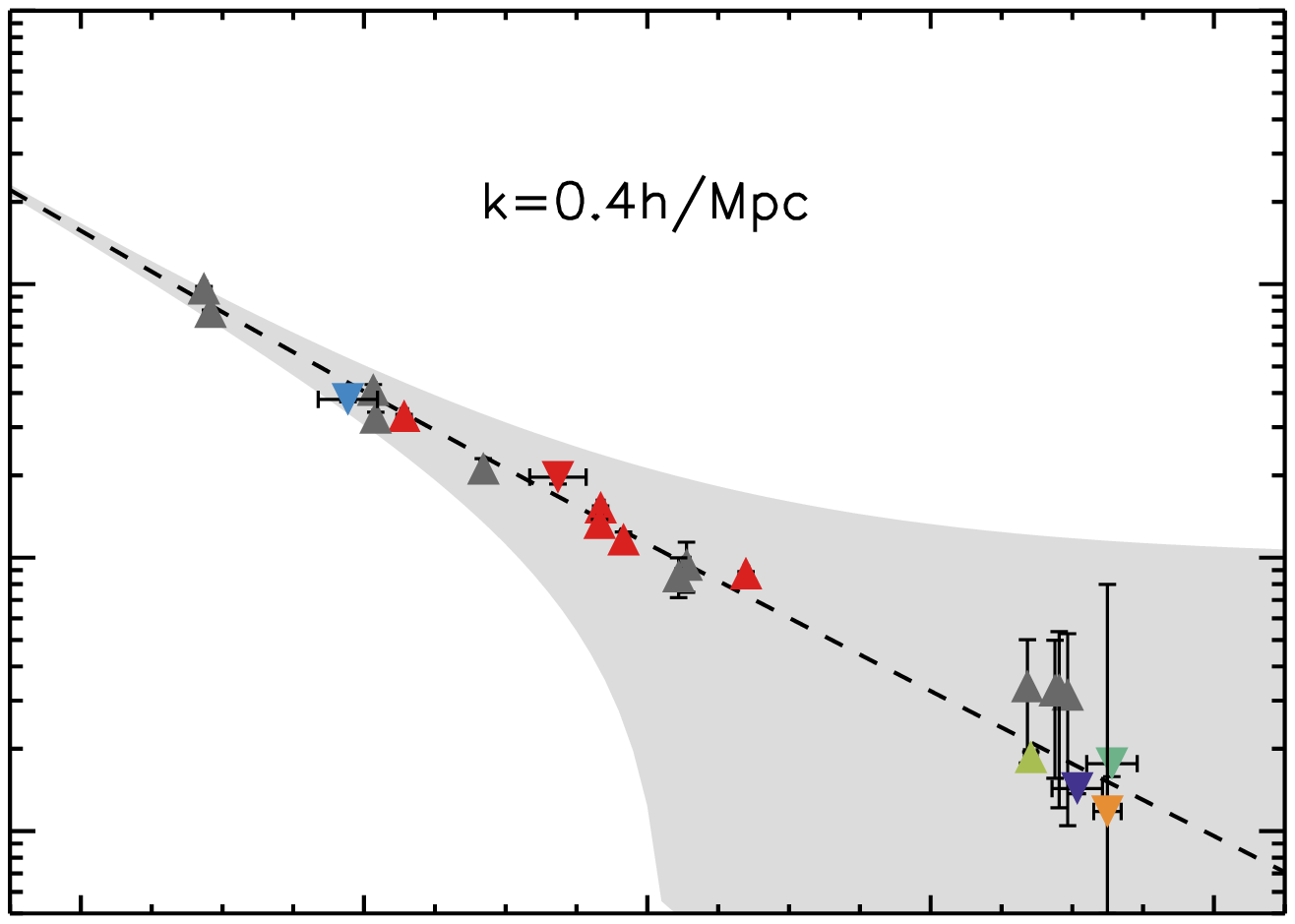}\\
\includegraphics[width=0.69\columnwidth, trim=16mm 18mm 25mm 17mm]{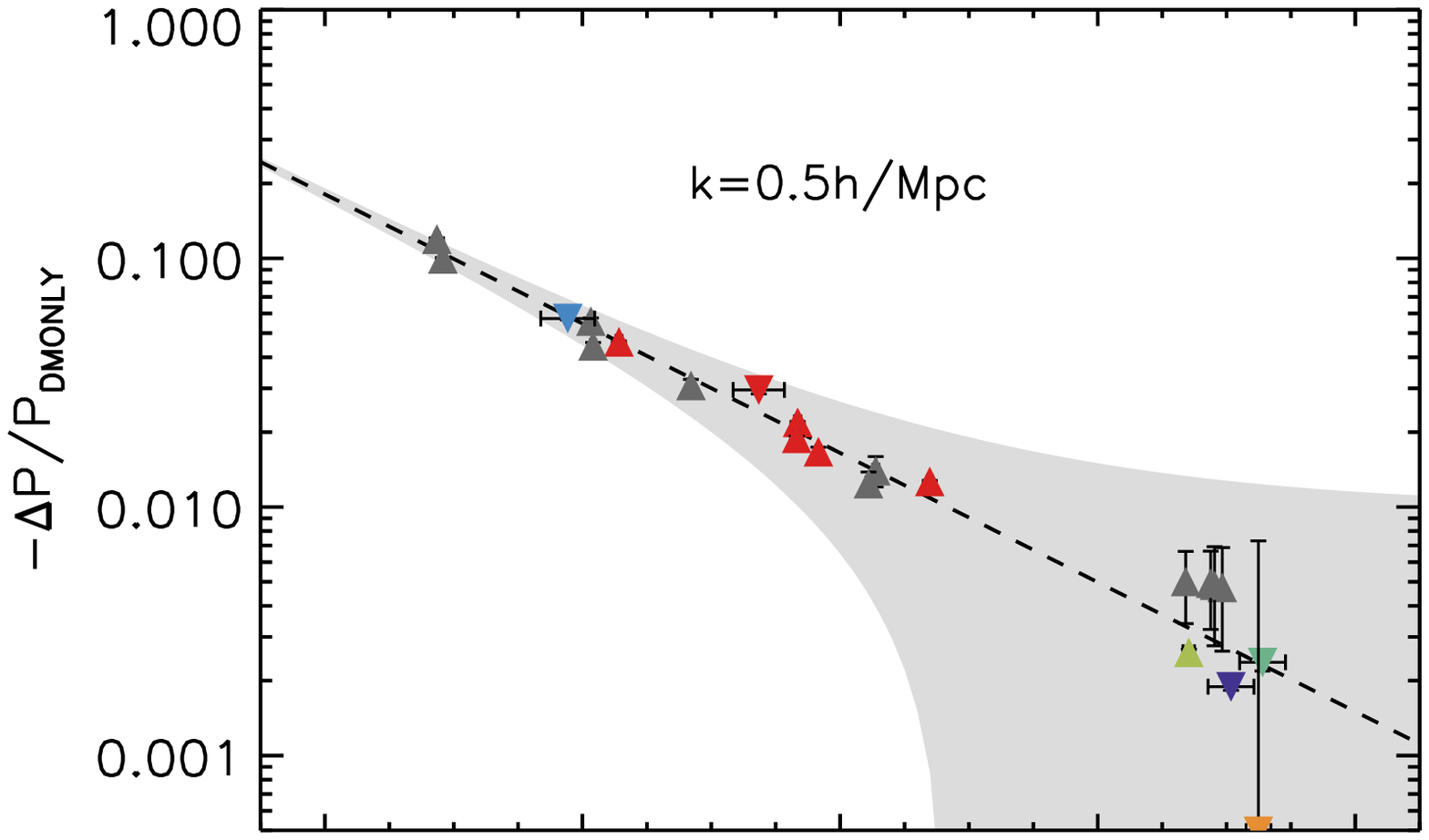}
\includegraphics[width=0.69\columnwidth, trim=24mm 18mm 17mm 17mm]{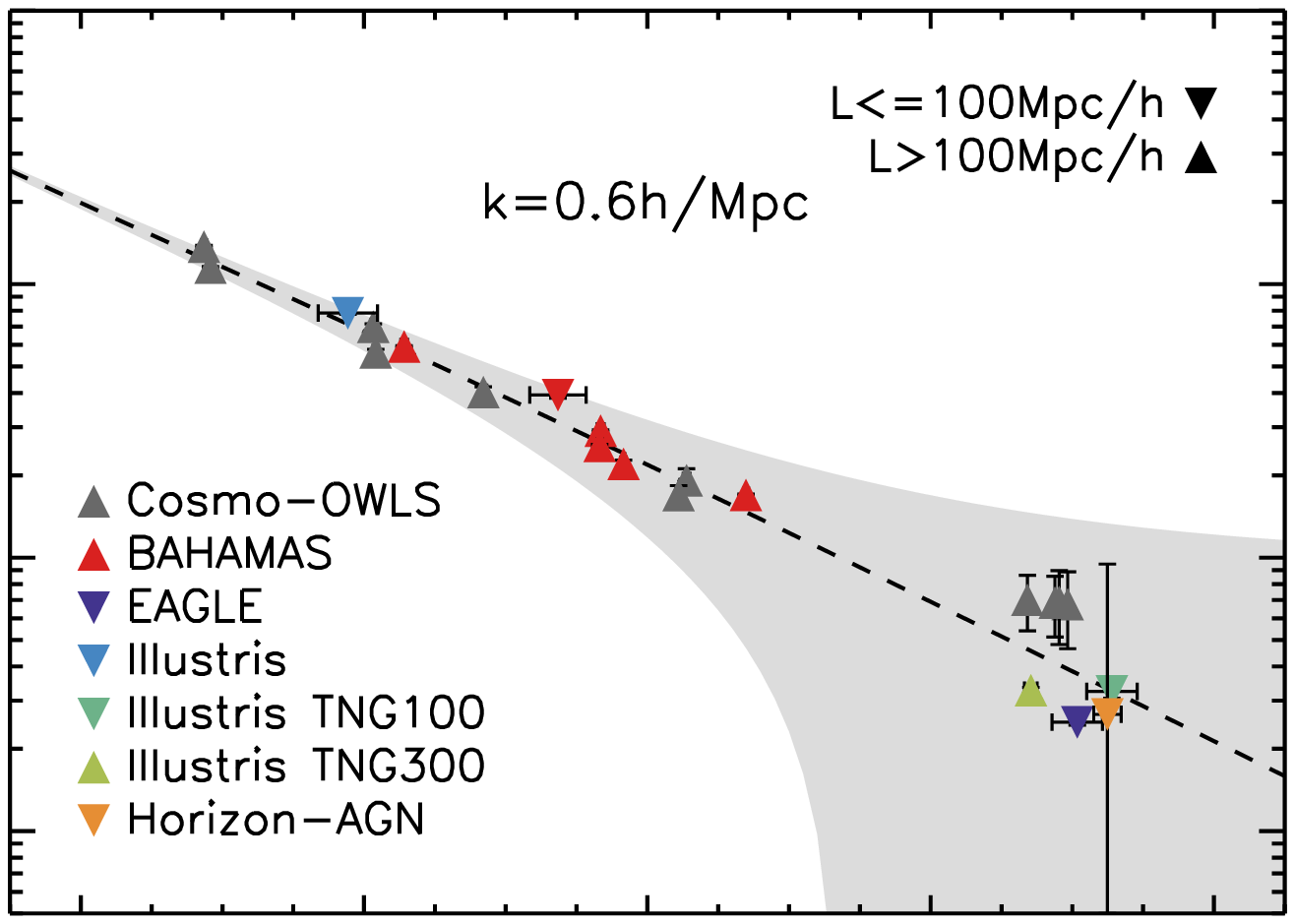}
\includegraphics[width=0.69\columnwidth, trim=32mm 18mm 9mm 17mm]{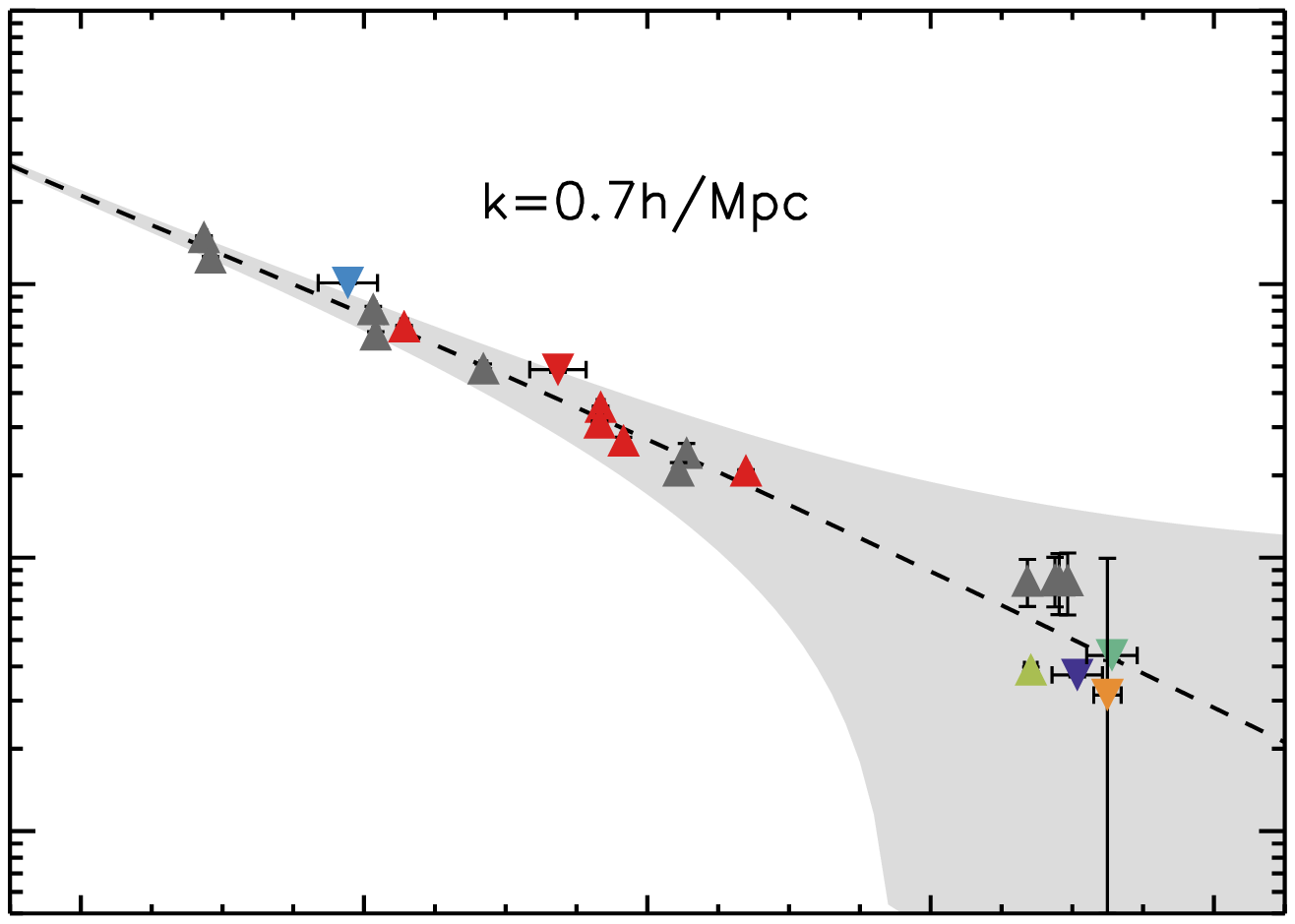}\\
\includegraphics[width=0.69\columnwidth, trim=16mm 8mm 25mm 17mm]{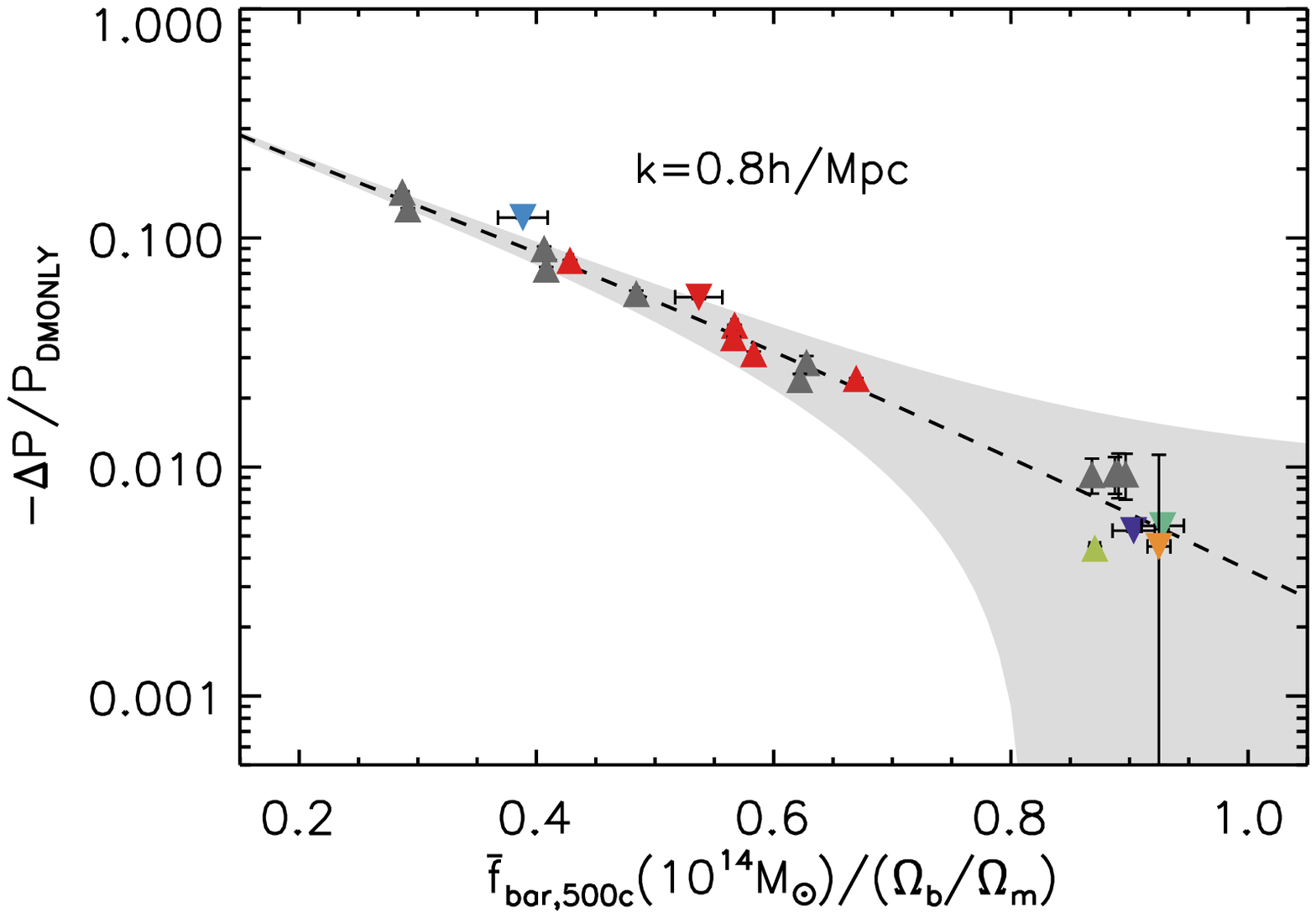}
\includegraphics[width=0.69\columnwidth, trim=24mm 8mm 17mm 17mm]{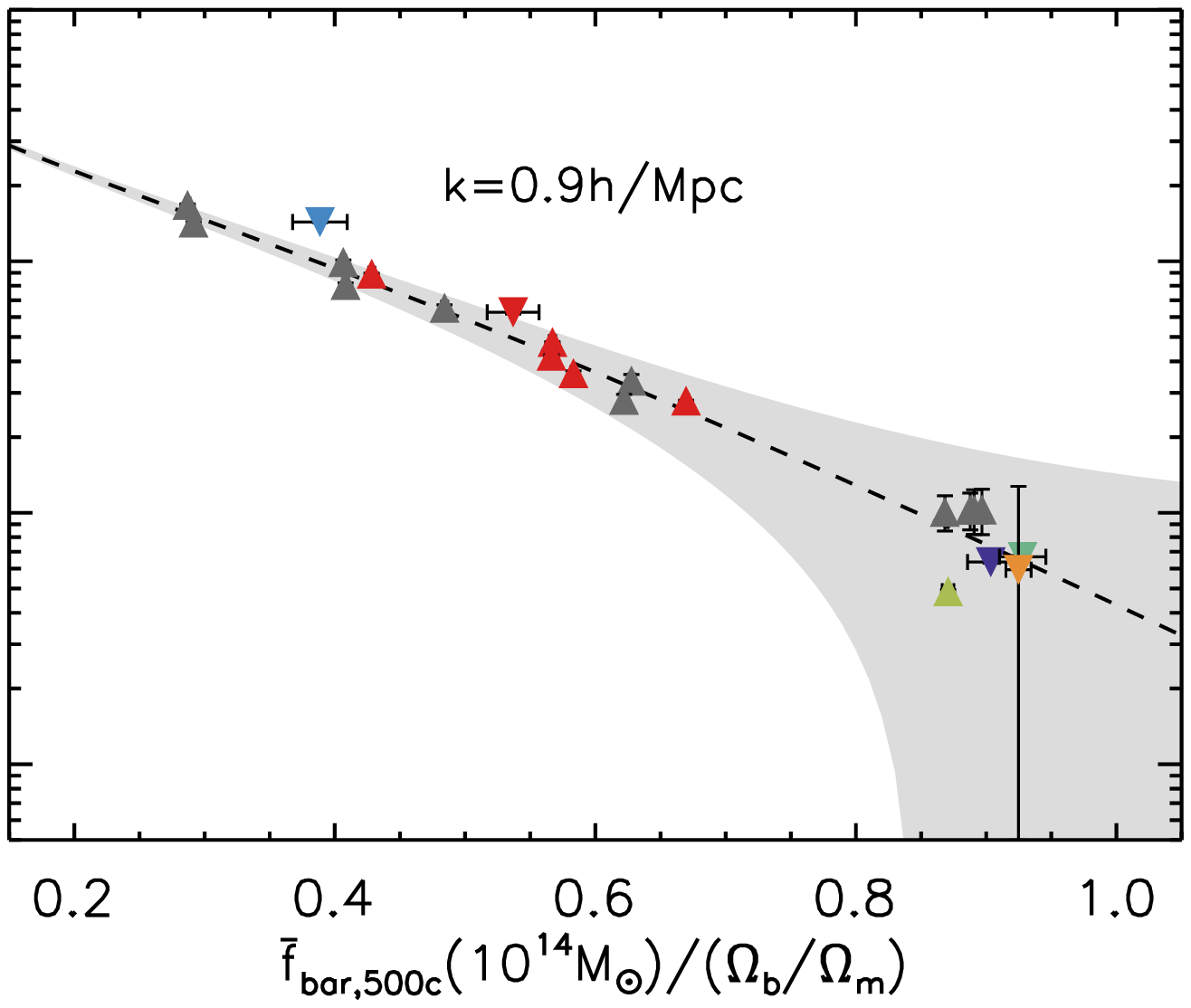}
\includegraphics[width=0.69\columnwidth, trim=32mm 8mm 9mm 17mm]{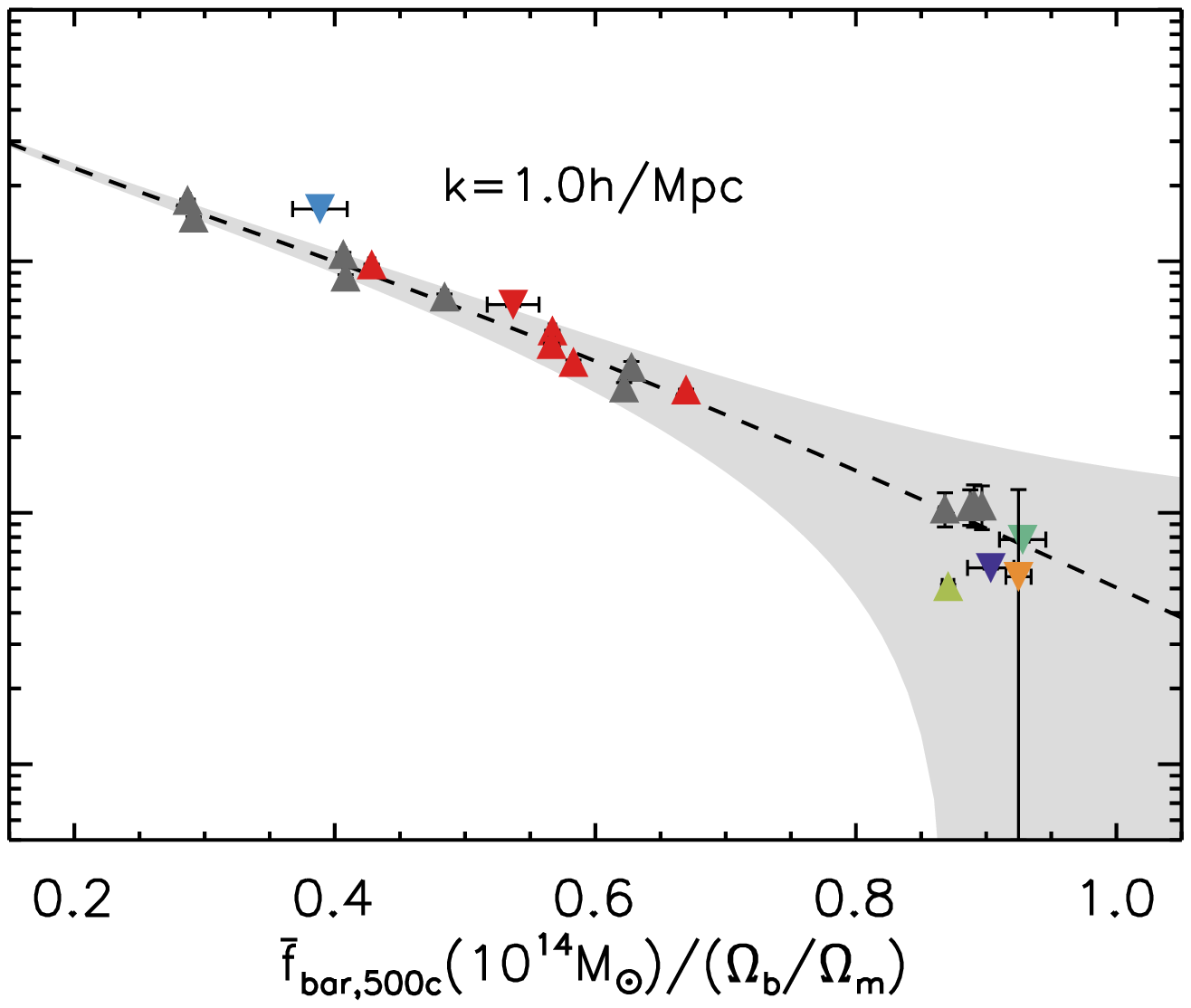}
\caption{As Figure~\ref{fig:fbar500pow_lin}, but with a logarithmic vertical axis. Each panel shows the relative change in total matter power due to galaxy formation on a different scale, from $k=0.2\kunit$ in the top left panel through $k=1\kunit$ in the bottom right panel. The dashed line shows the best-fit model, fit to the simulations with $400\runit$ boxes on all scales $k<1.1\kunit$ simultaneously (see equation~\eqref{eq:powbarmodel} and Table~\ref{tab:modelparams}), with the grey region showing where the absolute deviation from the fit is $\leq 1\%$.}
\label{fig:fbar500pow_panels}
\end{center}
\end{figure*}

The best-fit parameter values depend on the definition used for the rescaled baryon fraction, $\tilde{f}_\mathrm{bar}\equiv f_\mathrm{bar}/(\Omega_\mathrm{b}/\Omega_\mathrm{m})$. In Table~\ref{tab:modelparams}, we provide best-fit parameters for both baryon fractions measured within $R_\mathrm{500c}$ of haloes in the range $M_\mathrm{500c}=[6\times 10^{13},2\times 10^{14}]\munitnoh$ and within $R_\mathrm{200c}$ of haloes in the range $M_\mathrm{200c}=[6\times 10^{13},2\times 10^{14}]\munitnoh$. We find that the mean baryon fraction of these haloes is more strongly correlated with the suppression of power than that of haloes with masses around $10^{13}\munitnoh$ or $10^{15}\munitnoh$. Only power spectra of $400\runit$ cosmo-OWLS and \bahamas{} simulations for $k<1.1\kunit$ were included in the fit, but the best-fit model yields an excellent fit to the power ratio of other simulations as well, as seen in Figure~\ref{fig:fbar500pow_panels}. Using only the baryon fraction for $\sim 10^{14}\munitnoh$ haloes as a parameter, our best-fit model is accurate to within $1\%$ (absolute) for all simulations save the original Illustris run, C-OWLS\_AGN\_Theat8.7\_WMAP7 (which deviates up to $1.5\%$) and BAHAMAS\_nu0\_WMAP9\_L100N512 (red downward triangle) though shifting to a baryon fraction $\approx 0.7\sigma$ lower brings the latter into excellent agreement. Additionally, small boxes are significantly affected by cosmic variance and missing modes, which are not included in the vertical error bars.

\begin{figure*}
\begin{center}
\includegraphics[width=1.0\columnwidth, trim=20mm 8mm 21mm 9mm]{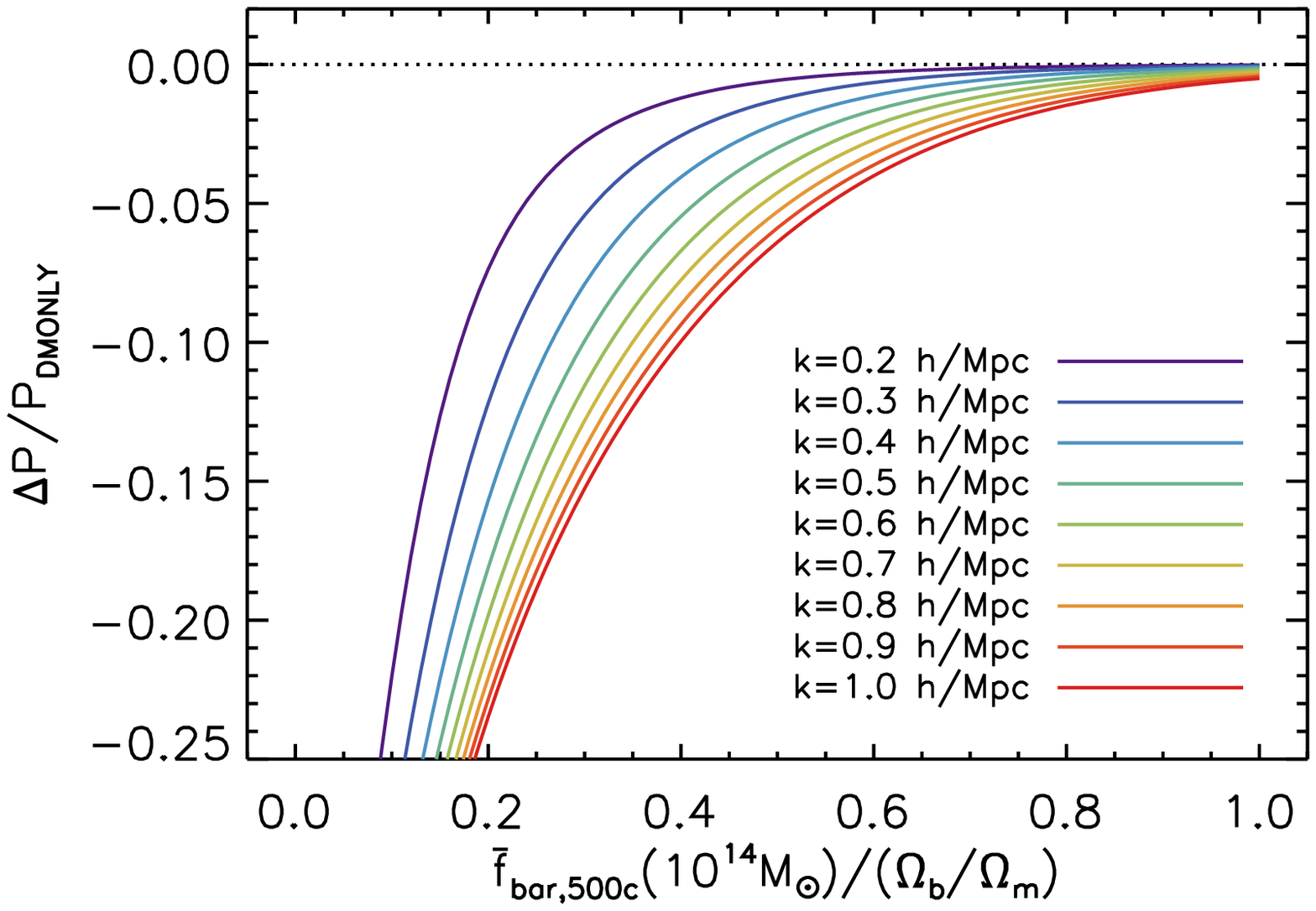}
\includegraphics[width=1.0\columnwidth, trim=27mm 8mm 14mm 9mm]{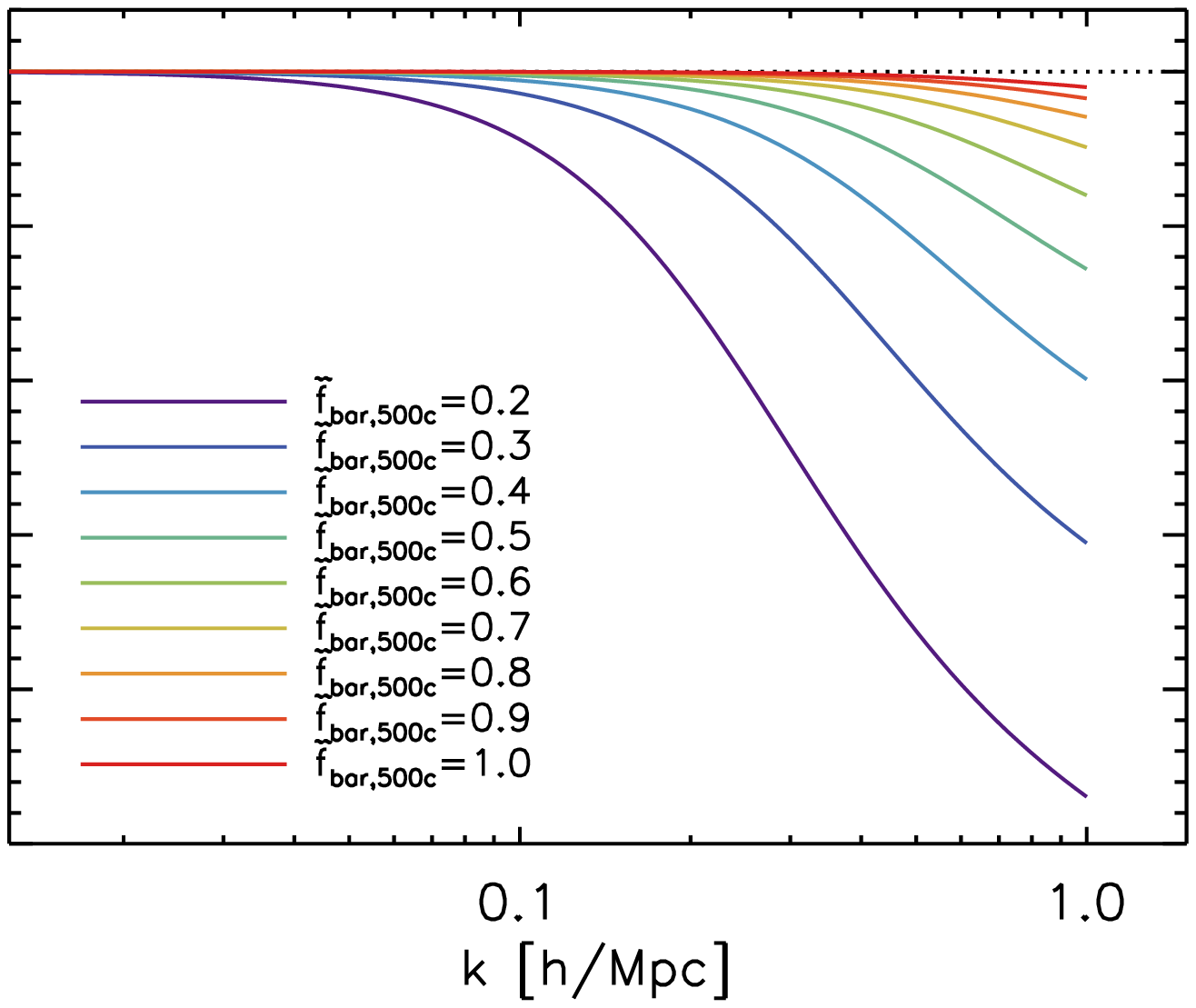}
\caption{The relative change in the matter power spectrum due to galaxy formation prediction by our model, equation~\eqref{eq:powbarmodel}, as a function of the rescaled mean baryon fraction in groups for different wavenumbers (left panel) and as a function of wavenumber for different rescaled mean baryon fractions in groups (right panel).}
\label{fig:fbar500pow_model}
\end{center}
\end{figure*}

For most simulations, the fit is accurate to within $0.5\%$ for $k\leq 1\kunit$, the outliers being simulations that, like Illustris, have unrealistically strong feedback. We expect that in these simulations AGN feedback is strong enough to significantly alter the density profiles of clusters, which will contribute to scales $k<1\kunit$. While both sets of parameter values shown in Table~\ref{tab:modelparams} provide an excellent fit, the parameters for $\tilde{f}_\mathrm{bar,200c}$ provide a marginally better fit for low baryon fractions. We note that the most realistic simulations in the current set have baryon fractions $\tilde{f}_\mathrm{bar,500c}\approx 0.59$ and $\tilde{f}_\mathrm{bar,200c}\approx 0.62$. Figure~\ref{fig:fbar500pow_model} shows the power suppression due to galaxy formation predicted by our best-fit model as a function of both scale and baryon fraction.

Since the predictions of our model depend only on the baryon fraction for groups of galaxies, this allows for accurate corrections of theoretical dark matter only power spectra up to $k=1\kunit$ using a single observable quantity. In a follow-up work, we attempt to extend this model to higher redshifts and smaller scales, which involves the changes feedback makes to the internal structure of group-sized haloes.

\section{Summary and discussion}
\label{sec:discussion}
The goal of this paper was two-fold: to greatly increase the number of matter power spectra from cosmological, hydrodynamical simulations available to the modelling community, and to increase our understanding of how both galaxy formation (particularly AGN feedback) and numerical choices influence these power spectra. To this end, we combined power spectra from the OWLS, cosmo-OWLS and \bahamas{} simulation suites. Relative to the other two sets, \bahamas{} best matches observational constraints, and is therefore expected to provide the most accurate predictions. Additionally, for \bahamas{} we have made improvements to the dark matter only counterpart simulations to ensure they agree with the hydrodynamical simulations on large scales (i.e.\ running them with two dark matter fluids, see \S\ref{subsec:dmonly}). Based on this large set of simulations, we were able to find a tight relation between the baryon fraction at the group scale and the suppression of power for $k<1\kunit$.

Our main findings can be summarized as follows:
\begin{itemize}
\item Increasing or decreasing the effectiveness of AGN feedback changes the matter power spectrum significantly on scales $k\gtrsim 0.1\kunit$ (\S\ref{subsec:varAGN}, Fig.~\ref{fig:diff_Theat}). More effective feedback causes a larger suppression of power.
\item Changes in cosmology at the level of the difference between WMAP and Planck have a small but significant impact on the relative change in the matter power spectrum due to baryonic effects, $\Delta P/P$, on scales $1\lesssim k\lesssim 10\kunit$ (relative shifts up to $30\%$, see \S\ref{subsec:varcosm}, Fig.~\ref{fig:diff_cosm}). The relatives change in the matter power spectrum is less sensitive to changes in the total neutrino mass (Fig.~\ref{fig:diff_nu}).
\item The back-reaction of the redistribution of baryons by cooling and feedback processes on the distribution of cold dark matter causes a mild relative enhancement of power on large scales ($k\lesssim 3\kunit$), but only if AGN feedback is not unrealistically strong (\S\ref{subsec:backreaction}, Fig.~\ref{fig:back_Theat}). On smaller scales, numerical resolution may play a larger role for the back-reaction than for the total matter power spectrum.
\item The suppression of the total matter power spectrum due to feedback generally increases down to redshift zero, the exception being scales $0.8\lesssim k\lesssim 8\kunit$ at late times ($z\leq 0.5$), where the suppression may slightly diminish, likely due to re-accretion of ejected gas (\S\ref{subsec:redshift}, Fig.~\ref{fig:diff_z}).
\item Having no or ineffective stellar feedback in the presence of AGN feedback greatly \emph{increases} the suppression of the matter power spectrum on large scales (\S\ref{subsec:SNAGN}, Fig.~\ref{fig:diff_SNAGN}). This is because more gas is available to accrete onto the supermassive black holes.
\item Comparison of different realizations of the $400\runit$ box of \bahamas{} suggests that it is large enough for cosmic variance to be negligible for the relative total matter power spectrum (\S\ref{subsec:cosvar}, Fig.~\ref{fig:diff_cosmicvar}). This would mean that the ratio of the power spectra from hydrodynamical and dark matter only \bahamas{} simulations can be used to accurately correct matter power spectra from large-volume dark matter only runs, emulator predictions or analytical power spectra up to $k\approx 10\kunit$, leaving only the uncertainty in galaxy formation to be accounted for. However, confirmation using larger volumes is required to rule out effects due to large-scale modes missing from all realizations.
\item We compared the predictions for the effects of galaxy formation on the matter power spectrum of simulations in our set with those of EAGLE, Illustris, IllustrisTNG and Horizon-AGN (\S\ref{subsec:complit}, Fig.~\ref{fig:diffback_literature}). While there is currently no consensus between different groups' simulations, all those with effective (AGN) feedback agree that the power is significantly suppressed relative to dark matter only simulations on all scales $k<10\kunit$, with a maximum suppression $>10\%$. The differences in predictions for large scales can be explained by differences in the effectiveness of feedback as probed by the predicted baryon fractions in massive haloes.
\item We used our large set of matter power spectra presented to show that the suppression of power on large scales is strongly correlated with the mean baryon fraction of $\sim 10^{14}\munitnoh$ haloes. We presented an empirical model capable of predicting the effect of galaxy formation on the matter power spectrum to within $1\%$ for all $k<1\kunit$ given only this baryon fraction (\S\ref{subsec:barfrac}, Fig.~\ref{fig:fbar500pow_panels}). This model also fits the results from the EAGLE, Illustris, IllustrisTNG and Horizon-AGN simulations, albeit with slightly poorer accuracy -- though we note that the uncertainties for these simulations are larger as well, due to their smaller volumes.
\item To avoid large-scale numerical artefacts in the relative difference between the matter power spectra of hydrodynamical and N-body simulations, it is important to consider differences in the initial conditions of the simulations being compared. These include differences in the transfer functions used and/or the number of particles (Appendix~\ref{app:2fluid}).
\end{itemize}

In line with our own findings (\S\ref{subsubsec:understandlitdiff} and \S\ref{subsec:barfrac}), \citet{Semboloni2013} and recently \citet{SchneiderTeyssier2019} have demonstrated that it is possible to predict the ratio of hydrodynamical to N-body matter power spectra quite well given the stellar and gas fractions of the hydrodynamical simulation as a function of halo mass. The gas fraction of groups and clusters correlates strongly with the suppression of power on scales $k\lesssim 10\kunit$, as expected given that the primary way in which the distribution of matter is altered on these scales is through ejection of gas by feedback. At the same time, stellar and AGN feedback also set the stellar fraction in haloes by heating/ejecting gas and quenching star formation. While the stellar fractions or galaxy stellar mass function are often used as a constraint for subgrid recipes in simulations, cluster gas fractions are not (\bahamas{} being one exception, \textsc{fable} (\citealt{Henden2018}) being another). Based on the results presented here, we argue that to accurately predict the clustering of matter, it is important to consider both.

In \S\ref{subsec:barfrac}, we demonstrated that the mean group-scale baryon fraction and the large-scale power suppression due to galaxy formation are strongly correlated. A comparison of cosmo-OWLS and \bahamas{} with simulations from the literature showed that Illustris predicts an unrealistically strong effect of feedback on the matter power spectrum for $k<1\kunit$ due to its too-low baryon fraction in $\sim 10^{14}\munitnoh$ haloes compared to observations, while EAGLE, IllustrisTNG and Horizon-AGN predict an unrealistically weak effect due to their too-high baryon fraction at the same mass scale. We conclude that to study the effect of baryonic processes on large scales, only simulations that reproduce the observed baryon fraction of large groups of galaxies are suitable. Of the simulations we compared to here, only the (fiducial) \bahamas{} simulations satisfy that requirement.

Improving observational constraints on gas fractions and profiles in massive haloes will be of tremendous importance in the near future. Based on the results of \S\ref{subsec:barfrac}, we argue that constraining the baryon fractions of group-scale haloes will be especially vital in the short term, as the model presented there may allow us to accurately correct dark matter only power spectra used to interpret weak lensing surveys \citep[e.g.][]{Potter2017,Knabenhans2019} for the effects of galaxy formation to the required $1\%$ precision down to $k=1\kunit$ at $z=0$ and possibly beyond.

While outside the scope of the current paper, a closer look at the predicted redshift evolution of the relative power spectrum would be of interest and provide a more detailed look into the origin of the differences seen between different simulations, e.g.\ whether gas is still being ejected at $z=0$ or whether this happened at early times and is now re-accreting. Ideally, measurements of the baryon fraction could be used to predict the large-scale suppression of power at any redshift.

Total and dark matter power spectra for the $92$ simulations listed in Table~\ref{tab:sims}, which includes $35$ simulations with AGN feedback, will be made publicly available at \texttt{powerlib.strw.leidenuniv.nl}. As stated in \S\ref{sec:introduction}, the underlying model in simulations with AGN is the same in each of the simulations presented here, though the parameter values differ.

\section*{Acknowledgements}
The authors thank Volker Springel for providing power spectra for Illustris and IllustrisTNG, and \citet{Chisari2018} for making those of Horizon-AGN publicly available. The authors additionally thank the Illustris and Illustris TNG teams for their publicly available group catalogues, and Volker Springel for providing data on the baryon content of massive haloes in Illustris TNG. MvD thanks Simeon Bird for useful discussions on the neutrino power spectrum.

MvD gratefully acknowledges support from the Netherlands Research Council NWO (VENI grant Nr. 639.041.748).

This project has received funding from the European Research Council (ERC) under the European Union's Horizon 2020 research and innovation programme (grant agreement No 769130).

This work used the DiRAC@Durham facility managed by the Institute for
Computational Cosmology on behalf of the STFC DiRAC HPC Facility
(\texttt{www.dirac.ac.uk}). The equipment was funded by BEIS capital funding
via STFC capital grants ST/K00042X/1, ST/P002293/1, ST/R002371/1 and
ST/S002502/1, Durham University and STFC operations grant
ST/R000832/1. DiRAC is part of the National e-Infrastructure.
\bibliographystyle{mn2e}
\setlength{\bibhang}{2.0em}
\setlength{\labelwidth}{0.0em}
\bibliography{powerlib}

\begin{thebibliography}{87}
\expandafter\ifx\csname natexlab\endcsname\relax\def\natexlab#1{#1}\fi

\bibitem[{{Ali-Ha{\"i}moud} \& {Bird}(2013)}]{Ali-HaimoudBird2013}
{Ali-Ha{\"i}moud} Y., {Bird} S., 2013, \mnras, 428, 3375

\bibitem[{{Angulo}, {Hahn} \& {Abel}(2013){Angulo}, {Hahn}, \&
  {Abel}}]{Angulo2013}
{Angulo} R.~E., {Hahn} O., {Abel} T., 2013, \mnras, 434, 1756

\bibitem[{{Barnes} {et~al}\mbox{.}(2018){Barnes}, {Vogelsberger}, {Kannan},
  {Marinacci}, {Weinberger}, {Springel}, {Torrey}, {Pillepich}, {Nelson},
  {Pakmor}, {Naiman}, {Hernquist}, \& {McDonald}}]{Barnes2018}
{Barnes} D.~J. {et~al.}, 2018, \mnras, 481, 1809

\bibitem[{{Booth} \& {Schaye}(2009)}]{BoothSchaye2009}
{Booth} C.~M., {Schaye} J., 2009, \mnras, 398, 53

\bibitem[{{Booth} \& {Schaye}(2013)}]{BoothSchaye2013}
---, 2013, Scientific Reports, 1738

\bibitem[{{Bourne}, {Zubovas} \& {Nayakshin}(2015){Bourne}, {Zubovas}, \&
  {Nayakshin}}]{Bourne2015}
{Bourne} M.~A., {Zubovas} K., {Nayakshin} S., 2015, \mnras, 453, 1829

\bibitem[{{Bower} {et~al}\mbox{.}(2017){Bower}, {Schaye}, {Frenk}, {Theuns},
  {Schaller}, {Crain}, \& {McAlpine}}]{Bower2017}
{Bower} R.~G., {Schaye} J., {Frenk} C.~S., {Theuns} T., {Schaller} M., {Crain}
  R.~A., {McAlpine} S., 2017, \mnras, 465, 32

\bibitem[{{Budzynski} {et~al}\mbox{.}(2014){Budzynski}, {Koposov}, {McCarthy},
  \& {Belokurov}}]{Budzynski2014}
{Budzynski} J.~M., {Koposov} S.~E., {McCarthy} I.~G., {Belokurov} V., 2014,
  \mnras, 437, 1362

\bibitem[{{Chabrier}(2003)}]{Chabrier2003}
{Chabrier} G., 2003, \pasp, 115, 763

\bibitem[{{Chisari} {et~al}\mbox{.}(2018){Chisari}, {Richardson}, {Devriendt},
  {Dubois}, {Schneider}, {Le Brun}, {Beckmann}, {Peirani}, {Slyz}, \&
  {Pichon}}]{Chisari2018}
{Chisari} N.~E. {et~al.}, 2018, \mnras, 480, 3962

\bibitem[{{Colombi} {et~al}\mbox{.}(2009){Colombi}, {Jaffe}, {Novikov}, \&
  {Pichon}}]{Colombi2009}
{Colombi} S., {Jaffe} A., {Novikov} D., {Pichon} C., 2009, \mnras, 393, 511

\bibitem[{{Copeland}, {Taylor} \& {Hall}(2018){Copeland}, {Taylor}, \&
  {Hall}}]{Copeland2018}
{Copeland} D., {Taylor} A., {Hall} A., 2018, \mnras, 480, 2247

\bibitem[{{Dalla Vecchia} \& {Schaye}(2012)}]{DallaVecchiaSchaye2012}
{Dalla Vecchia} C., {Schaye} J., 2012, \mnras, 426, 140

\bibitem[{{Debackere}, {Schaye} \& {Hoekstra}(2019){Debackere}, {Schaye}, \&
  {Hoekstra}}]{Debackere2019}
{Debackere} S.~N.~B., {Schaye} J., {Hoekstra} H., 2019, preprint
  (arXiv:1908.05765)

\bibitem[{{Dubois} {et~al}\mbox{.}(2012){Dubois}, {Devriendt}, {Slyz}, \&
  {Teyssier}}]{Dubois2012}
{Dubois} Y., {Devriendt} J., {Slyz} A., {Teyssier} R., 2012, \mnras, 420, 2662

\bibitem[{{Eifler} {et~al}\mbox{.}(2015){Eifler}, {Krause}, {Dodelson},
  {Zentner}, {Hearin}, \& {Gnedin}}]{Eifler2015}
{Eifler} T., {Krause} E., {Dodelson} S., {Zentner} A.~R., {Hearin} A.~P.,
  {Gnedin} N.~Y., 2015, \mnras, 454, 2451

\bibitem[{{Foreman}, {Becker} \& {Wechsler}(2016){Foreman}, {Becker}, \&
  {Wechsler}}]{Foreman2016}
{Foreman} S., {Becker} M.~R., {Wechsler} R.~H., 2016, \mnras, 463, 3326

\bibitem[{{Gonzalez} {et~al}\mbox{.}(2013){Gonzalez}, {Sivanandam},
  {Zabludoff}, \& {Zaritsky}}]{Gonzalez2013}
{Gonzalez} A.~H., {Sivanandam} S., {Zabludoff} A.~I., {Zaritsky} D., 2013,
  \apj, 778, 14

\bibitem[{{Hahn} {et~al}\mbox{.}(2017){Hahn}, {Martizzi}, {Wu}, {Evrard},
  {Teyssier}, \& {Wechsler}}]{Hahn2017}
{Hahn} O., {Martizzi} D., {Wu} H.-Y., {Evrard} A.~E., {Teyssier} R., {Wechsler}
  R.~H., 2017, \mnras, 470, 166

\bibitem[{{Harnois-D{\'e}raps} {et~al}\mbox{.}(2015){Harnois-D{\'e}raps}, {van
  Waerbeke}, {Viola}, \& {Heymans}}]{Harnois-Deraps2015}
{Harnois-D{\'e}raps} J., {van Waerbeke} L., {Viola} M., {Heymans} C., 2015,
  \mnras, 450, 1212

\bibitem[{{Hellwing} {et~al}\mbox{.}(2016){Hellwing}, {Schaller}, {Frenk},
  {Theuns}, {Schaye}, {Bower}, \& {Crain}}]{Hellwing2016}
{Hellwing} W.~A., {Schaller} M., {Frenk} C.~S., {Theuns} T., {Schaye} J.,
  {Bower} R.~G., {Crain} R.~A., 2016, \mnras, 461, L11

\bibitem[{{Henden} {et~al}\mbox{.}(2018){Henden}, {Puchwein}, {Shen}, \&
  {Sijacki}}]{Henden2018}
{Henden} N.~A., {Puchwein} E., {Shen} S., {Sijacki} D., 2018, \mnras, 479, 5385

\bibitem[{{Hildebrandt} {et~al}\mbox{.}(2017){Hildebrandt}, {Viola}, {Heymans},
  {Joudaki}, {Kuijken}, {Blake}, {Erben}, {Joachimi}, {Klaes}, {Miller},
  {Morrison}, {Nakajima}, {Verdoes Kleijn}, {Amon}, {Choi}, {Covone}, {de
  Jong}, {Dvornik}, {Fenech Conti}, {Grado}, {Harnois-D{\'e}raps}, {Herbonnet},
  {Hoekstra}, {K{\"o}hlinger}, {McFarland}, {Mead}, {Merten}, {Napolitano},
  {Peacock}, {Radovich}, {Schneider}, {Simon}, {Valentijn}, {van den Busch},
  {van Uitert}, \& {Van Waerbeke}}]{Hildebrandt2017}
{Hildebrandt} H. {et~al.}, 2017, \mnras, 465, 1454

\bibitem[{{Hinshaw} {et~al}\mbox{.}(2013){Hinshaw}, {Larson}, {Komatsu},
  {Spergel}, {Bennett}, {Dunkley}, {Nolta}, {Halpern}, {Hill}, {Odegard},
  {Page}, {Smith}, {Weiland}, {Gold}, {Jarosik}, {Kogut}, {Limon}, {Meyer},
  {Tucker}, {Wollack}, \& {Wright}}]{Hinshaw2013}
{Hinshaw} G. {et~al.}, 2013, \apjs, 208, 19

\bibitem[{{Huang} {et~al}\mbox{.}(2018){Huang}, {Eifler}, {Mandelbaum}, \&
  {Dodelson}}]{Huang2019}
{Huang} H.-J., {Eifler} T., {Mandelbaum} R., {Dodelson} S., 2018, ArXiv
  e-prints

\bibitem[{{Huterer} \& {Takada}(2005)}]{HutererTakada2005}
{Huterer} D., {Takada} M., 2005, Astroparticle Physics, 23, 369

\bibitem[{{Ivezi{\'c}} {et~al}\mbox{.}(2008){Ivezi{\'c}}, {Kahn}, {Tyson},
  {Abel}, {Acosta}, {Allsman}, {Alonso}, {AlSayyad}, {Anderson}, {Andrew}, \&
  et~al.}]{Ivezic2008}
{Ivezi{\'c}} {\v Z}. {et~al.}, 2008, preprint (arXiv:0805.2366)

\bibitem[{{Joudaki} {et~al}\mbox{.}(2017){Joudaki}, {Blake}, {Heymans}, {Choi},
  {Harnois-Deraps}, {Hildebrandt}, {Joachimi}, {Johnson}, {Mead}, {Parkinson},
  {Viola}, \& {van Waerbeke}}]{Joudaki2017}
{Joudaki} S. {et~al.}, 2017, \mnras, 465, 2033

\bibitem[{{Khandai} {et~al}\mbox{.}(2015){Khandai}, {Di Matteo}, {Croft},
  {Wilkins}, {Feng}, {Tucker}, {DeGraf}, \& {Liu}}]{Khandai2015}
{Khandai} N., {Di Matteo} T., {Croft} R., {Wilkins} S., {Feng} Y., {Tucker} E.,
  {DeGraf} C., {Liu} M.-S., 2015, \mnras, 450, 1349

\bibitem[{{Knabenhans} {et~al}\mbox{.}(2019){Knabenhans}, {Stadel}, {Marelli},
  {Potter}, {Teyssier}, {Legrand}, {Schneider}, {Sudret}, {Blot}, {Awan},
  {Burigana}, {Carvalho}, {Kurki-Suonio}, \& {Sirri}}]{Knabenhans2019}
{Knabenhans} M. {et~al.}, 2019, \mnras, 484, 5509

\bibitem[{{Komatsu} {et~al}\mbox{.}(2009){Komatsu}, {Dunkley}, {Nolta},
  {Bennett}, {Gold}, {Hinshaw}, {Jarosik}, {Larson}, {Limon}, {Page},
  {Spergel}, {Halpern}, {Hill}, {Kogut}, {Meyer}, {Tucker}, {Weiland},
  {Wollack}, \& {Wright}}]{Komatsu2009}
{Komatsu} E. {et~al.}, 2009, \apjs, 180, 330

\bibitem[{{Komatsu} {et~al}\mbox{.}(2011){Komatsu}, {Smith}, {Dunkley},
  {Bennett}, {Gold}, {Hinshaw}, {Jarosik}, {Larson}, {Nolta}, {Page},
  {Spergel}, {Halpern}, {Hill}, {Kogut}, {Limon}, {Meyer}, {Odegard}, {Tucker},
  {Weiland}, {Wollack}, \& {Wright}}]{Komatsu2011}
---, 2011, \apjs, 192, 18

\bibitem[{{Kravtsov}, {Vikhlinin} \& {Meshcheryakov}(2018){Kravtsov},
  {Vikhlinin}, \& {Meshcheryakov}}]{Kravtsov2018}
{Kravtsov} A.~V., {Vikhlinin} A.~A., {Meshcheryakov} A.~V., 2018, Astronomy
  Letters, 44, 8

\bibitem[{{Laureijs}(2009)}]{Laureijs2009}
{Laureijs} R., 2009, preprint (arXiv:0912.0914)

\bibitem[{{Le Brun} {et~al}\mbox{.}(2014){Le Brun}, {McCarthy}, {Schaye}, \&
  {Ponman}}]{LeBrun2014}
{Le Brun} A.~M.~C., {McCarthy} I.~G., {Schaye} J., {Ponman} T.~J., 2014,
  \mnras, 441, 1270

\bibitem[{{Lin} {et~al}\mbox{.}(2012){Lin}, {Stanford}, {Eisenhardt},
  {Vikhlinin}, {Maughan}, \& {Kravtsov}}]{Lin2012}
{Lin} Y.-T., {Stanford} S.~A., {Eisenhardt} P.~R.~M., {Vikhlinin} A., {Maughan}
  B.~J., {Kravtsov} A., 2012, \apjl, 745, L3

\bibitem[{{Lovisari}, {Reiprich} \& {Schellenberger}(2015){Lovisari},
  {Reiprich}, \& {Schellenberger}}]{Lovisari2015}
{Lovisari} L., {Reiprich} T.~H., {Schellenberger} G., 2015, \aap, 573, A118

\bibitem[{{Maughan} {et~al}\mbox{.}(2008){Maughan}, {Jones}, {Forman}, \& {Van
  Speybroeck}}]{Maughan2008}
{Maughan} B.~J., {Jones} C., {Forman} W., {Van Speybroeck} L., 2008, \apjs,
  174, 117

\bibitem[{{McCarthy} {et~al}\mbox{.}(2018){McCarthy}, {Bird}, {Schaye},
  {Harnois-Deraps}, {Font}, \& {van Waerbeke}}]{McCarthy2018}
{McCarthy} I.~G., {Bird} S., {Schaye} J., {Harnois-Deraps} J., {Font} A.~S.,
  {van Waerbeke} L., 2018, \mnras, 476, 2999

\bibitem[{{McCarthy} {et~al}\mbox{.}(2017){McCarthy}, {Schaye}, {Bird}, \& {Le
  Brun}}]{McCarthy2017}
{McCarthy} I.~G., {Schaye} J., {Bird} S., {Le Brun} A.~M.~C., 2017, \mnras,
  465, 2936

\bibitem[{{McCarthy} {et~al}\mbox{.}(2011){McCarthy}, {Schaye}, {Bower},
  {Ponman}, {Booth}, {Dalla Vecchia}, \& {Springel}}]{McCarthy2011}
{McCarthy} I.~G., {Schaye} J., {Bower} R.~G., {Ponman} T.~J., {Booth} C.~M.,
  {Dalla Vecchia} C., {Springel} V., 2011, \mnras, 412, 1965

\bibitem[{{McCarthy} {et~al}\mbox{.}(2010){McCarthy}, {Schaye}, {Ponman},
  {Bower}, {Booth}, {Dalla Vecchia}, {Crain}, {Springel}, {Theuns}, \&
  {Wiersma}}]{McCarthy2010}
{McCarthy} I.~G. {et~al.}, 2010, \mnras, 406, 822

\bibitem[{{Mead} {et~al}\mbox{.}(2016){Mead}, {Heymans}, {Lombriser},
  {Peacock}, {Steele}, \& {Winther}}]{Mead2016}
{Mead} A.~J., {Heymans} C., {Lombriser} L., {Peacock} J.~A., {Steele} O.~I.,
  {Winther} H.~A., 2016, \mnras, 459, 1468

\bibitem[{{Mead} {et~al}\mbox{.}(2015){Mead}, {Peacock}, {Heymans}, {Joudaki},
  \& {Heavens}}]{Mead2015}
{Mead} A.~J., {Peacock} J.~A., {Heymans} C., {Joudaki} S., {Heavens} A.~F.,
  2015, \mnras, 454, 1958

\bibitem[{{Mohammed} \& {Gnedin}(2018)}]{MohammedGnedin2018}
{Mohammed} I., {Gnedin} N.~Y., 2018, \apj, 863, 173

\bibitem[{{Mohammed} \& {Seljak}(2014)}]{MohammedSeljak2014}
{Mohammed} I., {Seljak} U., 2014, \mnras, 445, 3382

\bibitem[{{Mummery} {et~al}\mbox{.}(2017){Mummery}, {McCarthy}, {Bird}, \&
  {Schaye}}]{Mummery2017}
{Mummery} B.~O., {McCarthy} I.~G., {Bird} S., {Schaye} J., 2017, \mnras, 471,
  227

\bibitem[{{Nelson} {et~al}\mbox{.}(2015){Nelson}, {Pillepich}, {Genel},
  {Vogelsberger}, {Springel}, {Torrey}, {Rodriguez-Gomez}, {Sijacki}, {Snyder},
  {Griffen}, {Marinacci}, {Blecha}, {Sales}, {Xu}, \& {Hernquist}}]{Nelson2015}
{Nelson} D. {et~al.}, 2015, Astronomy and Computing, 13, 12

\bibitem[{{O'Leary} \& {McQuinn}(2012)}]{OLearyMcQuinn2012}
{O'Leary} R.~M., {McQuinn} M., 2012, \apj, 760, 4

\bibitem[{{Oppenheimer} \& {Dav{\'e}}(2006)}]{OppenheimerDave2006}
{Oppenheimer} B.~D., {Dav{\'e}} R., 2006, \mnras, 373, 1265

\bibitem[{{Pearson} {et~al}\mbox{.}(2017){Pearson}, {Ponman}, {Norberg},
  {Robotham}, {Babul}, {Bower}, {McCarthy}, {Brough}, {Driver}, \&
  {Pimbblet}}]{Pearson2017}
{Pearson} R.~J. {et~al.}, 2017, \mnras, 469, 3489

\bibitem[{{Peters} {et~al}\mbox{.}(2018){Peters}, {Brown}, {Kay}, \&
  {Barnes}}]{Peters2018}
{Peters} A., {Brown} M.~L., {Kay} S.~T., {Barnes} D.~J., 2018, \mnras, 474,
  3173

\bibitem[{{Pike} {et~al}\mbox{.}(2014){Pike}, {Kay}, {Newton}, {Thomas}, \&
  {Jenkins}}]{Pike2014}
{Pike} S.~R., {Kay} S.~T., {Newton} R.~D.~A., {Thomas} P.~A., {Jenkins} A.,
  2014, \mnras, 445, 1774

\bibitem[{{Pillepich} {et~al}\mbox{.}(2018){Pillepich}, {Springel}, {Nelson},
  {Genel}, {Naiman}, {Pakmor}, {Hernquist}, {Torrey}, {Vogelsberger},
  {Weinberger}, \& {Marinacci}}]{Pillepich2018a}
{Pillepich} A. {et~al.}, 2018, \mnras, 473, 4077

\bibitem[{{Planck Collaboration} {et~al}\mbox{.}(2014){Planck Collaboration},
  {Ade}, {Aghanim}, {Armitage-Caplan}, {Arnaud}, {Ashdown}, {Atrio-Barandela},
  {Aumont}, {Baccigalupi}, {Banday}, \& et~al.}]{Planck2014}
{Planck Collaboration} {et~al.}, 2014, \aap, 571, A16

\bibitem[{{Planck Collaboration} {et~al}\mbox{.}(2016){Planck Collaboration},
  {Ade}, {Aghanim}, {Arnaud}, {Ashdown}, {Aumont}, {Baccigalupi}, {Banday},
  {Barreiro}, {Bartlett}, \& et~al.}]{Planck2016}
---, 2016, \aap, 594, A13

\bibitem[{{Planck Collaboration} {et~al}\mbox{.}(2018){Planck Collaboration},
  {Aghanim}, {Akrami}, {Ashdown}, {Aumont}, {Baccigalupi}, {Ballardini},
  {Banday}, {Barreiro}, {Bartolo}, {Basak}, {Battye}, {Benabed}, {Bernard},
  {Bersanelli}, {Bielewicz}, {Bock}, {Bond}, {Borrill}, {Bouchet}, {Boulanger},
  {Bucher}, {Burigana}, {Butler}, {Calabrese}, {Cardoso}, {Carron},
  {Challinor}, {Chiang}, {Chluba}, {Colombo}, {Combet}, {Contreras}, {Crill},
  {Cuttaia}, {de Bernardis}, {de Zotti}, {Delabrouille}, {Delouis}, {Di
  Valentino}, {Diego}, {Dor{\'e}}, {Douspis}, {Ducout}, {Dupac}, {Dusini},
  {Efstathiou}, {Elsner}, {En{\ss}lin}, {Eriksen}, {Fantaye}, {Farhang},
  {Fergusson}, {Fernandez-Cobos}, {Finelli}, {Forastieri}, {Frailis},
  {Franceschi}, {Frolov}, {Galeotta}, {Galli}, {Ganga}, {G{\'e}nova-Santos},
  {Gerbino}, {Ghosh}, {Gonz{\'a}lez-Nuevo}, {G{\'o}rski}, {Gratton},
  {Gruppuso}, {Gudmundsson}, {Hamann}, {Handley}, {Herranz}, {Hivon}, {Huang},
  {Jaffe}, {Jones}, {Karakci}, {Keih{\"a}nen}, {Keskitalo}, {Kiiveri}, {Kim},
  {Kisner}, {Knox}, {Krachmalnicoff}, {Kunz}, {Kurki-Suonio}, {Lagache},
  {Lamarre}, {Lasenby}, {Lattanzi}, {Lawrence}, {Le Jeune}, {Lemos},
  {Lesgourgues}, {Levrier}, {Lewis}, {Liguori}, {Lilje}, {Lilley}, {Lindholm},
  {L{\'o}pez-Caniego}, {Lubin}, {Ma}, {Mac{\'{\i}}as-P{\'e}rez}, {Maggio},
  {Maino}, {Mandolesi}, {Mangilli}, {Marcos-Caballero}, {Maris}, {Martin},
  {Martinelli}, {Mart{\'{\i}}nez-Gonz{\'a}lez}, {Matarrese}, {Mauri}, {McEwen},
  {Meinhold}, {Melchiorri}, {Mennella}, {Migliaccio}, {Millea}, {Mitra},
  {Miville-Desch{\^e}nes}, {Molinari}, {Montier}, {Morgante}, {Moss}, {Natoli},
  {N{\o}rgaard-Nielsen}, {Pagano}, {Paoletti}, {Partridge}, {Patanchon},
  {Peiris}, {Perrotta}, {Pettorino}, {Piacentini}, {Polastri}, {Polenta},
  {Puget}, {Rachen}, {Reinecke}, {Remazeilles}, {Renzi}, {Rocha}, {Rosset},
  {Roudier}, {Rubi{\~n}o-Mart{\'{\i}}n}, {Ruiz-Granados}, {Salvati}, {Sandri},
  {Savelainen}, {Scott}, {Shellard}, {Sirignano}, {Sirri}, {Spencer},
  {Sunyaev}, {Suur-Uski}, {Tauber}, {Tavagnacco}, {Tenti}, {Toffolatti},
  {Tomasi}, {Trombetti}, {Valenziano}, {Valiviita}, {Van Tent}, {Vibert},
  {Vielva}, {Villa}, {Vittorio}, {Wandelt}, {Wehus}, {White}, {White},
  {Zacchei}, \& {Zonca}}]{Planck2018}
---, 2018, preprint (arXiv:1807.06209)

\bibitem[{{Potter}, {Stadel} \& {Teyssier}(2017){Potter}, {Stadel}, \&
  {Teyssier}}]{Potter2017}
{Potter} D., {Stadel} J., {Teyssier} R., 2017, Computational Astrophysics and
  Cosmology, 4, 2

\bibitem[{{Pratt} {et~al}\mbox{.}(2009){Pratt}, {Croston}, {Arnaud}, \&
  {B{\"o}hringer}}]{Pratt2009}
{Pratt} G.~W., {Croston} J.~H., {Arnaud} M., {B{\"o}hringer} H., 2009, \aap,
  498, 361

\bibitem[{{Rasmussen} \& {Ponman}(2009)}]{Rasmussen2009}
{Rasmussen} J., {Ponman} T.~J., 2009, \mnras, 399, 239

\bibitem[{{Salpeter}(1955)}]{Salpeter1955}
{Salpeter} E.~E., 1955, \apj, 121, 161

\bibitem[{{Sanderson} {et~al}\mbox{.}(2013){Sanderson}, {O'Sullivan}, {Ponman},
  {Gonzalez}, {Sivanandam}, {Zabludoff}, \& {Zaritsky}}]{Sanderson2013}
{Sanderson} A.~J.~R., {O'Sullivan} E., {Ponman} T.~J., {Gonzalez} A.~H.,
  {Sivanandam} S., {Zabludoff} A.~I., {Zaritsky} D., 2013, \mnras, 429, 3288

\bibitem[{{Schaye} {et~al}\mbox{.}(2015){Schaye}, {Crain}, {Bower}, {Furlong},
  {Schaller}, {Theuns}, {Dalla Vecchia}, {Frenk}, {McCarthy}, {Helly},
  {Jenkins}, {Rosas-Guevara}, {White}, {Baes}, {Booth}, {Camps}, {Navarro},
  {Qu}, {Rahmati}, {Sawala}, {Thomas}, \& {Trayford}}]{Schaye2015}
{Schaye} J. {et~al.}, 2015, \mnras, 446, 521

\bibitem[{{Schaye} {et~al}\mbox{.}(2010){Schaye}, {Dalla Vecchia}, {Booth},
  {Wiersma}, {Theuns}, {Haas}, {Bertone}, {Duffy}, {McCarthy}, \& {van de
  Voort}}]{Schaye2010}
---, 2010, \mnras, 402, 1536

\bibitem[{{Schneider} \& {Teyssier}(2015)}]{SchneiderTeyssier2015}
{Schneider} A., {Teyssier} R., 2015, \jcap, 12, 049

\bibitem[{{Schneider} {et~al}\mbox{.}(2016){Schneider}, {Teyssier}, {Potter},
  {Stadel}, {Onions}, {Reed}, {Smith}, {Springel}, {Pearce}, \&
  {Scoccimarro}}]{Schneider2016}
{Schneider} A. {et~al.}, 2016, \jcap, 4, 047

\bibitem[{{Schneider} {et~al}\mbox{.}(2019){Schneider}, {Teyssier}, {Stadel},
  {Chisari}, {Le Brun}, {Amara}, \& {Refregier}}]{SchneiderTeyssier2019}
{Schneider} A., {Teyssier} R., {Stadel} J., {Chisari} N.~E., {Le Brun}
  A.~M.~C., {Amara} A., {Refregier} A., 2019, \jcap, 3, 020

\bibitem[{{Semboloni}, {Hoekstra} \& {Schaye}(2013){Semboloni}, {Hoekstra}, \&
  {Schaye}}]{Semboloni2013}
{Semboloni} E., {Hoekstra} H., {Schaye} J., 2013, \mnras, 434, 148

\bibitem[{{Semboloni} {et~al}\mbox{.}(2011){Semboloni}, {Hoekstra}, {Schaye},
  {van Daalen}, \& {McCarthy}}]{Semboloni2011}
{Semboloni} E., {Hoekstra} H., {Schaye} J., {van Daalen} M.~P., {McCarthy}
  I.~G., 2011, \mnras, 417, 2020

\bibitem[{{Spergel} {et~al}\mbox{.}(2007){Spergel}, {Bean}, {Dor{\'e}},
  {Nolta}, {Bennett}, {Dunkley}, {Hinshaw}, {Jarosik}, {Komatsu}, {Page},
  {Peiris}, {Verde}, {Halpern}, {Hill}, {Kogut}, {Limon}, {Meyer}, {Odegard},
  {Tucker}, {Weiland}, {Wollack}, \& {Wright}}]{Spergel2007}
{Spergel} D.~N. {et~al.}, 2007, \apjs, 170, 377

\bibitem[{{Springel}, {Di Matteo} \& {Hernquist}(2005){Springel}, {Di Matteo},
  \& {Hernquist}}]{Springel2005}
{Springel} V., {Di Matteo} T., {Hernquist} L., 2005, \mnras, 361, 776

\bibitem[{{Springel} {et~al}\mbox{.}(2018){Springel}, {Pakmor}, {Pillepich},
  {Weinberger}, {Nelson}, {Hernquist}, {Vogelsberger}, {Genel}, {Torrey},
  {Marinacci}, \& {Naiman}}]{Springel2018}
{Springel} V. {et~al.}, 2018, \mnras, 475, 676

\bibitem[{{Stafford} {et~al}\mbox{.}(2019){Stafford}, {McCarthy}, {Crain},
  {Salcido}, {Schaye}, {Font}, {Kwan}, \& {Pfeifer}}]{Stafford2019}
{Stafford} S.~G., {McCarthy} I.~G., {Crain} R.~A., {Salcido} J., {Schaye} J.,
  {Font} A.~S., {Kwan} J., {Pfeifer} S., 2019, arXiv e-prints, arXiv:1907.09497

\bibitem[{{Sun} {et~al}\mbox{.}(2009){Sun}, {Voit}, {Donahue}, {Jones},
  {Forman}, \& {Vikhlinin}}]{Sun2009}
{Sun} M., {Voit} G.~M., {Donahue} M., {Jones} C., {Forman} W., {Vikhlinin} A.,
  2009, \apj, 693, 1142

\bibitem[{{Tenneti} {et~al}\mbox{.}(2015){Tenneti}, {Mandelbaum}, {Di Matteo},
  {Kiessling}, \& {Khandai}}]{Tenneti2015}
{Tenneti} A., {Mandelbaum} R., {Di Matteo} T., {Kiessling} A., {Khandai} N.,
  2015, \mnras, 453, 469

\bibitem[{{Valkenburg} \& {Villaescusa-Navarro}(2017)}]{Valkenburg2017}
{Valkenburg} W., {Villaescusa-Navarro} F., 2017, \mnras, 467, 4401

\bibitem[{{van Daalen} \& {Schaye}(2015)}]{vanDaalenSchaye2015}
{van Daalen} M.~P., {Schaye} J., 2015, \mnras, 452, 2247

\bibitem[{{van Daalen} {et~al}\mbox{.}(2011){van Daalen}, {Schaye}, {Booth}, \&
  {Dalla Vecchia}}]{vanDaalen2011}
{van Daalen} M.~P., {Schaye} J., {Booth} C.~M., {Dalla Vecchia} C., 2011,
  \mnras, 415, 3649

\bibitem[{{van Daalen} {et~al}\mbox{.}(2014){van Daalen}, {Schaye}, {McCarthy},
  {Booth}, \& {Dalla Vecchia}}]{vanDaalen2014}
{van Daalen} M.~P., {Schaye} J., {McCarthy} I.~G., {Booth} C.~M., {Dalla
  Vecchia} C., 2014, \mnras, 440, 2997

\bibitem[{{van Uitert} {et~al}\mbox{.}(2018){van Uitert}, {Joachimi},
  {Joudaki}, {Amon}, {Heymans}, {K{\"o}hlinger}, {Asgari}, {Blake}, {Choi},
  {Erben}, {Farrow}, {Harnois-D{\'e}raps}, {Hildebrandt}, {Hoekstra},
  {Kitching}, {Klaes}, {Kuijken}, {Merten}, {Miller}, {Nakajima}, {Schneider},
  {Valentijn}, \& {Viola}}]{vanUitert2018}
{van Uitert} E. {et~al.}, 2018, \mnras, 476, 4662

\bibitem[{{Vikhlinin} {et~al}\mbox{.}(2006){Vikhlinin}, {Kravtsov}, {Forman},
  {Jones}, {Markevitch}, {Murray}, \& {Van Speybroeck}}]{Vikhlinin2006}
{Vikhlinin} A., {Kravtsov} A., {Forman} W., {Jones} C., {Markevitch} M.,
  {Murray} S.~S., {Van Speybroeck} L., 2006, \apj, 640, 691

\bibitem[{{Vogelsberger} {et~al}\mbox{.}(2014{\natexlab{a}}){Vogelsberger},
  {Genel}, {Springel}, {Torrey}, {Sijacki}, {Xu}, {Snyder}, {Bird}, {Nelson},
  \& {Hernquist}}]{Vogelsberger2014a}
{Vogelsberger} M. {et~al.}, 2014{\natexlab{a}}, \nat, 509, 177

\bibitem[{{Vogelsberger} {et~al}\mbox{.}(2014{\natexlab{b}}){Vogelsberger},
  {Genel}, {Springel}, {Torrey}, {Sijacki}, {Xu}, {Snyder}, {Nelson}, \&
  {Hernquist}}]{Vogelsberger2014b}
---, 2014{\natexlab{b}}, \mnras, 444, 1518

\bibitem[{{Weinberger} {et~al}\mbox{.}(2017){Weinberger}, {Springel},
  {Hernquist}, {Pillepich}, {Marinacci}, {Pakmor}, {Nelson}, {Genel},
  {Vogelsberger}, {Naiman}, \& {Torrey}}]{Weinberger2017}
{Weinberger} R. {et~al.}, 2017, \mnras, 465, 3291

\bibitem[{{Wiersma} {et~al}\mbox{.}(2009){Wiersma}, {Schaye}, {Theuns}, {Dalla
  Vecchia}, \& {Tornatore}}]{Wiersma2009}
{Wiersma} R.~P.~C., {Schaye} J., {Theuns} T., {Dalla Vecchia} C., {Tornatore}
  L., 2009, \mnras, 399, 574

\bibitem[{{Yoon} {et~al}\mbox{.}(2019){Yoon}, {Jee}, {Tyson}, {Schmidt},
  {Wittman}, \& {Choi}}]{Yoon2019}
{Yoon} M., {Jee} M.~J., {Tyson} J.~A., {Schmidt} S., {Wittman} D., {Choi} A.,
  2019, \apj, 870, 111

\bibitem[{{Zentner} {et~al}\mbox{.}(2013){Zentner}, {Semboloni}, {Dodelson},
  {Eifler}, {Krause}, \& {Hearin}}]{Zentner2013}
{Zentner} A.~R., {Semboloni} E., {Dodelson} S., {Eifler} T., {Krause} E.,
  {Hearin} A.~P., 2013, \prd, 87, 043509

\end{thebibliography}

\appendix
\section{The role of box size and resolution, or the importance of calibration}
\label{app:restests}
The results of cosmological, hydrodynamical simulations often depend on the resolution at which they were run, particularly when subgrid recipes are involved. As subgrid recipes aim to model processes happening on unresolved scales, based on the information available on scales that are resolved, their outcomes are inherently dependent on the resolution. The prescription may give a very similar outcome in some range of resolved scales, but if the resolution changes significantly then its parameters will have to be adapted, or the prescription itself modified or removed.

Here, we test the effect of changing the resolution and/or the box size of the simulations on the relative effect that galaxy formation has on the matter power spectrum. We focus here on the cosmo-OWLS AGN model, for which appropriate simulations were available, but since the underlying subgrid prescriptions and resolutions are the same in OWLS and \bahamas{}, the results can confidently be extended to those AGN simulations as well.

\begin{figure}
\begin{center}
\includegraphics[width=1.0\columnwidth, trim=16mm 8mm 10mm -4mm]{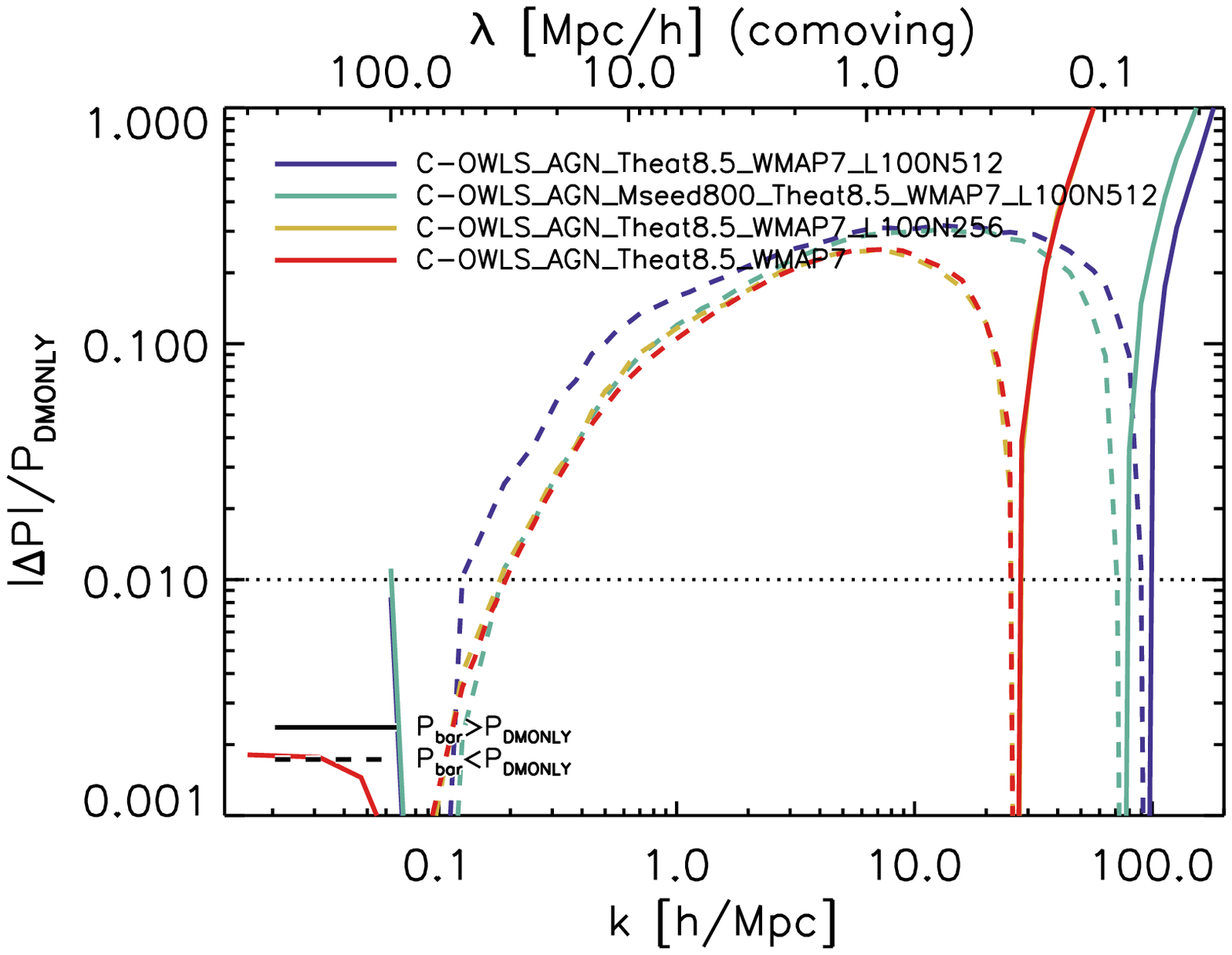}
\caption{The dependence of the effect of galaxy formation on box size and resolution. In blue we show the relative effect on the matter power spectrum for a WMAP7 cosmo-OWLS AGN simulation with a $100\runit$ box and $2\times 512^3$ particles and a relatively high heating temperature of $10^{8.5}\Tunit$. The green line shows the effects of increasing the minimum halo mass for black hole seeding, matching that of a simulation with an $8\times$ worse mass resolution. In this case, the effect AGN feedback has on large scales is diminished significantly. If we decrease the resolution at fixed box size, by reducing the particle number to $2\times 256^3$ (in yellow), we find an almost identical result as for the higher-resolution simulations with the same minimum seed mass on scales $k<10\kunit$, but on smaller scales the cross-over scale shifts by a factor of a few (see main text). A simulation with the same resolution but with a $64\times$ larger volume (in red) yields almost the same result, implying that the relative effect of galaxy formation on the power spectrum is insensitive to cosmic variance.}
\label{fig:diff_resalt}
\end{center}
\end{figure}

Figure~\ref{fig:diff_resalt} shows the relative resolution- and box size-dependence of a simulation including AGN feedback with a relatively high heating temperature of $10^{8.5}\Tunit$. In blue the effect on the power spectrum relative to dark matter only is shown for a simulation with $2\times 512^3$ particles in a $(100\runit)^3$ volume, making it a relatively high-resolution simulation for the set presented here. The AGN suppress the power spectrum by $1\%$ at $k\approx 0.1\kunit$, with the magnitude of the effect increasing to $k\approx 20\kunit$ before rapidly dropping when nearing the cross-over scale of the AGN and dark matter only power spectra around $k\approx 85\kunit$. This cross-over scale is most sensitive to changes in resolution, firstly because of changes in behaviour of subgrid recipes with resolution, which affect all scales, and secondly because it is small enough for the diminished particle clustering due to limited spatial resolution to become noticeable.\footnote{Note that the cross over itself is not a sign of having run into the unresolved regime: hot gas outflows suppress the power on large scales, relative to dark matter only, but baryons accreting onto galaxies increase the power on small scales. A physical cross-over scale of $\sim 0.1\runit$ is therefore expected to occur even at infinite resolution.} The latter primarily affects the (single-fluid) dark matter only simulations. Though not shown here, we find that for our medium-resolution dark matter only simulations (e.g.\ L100N256 or L400N1024) the matter clustering is significantly under-predicted for $k\gtrsim 10\kunit$.

Before directly comparing to a simulation with a different resolution, we first show the effect of changing the AGN parameter most directly related to it: the minimum halo mass for black hole seeding, equal to $100$ dark matter particles in the fiducial model. By changing the minimum mass to $800$, we seed only those haloes that would be sufficiently resolved in a simulation with an $8\times$ lower mass resolution. Changing only this single parameter has an immediate effect on the power spectrum, shown in green in Figure~\ref{fig:diff_resalt}. The large-scale power decrement seen for the simulation in blue, for which only the minimum halo mass for BH seeding is different, goes down significantly on large scales, and the cross-over scale moves to slightly larger scales as well.

When the resolution is now reduced to $2\times 256^3$ particles while keeping all other variables fixed (from green to yellow), the relative effect of galaxy formation on scales $k\lesssim 3\kunit$ is virtually unchanged. On smaller scales, particularly for $k\gtrsim 10\kunit$ the sensitivity to resolution is large, and the cross-over scale increases by a factor of a few. This should be kept in mind when trying to draw conclusions from our results at any scale $k>10\kunit$.

Finally, increasing the simulated volume by a factor 64 at fixed resolution, shown in red, barely affects the results on any scale, indicating that the effect of cosmic variance on the \emph{relative} power spectrum is quite low.

At this point, one may well wonder which simulations make more accurate predictions for the effect of galaxy formation on the matter power spectrum. The naive answer is the simulation with the highest resolution (given a volume at least $\sim(100\runit)^3$ in size), and on scales $k\gtrsim 10\kunit$ this will likely be correct for resolutions similar to the ones considered here -- as long as the subgrid parameters match the resolution. However, on larger scales, which are the main interest for weak lensing and other cosmological probes, the answer is not as straightforward. Given the uncertainties that currently exist in galaxy formation, both the medium- and high-resolution simulations give valid answers for $k<10\kunit$. What matters far more than the numerical resolution, is how these simulations compare with relevant observables, such as the stellar mass function and the hot gas fractions in groups and clusters. On well-resolved scales, a simulation calibrated to observables, like the \bahamas{} simulations are, is therefore expected to produce more accurate predictions than a higher-resolution model that does not reproduce these observables. Still, degeneracies may exist, and so it remains useful to cover a wide array of possible predictions when aiming to mitigate the effects of galaxy formation on weak lensing observables \citep[as argued by e.g.][]{MohammedGnedin2018}.

\begin{figure}
\begin{center}
\includegraphics[width=1.0\columnwidth, trim=16mm 8mm 10mm -4mm]{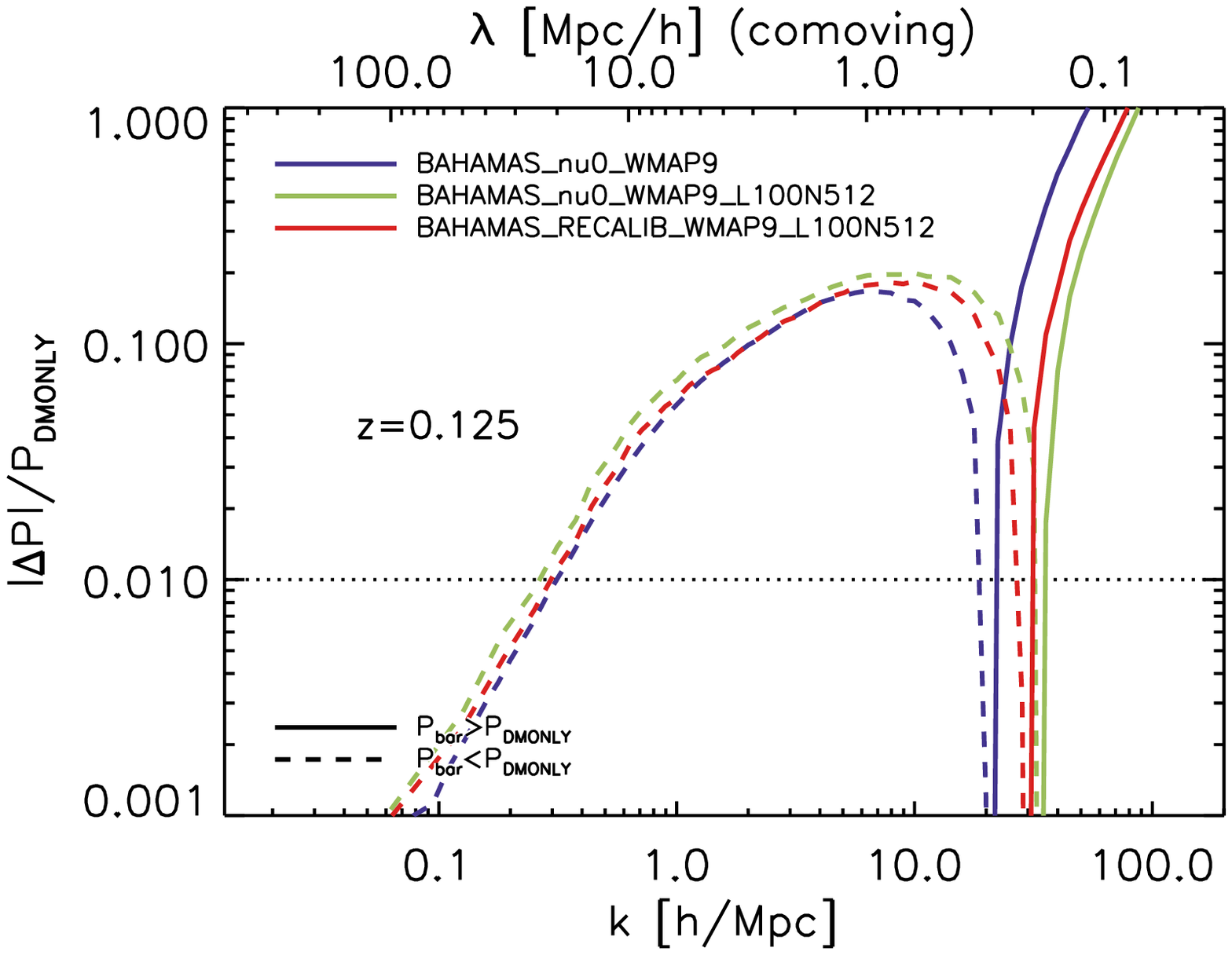}
\caption{Convergence of the effect of galaxy formation on the matter power spectrum with respect to resolution in \bahamas{}. Here we compare a fiducial \bahamas{} simulation ($400\runit$ box with $2\times 1024^3$ particles) to two higher-resolution runs ($100\runit$ box with $2\times 512^3$ particles), one using the same physical subgrid parameters while the other uses parameters that were recalibrated to observations. Note that we compare at $z=0.125$. Without recalibration, a higher-resolution simulation predicts increased suppression of power on all scales, and a significantly smaller cross-over scale. However, when the simulation is recalibrated to observations (i.e.\ its feedback parameters are adjusted to restore agreement with the galaxy stellar mass function and the gas fractions of groups and clusters), the predicted relative power spectrum is in excellent agreement for $k\lesssim 6\kunit$.}
\label{fig:diffback_convergence}
\end{center}
\end{figure}

To show that calibration is indeed more important than resolution, we compare independently calibrated \bahamas{} simulations with different box sizes and resolutions in Figure~\ref{fig:diffback_convergence}. We show the effect of galaxy formation on the matter power spectrum for three simulations: a standard $400\runit$ WMAP9 \bahamas{} simulation with $2\times 1024^3$ particles in blue, a higher-resolution $100\runit$ run with the same physical parameters (see \S\ref{subsubsec:AGN}) in green, and another high-resolution $100\runit$ run in which stellar and AGN feedback were recalibrated to the same observables as the standard \bahamas{} runs, in red. For more information on the latter two runs we refer to Appendix C of \citet{McCarthy2017}. Note that since the recalibrated simulation did not run down to $z=0$, we here compare at $z=0.125$ instead. 

When the standard \bahamas{} run (blue) is repeated in a smaller volume at a higher resolution (green), the suppression of power due to feedback increases on all scales. Additionally, the cross-over point of the hydrodynamic and dark matter only power spectra shifts towards smaller scales. However, when feedback in the higher-resolution run is recalibrated to the same observables (red), these differences diminish, and almost identical results are found for $k\lesssim 10\kunit$, with differences $<1\%$.

\section{2-fluid DMONLY resimulations and large-scale offsets in relative power}
\label{app:2fluid}
\begin{figure}
\begin{center}
\includegraphics[width=1.0\columnwidth, trim=16mm 8mm 10mm -4mm]{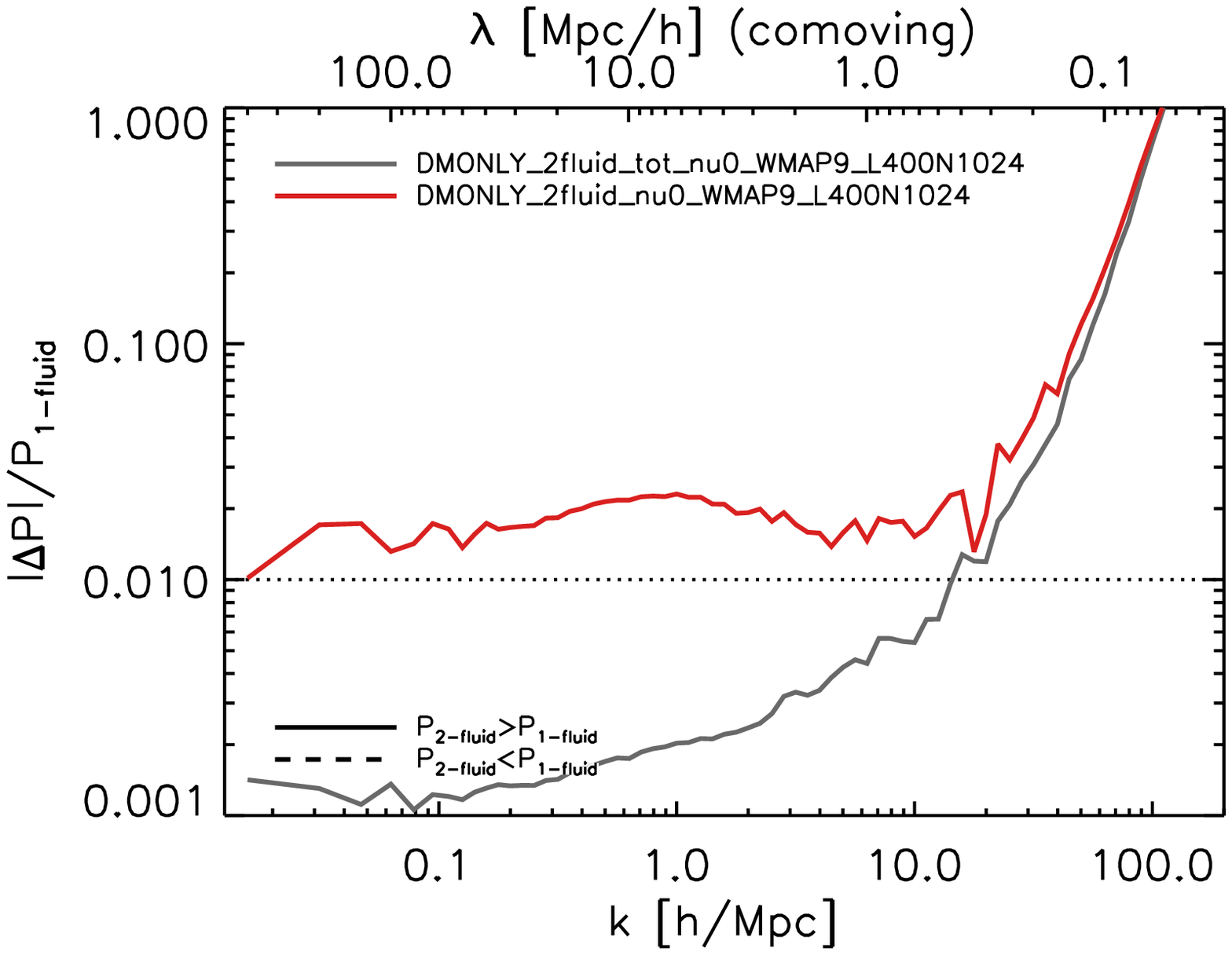}
\caption{Relative difference in the matter power spectra of dark matter only simulations using two CDM particle fluids (``2-fluid'') and a fiducial run using one particle fluid (``1-fluid''). In one of the 2-fluid runs, shown in red, we initialize each fluid with a separate transfer function (assuming CDM or baryons), as is done in \bahamas{} simulations containing baryons. Doing so increases the power by $>1\%$ on all scales, and explains why such large-scale offsets in the relative matter power spectrum can be seen when dark matter only simulations use a single fluid initialized with a total matter transfer function. The other 2-fluid run, shown in gray, still uses the same transfer function for both fluids, and therefore shows only the effect of adding a second particle species. Here, the increase in clustering is sub-percent for $k<10\kunit$. This explains the $\sim 0.1\%$ large-scale offsets seen for some (cosmo\discretionary{-)}{}{)}OWLS simulations.}
\label{fig:2fluid}
\end{center}
\end{figure}

As mentioned in \S\ref{subsec:dmonly}, most hydrodynamical and dark matter only simulations run with the same code will not show identical clustering on large scales at redshift zero, even when seeded with identical phases. This is due to differences in how these simulations are typically performed -- specifically, the transfer functions and number of particles often differ. Since the combined effects can lead to significant large-scale offsets, we have re-run some of our dark matter only simulations in an attempt to mitigate this. For \bahamas{}, we performed 2-fluid DMONLY runs that have two CDM particle types, each represented by $1024^3$ particles (instead of the fiducial $1024^3$ particles of a single species). Each set of particles has a different mass and is initialized with a different matter transfer function in such a way as to exactly mimic the initial conditions of the hydrodynamical simulations. In Figure~\ref{fig:2fluid} we show the change in the $z=0$ matter power spectrum for such a simulation relative to a fiducial dark matter only run that is otherwise identical, in red. Due mainly to initializing the particles with two different transfer functions instead of the fiducial combined transfer function, the clustering changes by $>1\%$ on all relevant scales, $k\lesssim 10\kunit$. This effect was previously explored by \citet{Valkenburg2017}. As this offset is significant compared to the effects we are interested in, we therefore compare hydrodynamical \bahamas{} simulations to 2-fluid DMONLY runs initialized in this way throughout this paper. On smaller scales, the difference is much larger, due to spurious clustering between the two different particle types in the 2-fluid run. This would likely be mitigated by including a phase shift in one of the particle species when splitting the original $1024^3$ particle distribution in two, as described in \citet{Angulo2013} -- however, as this has not been done for the baryons either, we choose not to do so here.

To better consider the effect of adding another particle species, separate from changing the transfer functions, we also ran a 2-fluid DMONLY simulation that uses the total transfer function for both species, in that way mimicking the initial conditions of OWLS and cosmo-OWLS. The clustering results relative to the fiducial 1-fluid dark matter only simulation are shown in Figure~\ref{fig:2fluid} as well, in grey. Here, too, the 2-fluid run shows increased clustering on all scales, though the effect is sub-percent on the scales most relevant for our study, $k<10\kunit$. We have therefore not performed additional runs of this kind to replace the fiducial OWLS and cosmo-OWLS DMONLY simulations, and $\sim 0.1\%$ large-scale offsets can therefore still be seen for some relative power spectra. If compensated for, the relative power would be suppressed by roughly $0.2\%$ on large scales for WMAP7 cosmo-OWLS simulations (see e.g.\ the left-hand panels of Figures~\ref{fig:diff_Theat} and \ref{fig:back_Theat}).

We note that simulations from the literature show similar offsets between the very large-scale power predicted by hydrodynamical and dark matter only runs. In particular, the power spectra from Horizon-AGN \citep{Chisari2018} show a large-scale offset of $0.5-0.7\%$ (see Figure~\ref{fig:diffback_literature}), which is likely due to the effects described above. Given that Figure~\ref{fig:2fluid} shows that this offset is expected to be roughly constant on large scales, we expect that compensating for this offset would increase the agreement between different models on large scales.

Finally, we have checked that when the CDM power of a hydrodynamical \bahamas{} simulation is compared to that of only one component of the 2-fluid DMONLY simulation (i.e. comparing the power in components with the same number of particles and the same mass), rather than to the total power of dark matter (scaled by the mass ratio as described in \S\ref{subsec:backreaction}), the back-reactions are virtually identical.

\bsp
\label{lastpage}
\end{document}